\documentclass[useAMS,usenatbib]{mn2e}

\usepackage{graphicx}
\usepackage{natbib}
\usepackage{longtable,lscape}
\usepackage{epstopdf}
\newcommand{\kms}{km~s$^{-1}$}

\newcommand{\msun}{$\rm M_{\odot}$}

\newcommand{\be}{\begin{equation}}
\newcommand{\ee}{\end{equation}}
\newcommand{\bd}{\begin{displaymath}}
\newcommand{\ed}{\end{displaymath}}
\newcommand{\bi}{\begin{itemize}}
\newcommand{\ei}{\end{itemize}}
\newcommand{\bfig}{\begin{figure}}
\newcommand{\efig}{\end{figure}}
\newcommand{\bc}{\begin{center}}
\newcommand{\ec}{\end{center}}
\newcommand{\hii}{{H\scriptsize{II}}}
\newcommand{\ai}{$\alpha_{\textrm{\tiny{IRAC}}}$}

\newcommand{\vlsr}{V$_{\mathrm{LSR}}$}
\newcommand{\coa}{$^{12}\mathrm{CO}$}
\newcommand{\cob}{$^{13}\mathrm{CO}$}
\newcommand{\lsun}{L$_{\odot}$}
\newcommand{\lfir}{L$_{\textrm{\tiny{FIR}}}$}
\newcommand{\ghm}{$\Gamma_{\textrm{\tiny{HM}}}$}
\newcommand{\Q}{$\mathcal{Q}$}

\title[Study of embedded clusters]{A multiwavelength study of embedded clusters in \\W5-east, NGC7538, S235, S252 and S254-S258}
\author[L. Chavarr\'{\i}a et al.]{L. Chavarr\'{\i}a$^{1,2,3,4}$\thanks{E-mail: luisagustinchavarria@gmail.com}, L. Allen$^{5}$, C. Brunt$^{6}$, J. L. Hora$^{2}$, A. Muench$^{2}$ and G. Fazio$^{2}$\\
$^{1}$Universidad de Chile, Camino del Observatorio 1515, Santiago, Chile\\
$^{2}$Harvard-Smithsonian Center for Astrophysics, 60 Garden Street, Cambridge, MA02138, USA\\
$^{3}$Laboratoire d'Astrophysique de Bordeaux, 2 rue de l'Observatoire, 33271 Floirac Cedex, France\\
$^{4}$Centro de Astrobiolog\'{\i}a (CSIC/INTA), Ctra. de Torrej\'on a Ajalvir, km 4 28850, Torrej\'on de Ardoz, Madrid, Spain\\
$^{5}$National Optical Astronomical Observatory, 950 North Cherry Avenue, Tucson, AZ 85719, USA\\
$^{6}$The School of Physics, University of Exeter, The Queens Drive, Exeter, Devon, UK EX4 4QL
}

\begin{document}

\date{Received June 2013}
\pagerange{\pageref{firstpage}--\pageref{lastpage}} \pubyear{2002}
\maketitle
\label{firstpage}

\begin{abstract}
We present Spitzer, NIR and millimeter observations of the massive star forming regions W5-east, S235, S252, S254-S258 and NGC7538. Spitzer data is combined with near-IR observations to identify and classify the young population while \coa~and \cob~observations are used to examine the parental molecular cloud. We detect in total 3021 young stellar objects (YSOs). Of those, 539 are classified as Class~I, and 1186 as Class~II sources. YSOs are distributed in groups surrounded by a more scattered population. Class I sources are more hierarchically organized than Class II and associated with the most dense molecular material. We identify in total 41 embedded clusters containing between 52 and 73\% of the YSOs. Clusters are in general non-virialized, turbulent and have star formation efficiencies between 5 and 50\%. We compare the physical properties of embedded clusters harboring massive stars (MEC) and low-mass embedded clusters (LEC) and find that both groups follow similar correlations where the MEC are an extrapolation of the LEC. The mean separation between MEC members is smaller compared to the cluster Jeans length than for LEC members. These results are in agreement with a scenario where stars are formed in hierarchically distributed dusty filaments where fragmentation is mainly driven by turbulence for the more massive clusters. We find several young OB-type stars having IR-excess emission which may be due to the presence of an accretion disk.

%The spatial distribution of Class I and Class II sources is studied by means of the structural parameter \Q~and Kolmogorov-Smirnov (K-S) tests.
%We investigate the high-mass end of the initial mass function (IMF) for the most numerous clusters in each region and find IMF slopes with values between $-1.0$ and $-2.0$.
%We find a linear correlation between the dense molecular mass and the number of cluster members which is in agreement with previous findings. 
\end{abstract}

%%%%%%%%%%%%%%%%%%%%%%%%%%%%%%%%%%%%%%%%%%%%%%%%

\section{Introduction}
Embedded clusters are truly stellar nurseries, more than 90\% of the stars in our Galaxy are formed in such associations \citep{zin07}. Since they are young (with ages of less than $2-3$~Myr), they still contain the imprints of the parental molecular cloud. Moreover, the wide range of number stars \citep[10 to $10^4$,][]{lad03} and high density of members \citep[more than 20~stars$/$pc$^{-2}$,][]{lad03} makes embedded clusters perfect laboratories to study cluster dynamics, stellar evolution and star formation theories.

Among embedded clusters, those harboring massive stars (hereafter massive embedded clusters) are particularly important since both the formation of massive stars and the impact of massive stars feedback on the other cluster members and the parental molecular cloud are still not well understood. Massive stars begin hydrogen burning while they are still accreting material and the strong stellar winds and ultraviolet (UV) photons emitted will eventually stop the accretion before the star reaches its final mass. In addition, the emitted UV photons ionize the surrounding cloud and create an expanding \hii~region that disrupts and compress the natal molecular cloud. The feedback effects of massive stars over, for example, their disk life-times (in case they have disk) and/or over other cluster members are unclear. It is also unclear under what conditions the \hii~regions and shock waves will either destroy the molecular cloud or trigger star formation \citep[there are several examples showing molecular gas that has been swept up by expanding \hii~regions and that contains young stars, eg.][]{deh08,cha08a,wan11}. This feedback into the interstellar medium is absent in the case of low-mass stars and it may play an important role in the star formation rate and evolution of the Galaxy. 

Of the three models proposed to explain the formation of massive stars \citep*[competitive accretion in a protocluster environment, monolithic collapse in turbulent cores and stellar collisions and mergers in very dense systems,][]{bon04,mck03,bon08}, competitive accretion and turbulent cores are somehow a scaled-up version of low-mass star formation. Competitive accretion requires that massive stars form at the center of the cluster (known as primordial mass segregation). This has been observed in some young clusters. However, it can be also achieved by the dynamical interaction between the cluster members and the gas they are embedded in \citep[e.g.][]{cha10}. The turbulent core model proposes that density enhancements created by turbulent motions allows the high-accretion rates necessary for high-mass stars to form. This requires massive cores highly turbulent which have been lately reported \citep[e.g.][]{her12}. In addition, rotating toroid-like structures and outflows (which are star formation indicators via gravitational collapse) have been observed for sources with masses up to 25 \msun~\citep[eg.][]{ces05,gar07}. However, there is still no evidence of disks in O-type stars. This suggests that coalescence may be an alternative theory of formation for stars with masses of more than 30 \msun~\citep{zin07}. However, the high star densities necessary for coalescence to occur have not been yet reported.

Since embedded clusters are located deep inside their natal molecular cloud, they can be observed  only at infrared (IR), and millimeter wavelengths. In the last years, several authors have studied embedded clusters using a combination of Spitzer-IRAC \citep{faz04,all04} and near-IR (NIR) data, which is proven to be a powerful tool to identify and classify YSOs in regions of star formation \citep*[eg. Ophiuchus, Serpens, Perseus, Taurus and NGC1333,][]{gut08,win07,sch08}, most of them in the low-mass range. \citet{sch08} studied the spatial distribution of different class YSOs in embedded clusters and found that they mostly evolve from a hierarchical to a more centrally concentrated distribution. \citet{gut09} analyzed 36 low-mass embedded clusters and found that YSOs are likely formed by Jeans fragmentation of parsec-scale clumps, in agreement with the accretion scenario. Massive embedded clusters, on the other hand, have been more elusive to scrutinize mainly due to two reasons; a) they are less abundant than low-mass clusters  and hence usually located at several kpc from the Sun and, b) the early stages of massive star formation last only a few million years. Because of this, only a few massive embedded clusters have been evenly studied until date using a combination of near-IR and Spitzer data \citep[eg.][]{koe08,cha08a,kir08,dew11,ojh11}.  Those studies have been carried out by several authors using different data sets and analysis. As a consequence, their results are difficult to compare between each other and the available statistics is still poor.

We present an homogeneous Spitzer-IRAC, NIR and molecular data study on the young stellar population in five high-mass star forming regions: W5-east, S235, S252, S254-S258 and NGC7538. Our study aims to address the following questions: What are the physical properties of YSOs in massive embedded clusters? Are those properties similar to the low-mass case? What are the implications of those properties on the massive star formation scenario? Also, we provide a set of  physical quantities with a reasonable statistical weight that will help to constrain theoretical models of star formation and cluster dynamic.

Region S254-S258 was presented by \citet{cha08a}. In this work, we use their results for a more recent distance estimate derived from trigonometric parallax of methanol masers \citep[1.6 kpc,][]{ryg10}. 

The position and distance to the studied regions are summarized in Table~\ref{table_sources}. Following there is a brief description of the regions \citep[see][for a description of region S254-S258]{cha08a}. In \S~\ref{section_observations3} we explain our observations and the data reduction process. Results, including the identification of YSOs, analysis of their spatial distribution and the study of the molecular cloud structure are presented in \S~\ref{section_results3}. In \S~\ref{section_discussion3} discuss and compare the physical properties of YSOs for low-mass and massive embedded clusters. Our conclusions are presented in \S~\ref{section_conclusions3}. The estimation of background contamination, a comparison with previous observations, non-detection estimate and Gaussian decomposition of molecular spectrum are explained in Appendices~\ref{background} through~\ref{section_gaussian}.

\begin{table}
\caption{List of observed regions}
\label{table_sources}
\centering                         
\begin{tabular}{lccc}
\hline
Name & RA (J2000) & Dec (J2000) & D$_{\odot}$ \\
& hh mm ss & dd mm ss & [kpc] \\
\hline
W5-east         & 03  01  31.20   & 60  29  13.0    &  2.0 \\
S235             & 05  40  52.00   & 35  42  20.0    &  1.8 \\
S252             & 06  09  04.70   & 20  35  09.0    &  2.1 \\
S254-S258   & 06  12  46.00   & 18  00  38.0    &  1.6 \\
NGC7538      & 23  13  42.00   & 61  30  10.0    &  2.7 \\
\hline
\end{tabular}
\end{table}
%%%%%%%%%%%%%%%%%%%%%%%%%%%%%%%%%%%%%
%%%%%%%%%%%%%%%%%%%%%%%%%%%%%%%%%%%%%
  
%\begin{table*}
%\vbox to220mm{\vfil lanscape.tex
%    \caption{}
%\vfil}
%\label{landfig}
%\end{table*}
  
\section{Description of the studied regions}\label{section_description3}
The luminosities given in this section are normalized by each region assumed distance (see Table~\ref{table_sources}).
\subsection{W5-east}
%%%%%%%%%%%%%%%%%%%%%%%%%%%%%%%%%%%%%
%%%%%%%%%%%%%%%%%%%%%%%%%%%%%%%%%%%%%
\begin{figure*}
\begin{center}
\includegraphics{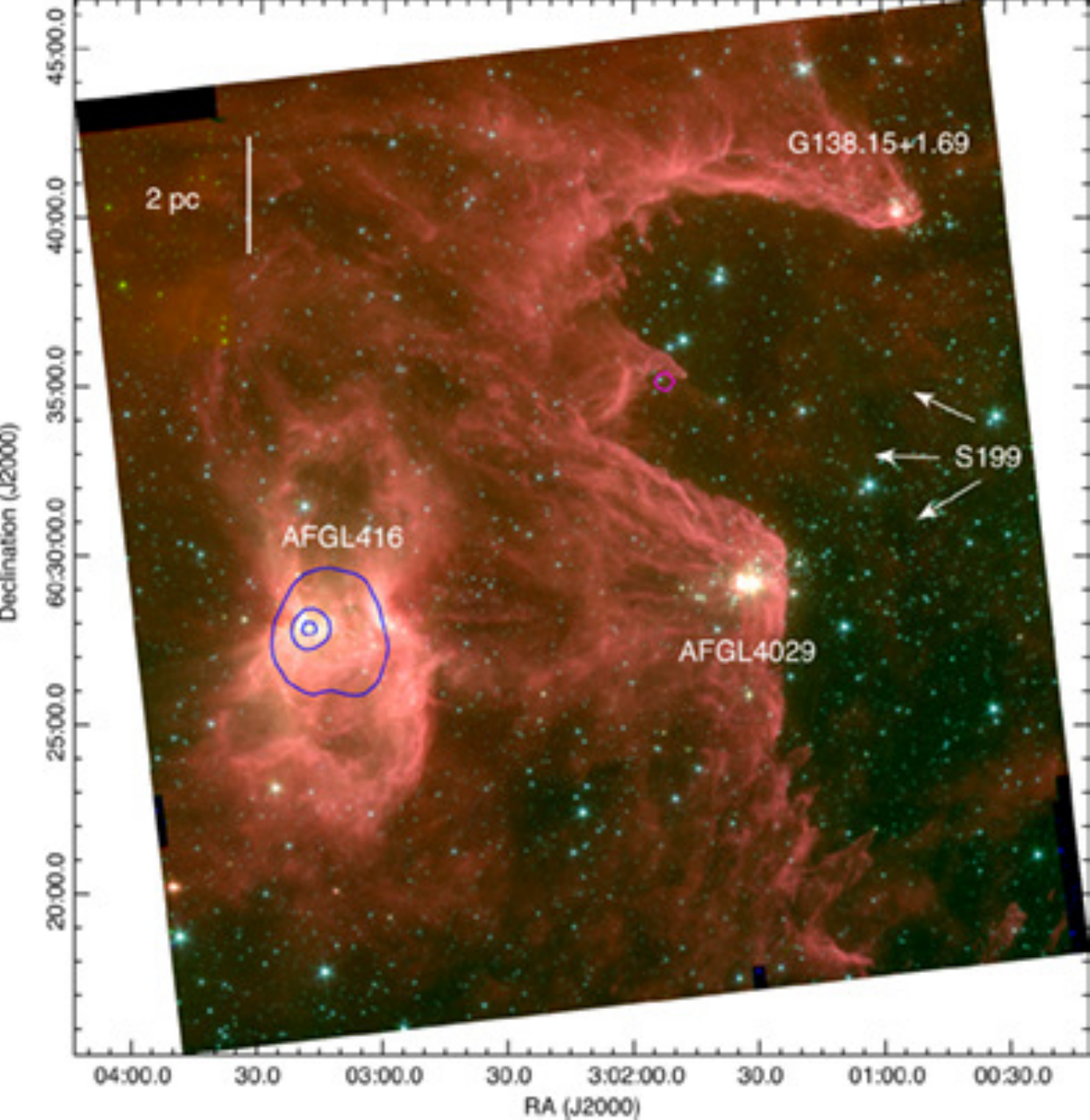}
\caption{IRAC 3-color image of region W5-east (blue: 3.6~$\mu$m, green: 4.5~$\mu$m, red: 8.0~$\mu$m). Blue and magenta contours correspond to the 1.4~GHz (21~cm) emission from the NRAO VLA Sky Survey \citep{con98}. Contours are at 10, 50 and 90\% of peak emission. Blue contours are associated with W5-east while magenta contours are likely background sources. The white arrows show the ionizing front advancing direction. The ionizing star HD18326 is out of the field of view (FOV).}\label{afgl4029_124}
\end{center}
\end{figure*}
%%%%%%%%%%%%%%%%%%%%%%%%%%%%%%%%%%%%%
%%%%%%%%%%%%%%%%%%%%%%%%%%%%%%%%%%%%%
W5-east is located on the west side of \hii~region Sharpless 199 \citep[S199, also called IC 1848,][]{sha59,koe08}. The \hii~region is powered by an O7V type star (HD18326) and it harbors three young star clusters: AFGL4029 \citep{pri76,car93,deh97}, AFGL416 or Sh 2-201 \citep{car93,ojh04} and G138.15+1.69 \citep*{bic03a}. The two former contain young massive stars. 

AFGL4029 has a far-IR luminosity (\lfir) of 1.7$\times 10^4$~\lsun~\citep{sne88}. It is associated with the IRAS source 02575+6017 and sub-millimeter emission \citep{mor08}. The cluster has an estimated number of members of 240 \citep*{car00} and around 80 sources with H$\alpha$ emission \citep*{ogu02,nak08}. It also contains UC\hii~regions \citep*{kur94,zap01} associated with the infrared sources IRS1 and IRS2 \citep{bei79}. IRS1 is believed to be the exciting source of a molecular outflow \citep{sne88} detected also at optical wavelengths \citep{ray90}. G138.15+1.69 is located 12 arc-minutes north-west of AFGL4029 \citep{bic03a}. It is associated with the IRAS source 02570+6028 (\lfir~$=380$~\lsun) and contains at least 80 members \citep{car00} and about 30 stars with H$\alpha$ emission \citep{ogu02,nak08}. Both G138.15+1.69 and AFGL4029 are immerse in pillar-like structures located along a dusty rim and suggesting a triggered star formation scenario \citep{koe08,niw09,deh12}. This is also supported by the younger ages associated to YSOs inside the rim compared to the ages of YSOs located outside the rim in direction to the ionizing star \citep{cha11}. AFGL416 is associated with the \hii~region Sharpless 201 (S201), the IRAS source 02593+6016 and sub-millimeter emission \citep{mor08}. The cluster luminosity (\lfir) is $4.2\times 10^4$~\lsun~and it harbors at least 90 members \citep{car00}. 

The dust content in the region was investigated by \citet{deh12} using Herschel observations in the far-IR. They find that the dust is being collected by the expansion of the ionizing front. In addition to this, they identify around 40 point sources at 100 microns over the same FOV as in our observations. Those sources are associated with the various clusters in the region as well as with the ionizing front edge and are presumably very young stellar objects.

The distance to cluster AFGL4029 is estimated between 1.9 and 3.8~kpc \citep*{hil06,chu90}. Since AFGL4029 and AFGL416 are both associated with the \hii~region S199, and G138.15+1.69 has a similar \vlsr~as AFGL4029 and AFGL416 \citep{mam87}, we adopt a distance of 2~kpc \citep{bec71,mam87} to the whole region W5-east.

\subsection{Sharpless 235}
%%%%%%%%%%%%%%%%%%%%%%%%%%%%%%%%%%%%%
%%%%%%%%%%%%%%%%%%%%%%%%%%%%%%%%%%%%%
\begin{figure*}
\begin{center}
\includegraphics[width=10cm]{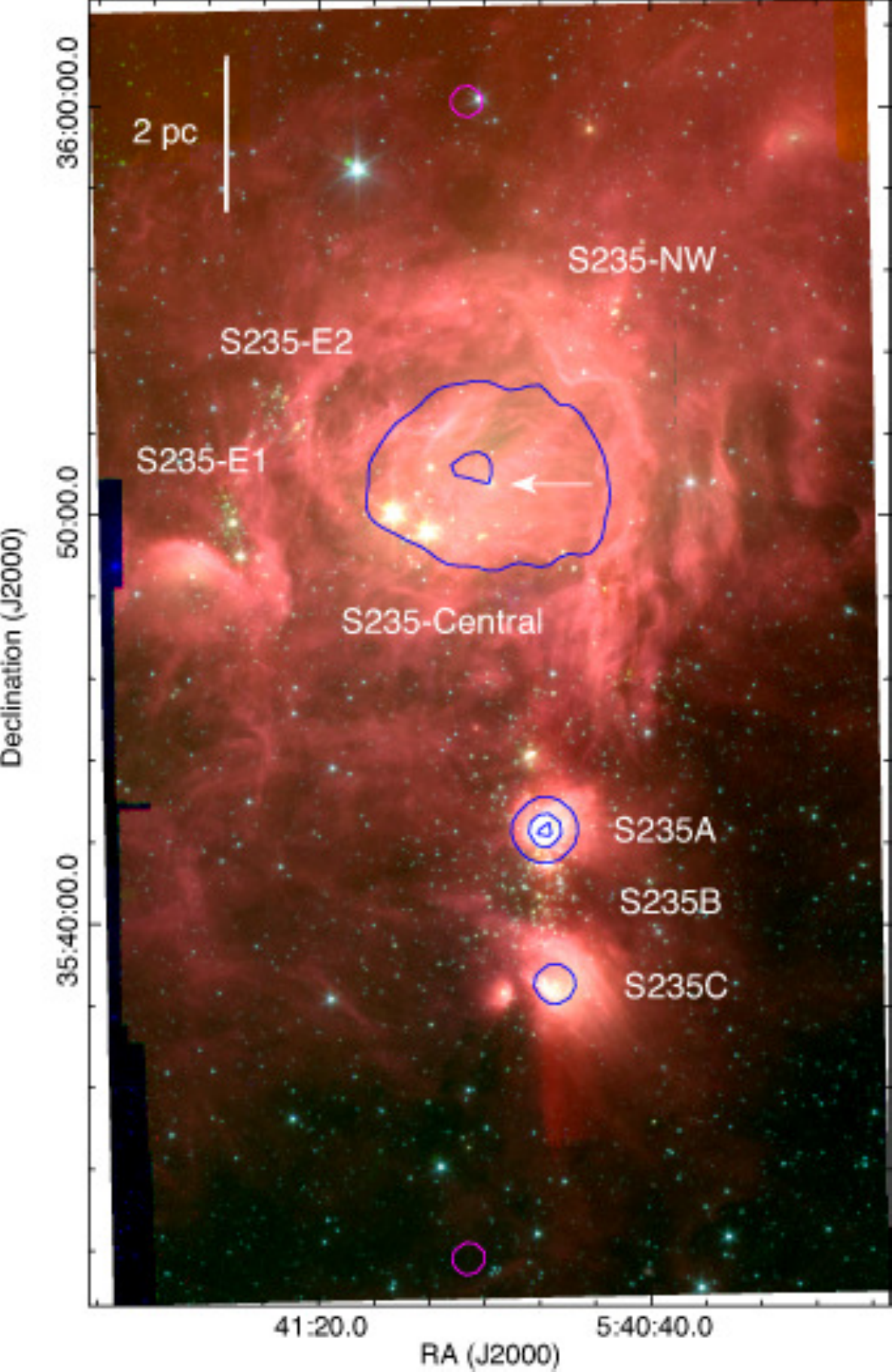}
\caption{IRAC 3-color image of region S235 (blue: 3.6~$\mu$m, green: 4.5~$\mu$m, red: 8.0~$\mu$m). The contours are the same as for Figure~\ref{afgl4029_124}. The white arrow indicates the ionizing star BD+351201.}
\label{s235_124}
\end{center}
\end{figure*}
%%%%%%%%%%%%%%%%%%%%%%%%%%%%%%%%%%%%%
%%%%%%%%%%%%%%%%%%%%%%%%%%%%%%%%%%%%%
Sharpless 235 (S235) is an \hii~region from the \citet{sha59} catalog first reported as an emission nebula by \citet{min46}. The \hii~region is powered by an O9.5V type star \citep*[BD+351201,][]{geo73} and also two bright infrared sources (IRS1 and IRS2) with estimated spectral types of early B stars \citep*{eva81,tho83}. Those are part of the young cluster S235 \citep{car93,bic03a}. 

There are at least 270 YSOs in this region \citep{dew11}. Of those, around 70\% belong to different clusters: S235, S235 East1, S235 East2, S235 NW and S235AB \citep[Camargo, Bonatto \& Bica 2011;][]{bic03a, kir08, dew11}. Cluster S235AB harbors three compact \hii~regions: S235A, S235B and S235C \citep{isr78}. \hii~regions S235A and S235B are powered by zero age main sequence stars (ZAMS) with spectral type between B0 and O9.5 \citep*{kra79,olo83}. They are associated with the infrared sources IRS3 and IRS4 \citep{eva81} as well as water and methanol masers \citep{fel06}.  
%A detailed study of S235A, S235B and S235C and their interaction with the environment is presented by \citet{fel04} and \citet{fel06}. 

Estimate distances to region S235 are between 1.6 and 2.5~kpc \citep{isr78,hun90}. In this paper, we adopt a distance of 1.8~kpc \citep{eva81}.

\subsection{Sharpless 252}
%%%%%%%%%%%%%%%%%%%%%%%%%%%%%%%%%%%%%
%%%%%%%%%%%%%%%%%%%%%%%%%%%%%%%%%%%%%
\begin{figure*}
\begin{center}
\includegraphics[width=16cm]{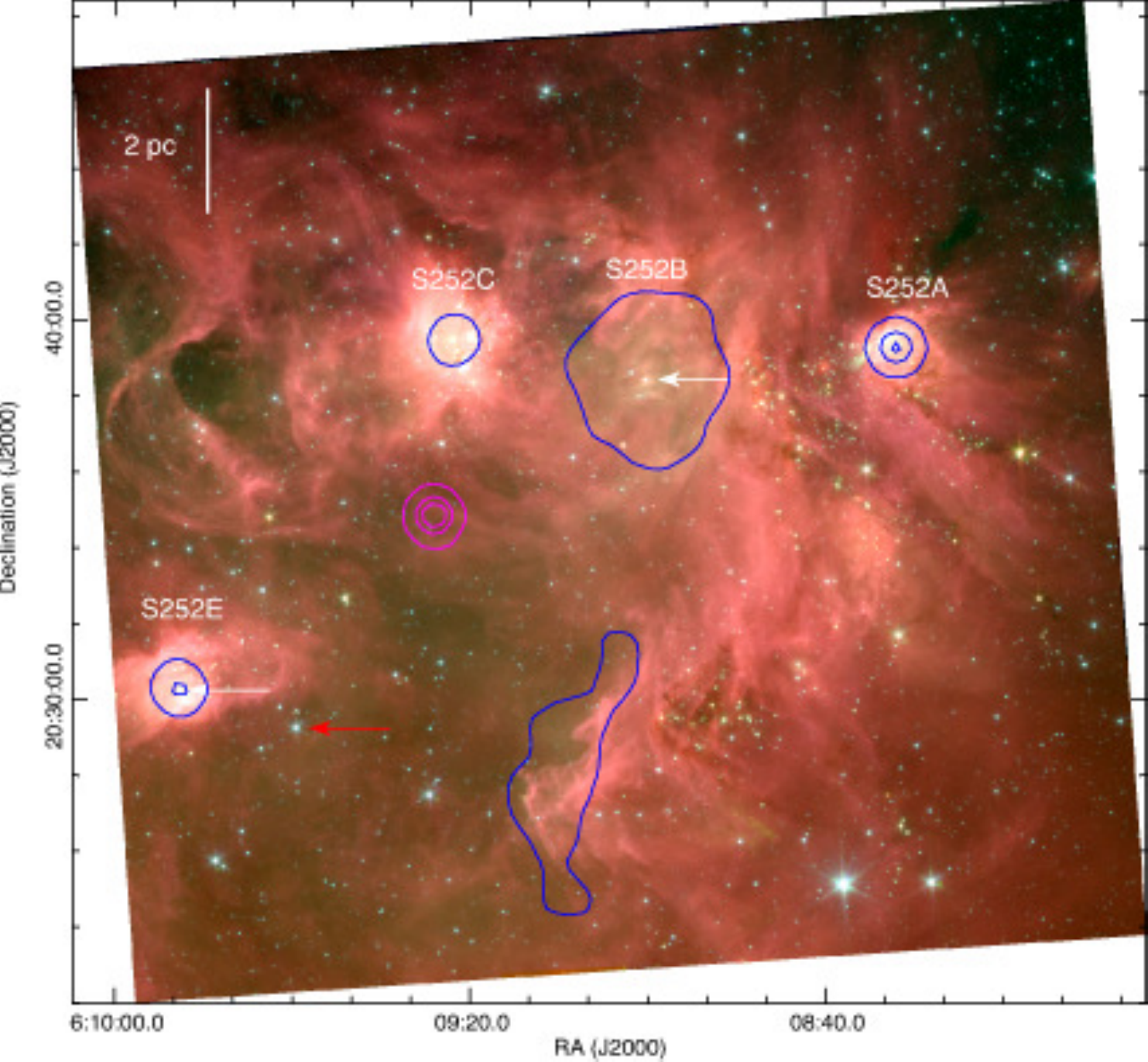}
\caption{IRAC 3-color image of region S252 (blue: 3.6~$\mu$m, green: 4.5~$\mu$m, red: 8.0~$\mu$m). The contours are the same as for Figure~\ref{afgl4029_124}. The ionizing star HD42088 is indicated by a red arrow. Other known ionizing stars are indicated by white arrows.}
\label{s252_124}
\end{center}
\end{figure*}
%%%%%%%%%%%%%%%%%%%%%%%%%%%%%%%%%%%%%
%%%%%%%%%%%%%%%%%%%%%%%%%%%%%%%%%%%%%

Sharpless 252 (S252, also called NGC2175) is an extended \hii~region (size of $\sim1$ degree) from the \citet{sha59} catalog. The \hii~region is powered by an O6.5V type star \citep[HD42088,][]{gra75}, and it contains four compact \hii~regions associated with a young population of stars \citep*[S252A, S252B, S252C and S252E from][]{fel77}.

The \hii~region S252A (also named AFGL5179) is powered by a B1V-O9.5V star \citep{hil56,jos12} and it contains an infrared cluster also named S252A with around 80 members \citep{bic03a,tej06}. S252A is also associated with the source IRAS 06055+2039 \citep[][\lfir$= 4.3\times 10^3$~\lsun]{car95}, a water maser \citep{lad79} and outflow activity \citep{xu06}. The \hii~region S252B is powered by a B1.0V type star \citep{jos12} which seems to be somehow isolated. The \hii~region S252C is powered by a B0.5 type star \citep{jos12} and is associated with the infrared cluster S252C \citep{cha89,bic03a}. The \hii~region S252E (also NGC2175s or AFGL5184) is powered by a B0V type star \citep{jos12} and is associated with an homonymous infrared cluster and the source IRAS 06068+2030 \citep[\lfir$= 2.9\times 10^3$~\lsun,][]{car95}. 

The distance estimate for S252 are between 1.25 and 2.9~kpc \citep{mir87,wou89}. We adopt a distance of 2.1~kpc, derived from trigonometric parallax of methanol masers \citep{rei09}.

\subsection{NGC7538}
%%%%%%%%%%%%%%%%%%%%%%%%%%%%%%%%%%%%%
%%%%%%%%%%%%%%%%%%%%%%%%%%%%%%%%%%%%%
\begin{figure*}
\begin{center}
\includegraphics[width=16cm]{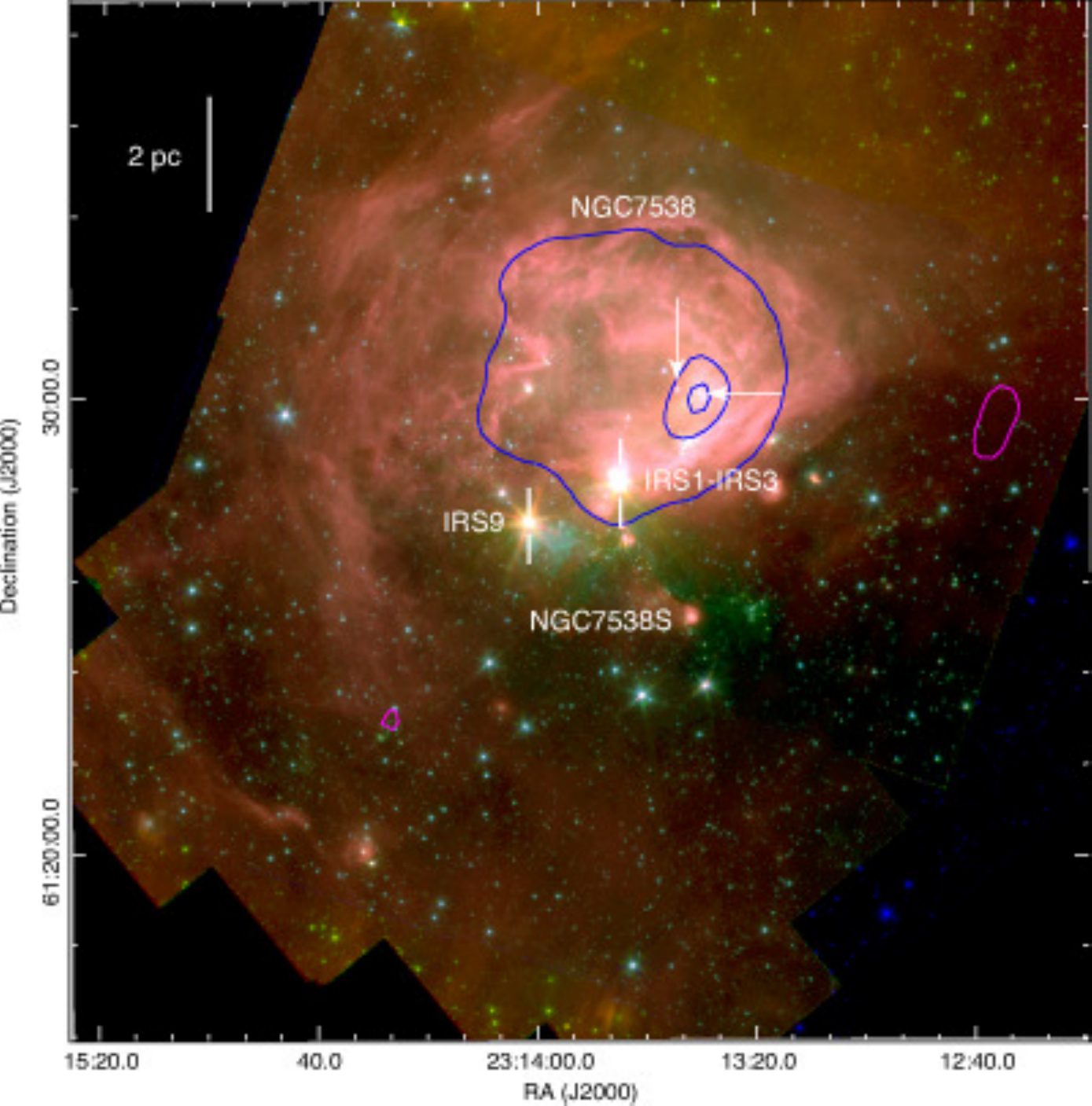}
\caption{IRAC 3-color image of region NGC7538 (blue: 3.6~$\mu$m, green: 4.5~$\mu$m, red: 8.0~$\mu$m). The contours are the same as for Figure~\ref{afgl4029_124}. The white arrows indicate the known ionizing stars. Infrared sources IRS1-IRS3 and IRS9 are indicated by bars.}
\label{ngc7538_124}
\end{center}
\end{figure*}
%%%%%%%%%%%%%%%%%%%%%%%%%%%%%%%%%%%%%
%%%%%%%%%%%%%%%%%%%%%%%%%%%%%%%%%%%%%
NGC7538 is an active site of star formation located in the Perseus spiral arm. It was first documented as an emission nebula (\hii~region S158) by \citet{sha59} and it contains massive stars in different evolutionary stages; main sequence stars which ionize the \hii~region NGC7538, with spectral types between O3 and O9 \citep[IRS5 and IRS6,][]{pug10}; infrared sources IRS1 \citep*[associated with a disk and an outflow;][]{pes09,san09}, IRS2 and IRS3 \citep*{wyn74}, located south of NGC7538 and associated with UC\hii~regions and with the infrared cluster NGC7538S \citep{car93,bic03a,san10}; young stellar objects like IRS9 and IRS11 \citep{wer79} and submillimeter clumps without associated IR emission \citep{rei05}. 

The spatial location of sources at different evolutionary stages suggest a sequence of star formation from north-east to south-west \citep{pug10}. \citet{bal04} detected 238 sources with NIR excess in NGC7538. More recent studies in the region have been done by \citet{rei05}, \citet{kra06} and \citet{bar07}. We suggest these authors and references therein for a more detailed description of this region. 

Distance estimations for NGC7538 are between 2.1~kpc \citep{bal04} and 3.5~kpc \citep{isr77}. The most used distance to the region is 2.8~kpc, corresponding to the photometric distance calculated by \citet*{cra78}. In this paper, we adopt a distance of 2.7~kpc, derived by \citet{mos09} using trigonometric parallax.

\section{Observations and Data Reduction}\label{section_observations3}
\subsection{Mid-IR imaging}
All regions were observed with the Infrared Array Camera (IRAC) on the \emph{Spitzer Space Telescope} between December 2004 and October 2007. We used an integration time of 10.4 seconds per dither in the High Dynamic Range (HDR) mode, with 3 dithers per map position. HDR mode also acquires 0.4 seconds integration time frames for the recovery of bright sources which are saturated in longer exposures. We used S. Carey's artifact correction scripts to remove column pull-down and some of the banding and muxbleed artifacts \citep{hor04}. IRAC mosaics were constructed using the Basic Calibrated Data (BCD) frames (S14.0.0, and S15.3.0) with IRACproc\footnote{https://www.cfa.harvard.edu/twiki/bin/view/Main/IracProc} \citep{sch06}. The final image scale is 0.6 arc-seconds per pixel.

We used IRAF DAOPHOT packages to extract sources and perform photometry in each band. The photometry was done using an aperture radius of 1.8 arc-seconds (3 pixels) for the 3.6 and 4.5~$\mu$m bands and 2.4 arc-seconds (4 pixels) for the 5.8 and 8.0~$\mu$m bands. Inner and outer sky annuli of 4.8 and 6 arc-seconds radius respectively were used in each IRAC band. We calculated the zero point magnitudes in each band using Vega fluxes. Their values are 18.443, 17.879, 17.234 and 16.477 magnitudes for the 3.6, 4.5, 5.8 and 8.0 $\mu$m bands respectively (aperture corrections are included). 

Spitzer-MIPS mosaics for all regions were downloaded from the NASA / IPAC Infrared Science Archive \footnote{http://sha.ipac.caltech.edu}. MIPS mosaics were constructed using the BCD frames (S18.12.0). The program ID for region W5-east is 20300 (PI is Lori Allen), for the other regions the program ID is 40005 (PI is Giovani Fazio).

\subsection{Near-IR imaging}
Observations with Flamingos were performed at the 2.1 meter telescope located at Kitt Peak National Observatory in December 2004 and January 2006. We observed J, H and K bands (centered at 1.24, 1.65 and 2.21 $\mu$m respectively) for W5-east and S252, J and K bands for NGC7538 and K-band for S235. Flamingos has a $2048\times 2048$ pixel Hawaii II HgCdTe detector array with a plate-scale of 0.611 arc-seconds per pixel which gives a field of view (FOV) of 20$\times$20 arc-minutes. The seeing during the observations was approximately 1.0 arc-second. The observations were done in dithering mode with 15-30 arc-seconds shifts for a total integration time of approximately 1000 seconds per band.

The near-IR data reduction was performed using IDL\footnote{Linearization, developed by Robert A. Gutermuth (http://www.astro.umass.edu/\~rguter/ Rob\_Gutermuth\_Astronomy/IDL\_Page.html)}, IRAF\footnote{Darks, Flat-field, Bad-pixel mask and background frame creation and application.} and WCS tools\footnote{Distortion and astrometry correction, developed by Doug Mink (http://tdc-www.harvard.edu/wcstools/).}. Image distortion at the edge of Flamingos FOV were corrected by applying a second order transformation in the X and Y axis using the position of matching sources from the 2MASS catalogue. This has an impact of less than 1 arc-second in the source position at the edges of the detector which produces a variation in the point spread function (PSF) over the mosaic. To avoid systemic errors in the flux estimation of detected sources, we performed aperture photometry with the following parameters: threshold of 3 sigma for detection and 1.8, 4.8 and 6 arc-seconds in radii for aperture, inner annulus and inner plus outer annulus respectively. Calibration was performed by minimizing residuals to corresponding 2MASS detections. No color terms were assumed in the zeropoint minimization. The RMS for residuals between the data sets used was less than 0.08 magnitudes in all bands.\\

We also obtained NIR data with SWIRC at the 6.5 meter MMT telescope located at Fred Lawrence Whipple Observatory in 2005. We acquired data in the J and H bands for S235 and in the H band for NGC7538. SWIRC has a $2048\times2048$ pixel Hawaii II detector array with a plate-scale of 0.15 arc-seconds per pixel which gives a FOV of 5$\times$5 arc-minutes. For NGC7538, the observations were performed in dithering mode with 15-20 arc-seconds shifts for a total integration time of 100 seconds in the H band. For S235 we observed in dithering mode for a total integration time of 30 seconds in H and 60 seconds in J. We performed aperture photometry of 0.9 arc-seconds in radii for both bands, using a detection threshold of 3 sigma, and 1.0 and 2.0 arc-seconds radii for inner annulus and inner plus outer annulus respectively. As for Flamingos, we calibrated SWIRC photometry by minimizing residuals to corresponding 2MASS detections.\\

Finally, we combined IRAC, Flamingos and SWIRC data by merging the photometry of the seven bands. The maximum tolerance in positional offsets between bands was 2.0 arc-seconds. We use detections with error of less than 0.2 magnitudes in our analysis.

\subsection{Millimeter observations}\label{section_CO}
Millimeter wavelength spectral line observations of the target regions were conducted at the Five College Radio Astronomy Observatory (FCRAO) 14~meter telescope in New Salem, Massachusetts between November 2003 and April 2006. We observed the J=1--0 lines of $^{12}$CO and $^{13}$CO simultaneously in on-the-fly mapping mode, using the 32 pixel SEQUOIA focal plane array \citep{eri99} and the dual channel correlator (DCC). Regions S252 and S235 were observed as part of the Extended Outer Galaxy Survey \citep[E-OGS,][]{bru04}, that extends the coverage of the FCRAO Outer Galaxy Survey (OGS) \citep{hey98} to Galactic longitude $l$~=~193, over a the latitude range --3.5$\leq$~$b$~$\leq$+5.5. Observations of W5-east were conducted in May-June of 2004 and 2005, and observations of NGC7538 were conducted in November 2003.

For all observations, pointing and focus checks were carried out every 3--4 hours,
shortly after dawn-dusk or after a significant change in source coordinates. The data were initially converted to the $T_{A}^{*}$ scale using the standard chopper wheel method \citep{kut81}. We used the OTFTOOL software, written by M. Heyer, G. Narayanan, and M. Brewer, to place the spectra on a regular 22.5" grid in Galactic $l, b$ coordinates. Individual spectra contributing to a given coordinate were assigned a $1/\sigma^{2}$ weighting during the griding, after first fitting and subtracting a first order baseline from signal-free regions of each spectrum.
The gridded data were scaled to the main beam temperature scale by dividing by the
main beam efficiency of 0.45 ($^{12}$CO) or 0.48 ($^{13}$CO). The FCRAO beam size is 45" for $^{12}$CO and 46" for $^{13}$CO. For S252, S235, and W5-east the 1024-channel DCC was configured with a total bandwidth of 50~MHz, yielding a channel separation of  0.126~km~s$^{-1}$ ($^{12}$CO) or 0.132~km~s$^{-1}$ ($^{13}$CO). For NGC7538, the total bandwidth was 25~MHz and the channel separation was 0.063~km~s$^{-1}$ ($^{12}$CO) or 0.066~km~s$^{-1}$ ($^{13}$CO). The velocity resolution is 1.21 times the channel spacing.

%\subsection{Archive data}
%We have also made use of archival Spitzer-MIPS (24~$\mu$m) and VLA (1.4~GHz) observations. The MIPS observations were obtained from the ...

\section{Results}\label{section_results3}
Spitzer-IRAC color images for W5-east, S235, S252 and NGC7538 are shown in Figures~\ref{afgl4029_124} to~\ref{ngc7538_124} respectively \citep[for region S254-S258, see Fig.~1 from][]{cha08a}. All regions show abundant extended emission in reddish color corresponding mainly to fluorescence from polycyclic aromatic hydrocarbon (PAH) molecules. Regions W5-east and S252 show PAHs ridges (more prominent in W5-east) at the interface between the expanding \hii~region and the neutral molecular material. It is possible to see several pillar-like structures along the ridges pointing towards the ionizing source. Some of them seem to have a single source at their tip (e.g. between AFGL4029 and G138.15$+$1.69) and others have several (e.g. clusters AFGL4029 and G138.15$+$1.69). The dust around the \hii~regions (as traced by PAHs emission) shows at least two different morphologies: single and bipolar cavity. Single cavities range from less than 1 (e.g. \hii~region S235C) to around 5 parsecs in size (e.g. \hii~region S254) and they have the ionizing source(s) located in their interior. On the other hand, bipolar cavities (e.g. \hii~regions AFGL416, S252E and S252C) have lobe sizes of a couple of parsecs and the ionizing source located between lobes. It is possible that bipolar cavities look like single cavities when seen through the end of one of their lobes. Another feature seen in the IRAC color images are several greenish blobs (e.g. around IRS9 and IRS1-3 in NGC7538, east side of S252A, S235-E1 and S235-E2) known as extended green objects (EGOs). Those correspond mainly to shocked H$_2$ detected in the IRAC 4.5~$\mu$m band and are used as tracers for outflow activity \citep{nor04,cyg08}. Finally, it is possible to identify several areas with a high density of sources, some of them associated with known embedded clusters (e.g. AFGL4029, S235A-C, S252A). In the following section, we explain how we identify the young population among the studied regions.

\subsection{Identifying and classifying young stars}\label{section_classification}
We combine NIR, IRAC and MIPS data to identify and classify YSOs in the observed regions. First, we identify YSOs from sources detected in the IRAC and NIR+IRAC bands. Then, we identify more YSOs  from sources detected only in the NIR bands by their color excess in the [J--H] vs. [H--K] color-color diagram. Finally, sources detected in the IRAC-8 microns and MIPS-24 microns bands but not detected in the IRAC-3.6 microns band are also identified as YSOs. %The background decontamination algorithms used for different identification schemes are described in Appendix~\ref{background}.

\subsubsection{Sources with IRAC 4-bands detections}\label{irac_detections}
\begin{figure}
\begin{center}
\includegraphics[width=7cm]{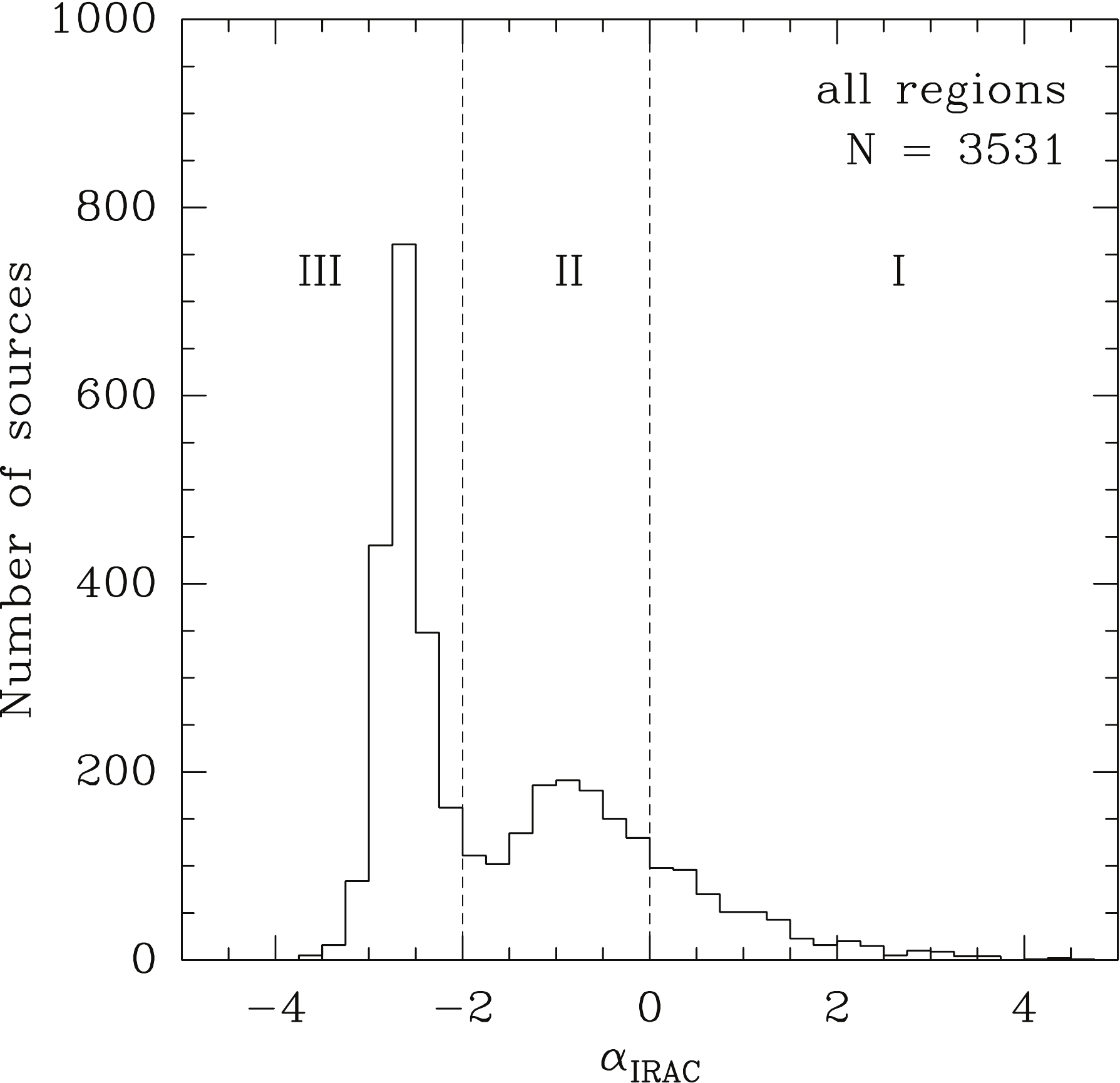}
\caption{Observed spectral indices distribution for sources detected in the four IRAC bands (see \S~\ref{irac_detections}). Dashed lines separate Class~I and Class~II sources. The minimum around \ai$=-2$ is also seen in other studies \citep{kum07,mue07}.}
\label{spectral_histograms}
\end{center}
\end{figure}

Sources with detections in all IRAC bands are classified using their observed IRAC spectral energy distribution slope \ai~as Class I (\ai~$> 0$), Class II ($-2 <$ \ai~$< 0$) and stellar photospheres or Class III (\ai~$< -2$) \citep{lad87}. This classification scheme follows an evolutionary trend that begins with a Class 0 (collapsing starless envelope), continues with Class I (star plus collapsing envelope), then Class II (star plus a disk) and Class III (star with a thin disk). If well there may be a few Class 0 sources among our Class I sample, this has no implications in our analysis since they are less evolved than the Class II sources as well. Among the sources detected in the IRAC bands, those detected also in the NIR bands are classified using their derreddened \ai~value. The percentage of sources that changed their classification between the observed and derreddened \ai~value is of less than 5\% in each region. After background contaminants subtraction, we identify 495 Class I and 1186 Class II out of the 3531 detected sources. The distribution of observed \ai~values is shown in Figure~\ref{spectral_histograms}. 
%IRAC color-color diagrams for each region are shown in Figure~\ref{irac_colcol}.

%%%%%%%%%%%%%%%%%%%%%%%%%%%%%%%%%%%%%
%%%%%%%%%%%%%%%%%%%%%%%%%%%%%%%%%%%%%
%%%%%%%%%%%%%%%%%%%%%%%%%%%%%%%%%%%%%
%%%%%%%%%%%%%%%%%%%%%%%%%%%%%%%%%%%%%

%%%%%%%%%%%%%%%%%%%%%%%%%%%%%%%%%%%%%
%%%%%%%%%%%%%%%%%%%%%%%%%%%%%%%%%%%%%
\begin{figure*}
\begin{center}
\includegraphics[width=7cm]{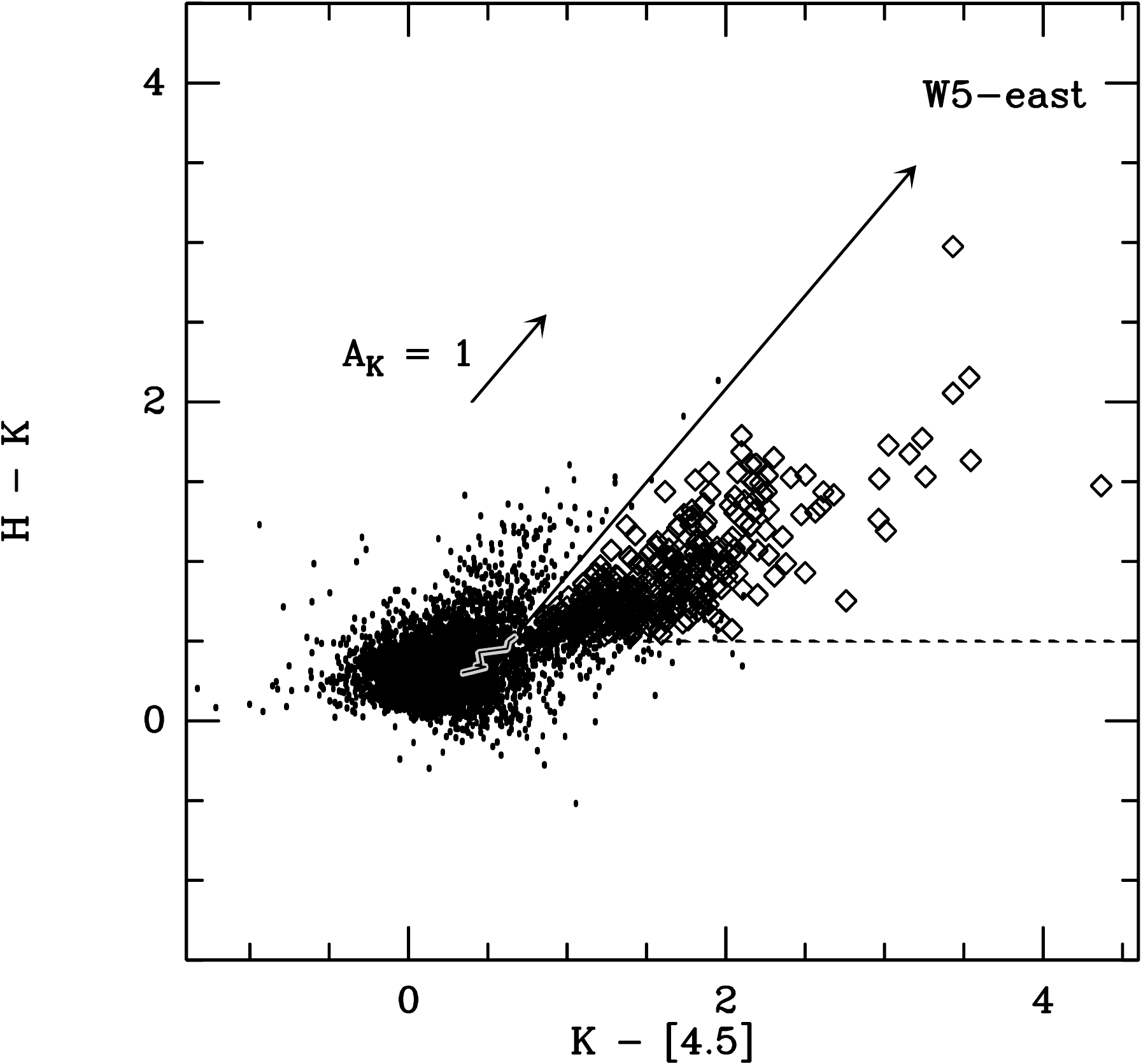}
\includegraphics[width=7cm]{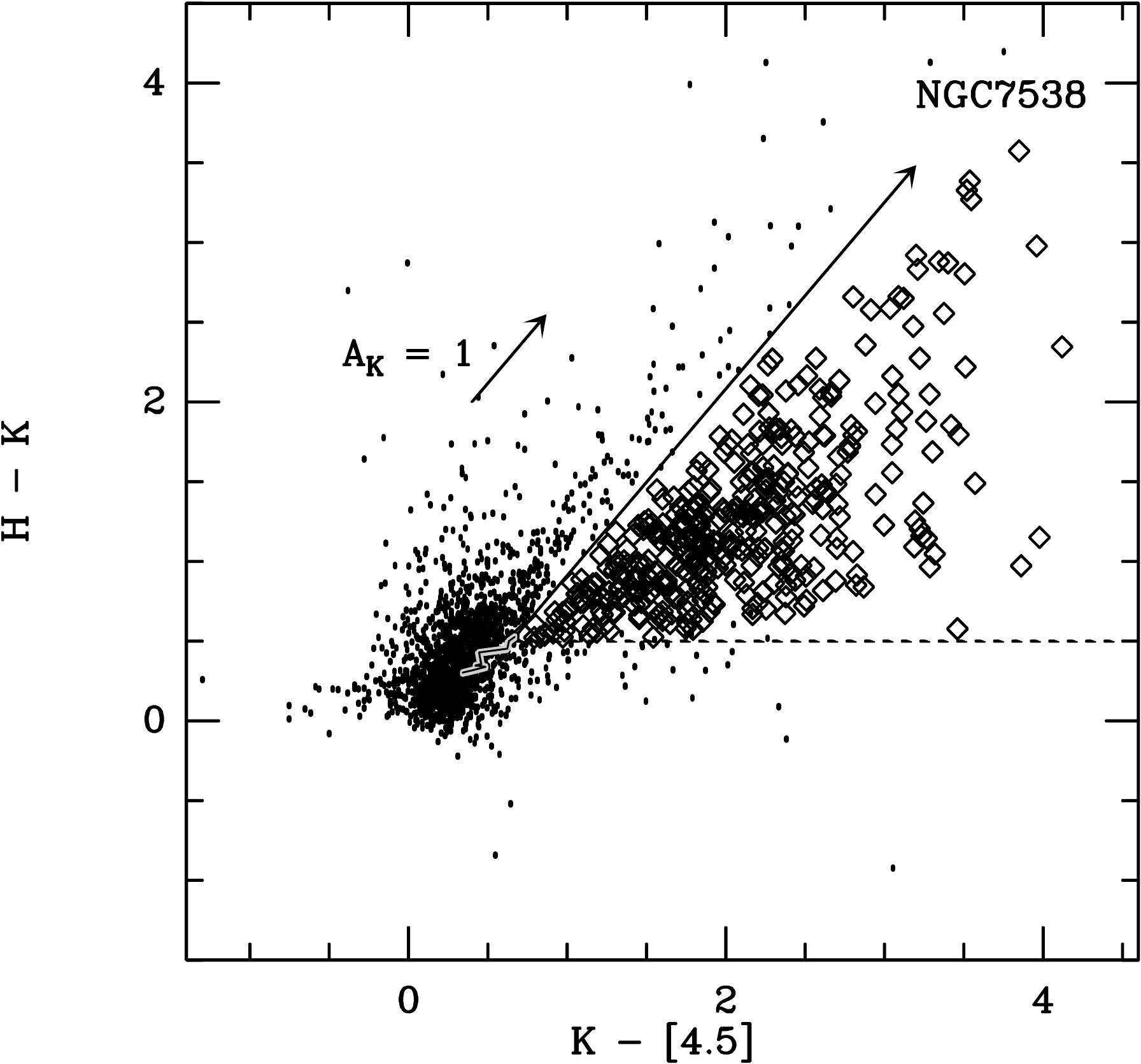}
\includegraphics[width=7cm]{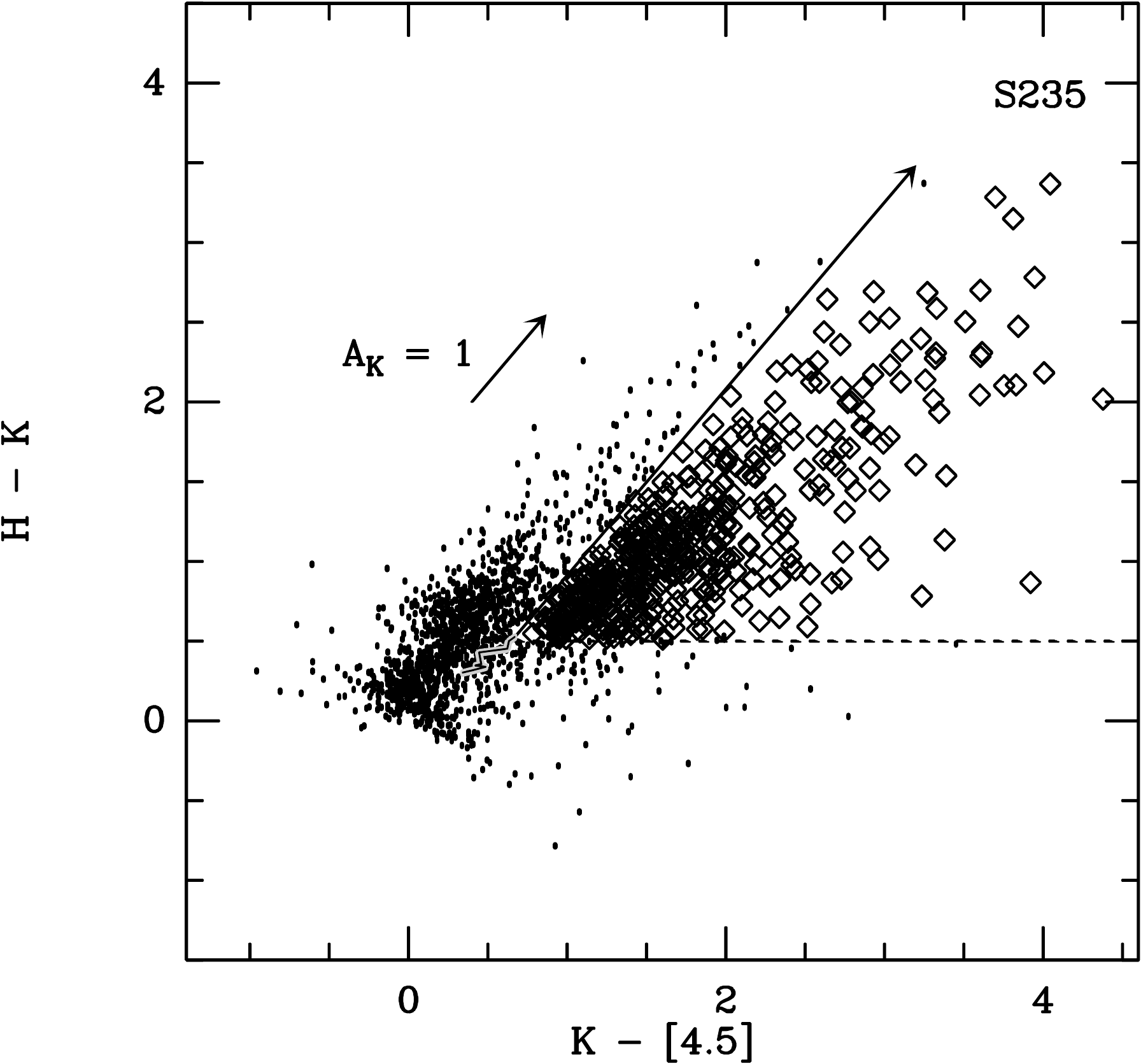}
\includegraphics[width=7cm]{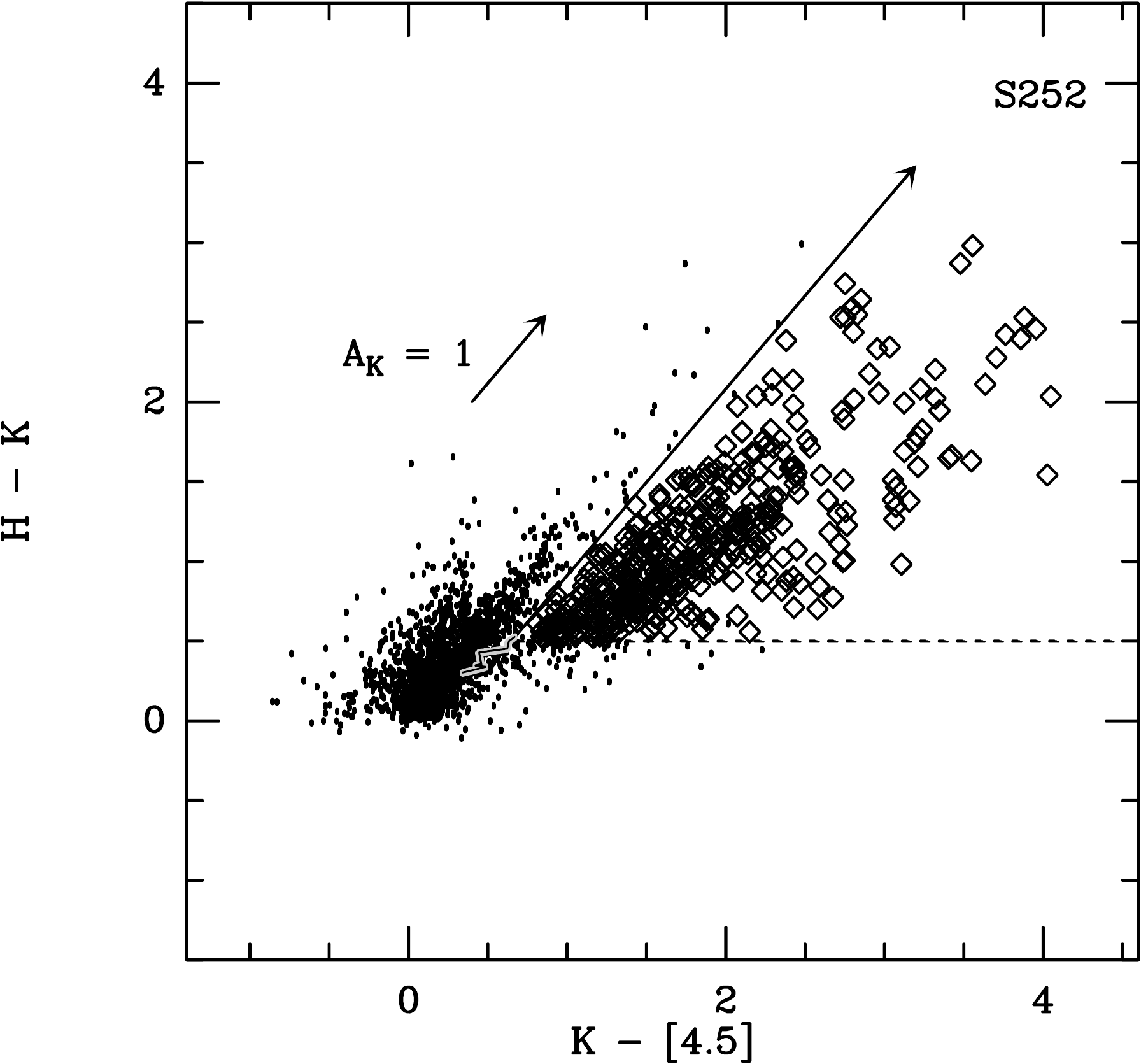}
\caption{H, K and IRAC 4.5 $\mu$m band color-color diagrams for each region. The long arrow corresponds to the reddening vector \citep{fla07}. Stars with infrared excess, shown as diamonds, lay between the reddening vector and the dashed horizontal line. The black line near (0,0), close to the reddening vector base shows the main sequence locus of a late type M dwarf star \citep{pat06}.}
\label{nir_irac_colcol}
\end{center}
\end{figure*}
%%%%%%%%%%%%%%%%%%%%%%%%%%%%%%%%%%%%%
%%%%%%%%%%%%%%%%%%%%%%%%%%%%%%%%%%%%%

\subsubsection{Adding MIPS detections}\label{section_MIPS_detections}
Some sources in our sample that are detected only at the longer IRAC wavelengths may correspond to embedded Class I or Class 0 sources. We include those YSOs in our Class I sample by adding sources detected at 8 with a 24~$\mu$m counterpart but not detected at 3.6~$\mu$m. This way we identify 5 additional Class I sources in region W5-east, 8 in S235, 18 in S252, 6 in S254-S258 and 7 in NGC7538.

\subsubsection{Sources with IRAC and NIR detections}\label{section_HK2_detections}
Additional YSOs are identified by combining the IRAC-4.5$\mu$m and the H and K bands \citep[e.g.][]{win07,gut08}. Sources with IR-excess due to presence of a dusty disk around them are located to the right of the reddening vector in the [H--K] vs.~[K--4.5] color-color diagram (Fig.~\ref{nir_irac_colcol}). From this diagram, we find 285 sources with IR-excess in region W5-east, 439 in S235, 419 in S252, 364 in S254-S258 and 387 in NGC7538. Of those, many were already classified as YSOs from their \ai~value (see \S~\ref{irac_detections}). Therefore, the number of additional YSOs identified per region are 135, 230, 237, 229 and 291 for W5-east, S235, S252, S254-S258 and NGC7538 respectively. Since the additional YSOs are not detected in all IRAC bands, those are not classified in Class I or Class II. The same happens with the YSOs detected only in the near-IR (see \S~\ref{nirexcess_section}).

\subsubsection{Sources detected only in the NIR}\label{nirexcess_section}
The NIR data is used to identify YSOs which are not detected in the IRAC bands due to bright PAHs emission and/or saturation in the IRAC bands \citep[eg. cluster S255-2 in region S254-S258,][]{cha08a}. We apply the method from \citet{cha10} to identify YSOs by their NIR color excess in the [H--K] vs.~[J--H] color-color diagram. First, we use the color-color diagram to estimate the reddening law $E_{J-H}/E_{H-K}$ towards each region and then we classify sources as YSOs if they are located at least 1 sigma to the right of the reddening vector. We identify 39 additional YSOs in W5-east, 59 in S235, 34 in S254-S258 and 42 in NGC7538. The derived $E_{J-H}/E_{H-K}$ values are 1.38 for W5-east, 1.50 for S235, 1.69 for S252, 1.60 for S254-S258 and 1.68 for NGC7538, all with errors up to 0.1.

%%%%%%%%%%%%%%%%%%%%%%%%%%%%%%%%%%%%%%%%%%%%%%%%
%%%%%%%%%%%%%%%%%%%%%%%%%%%%%%%%%%%%%%%%%%%%%%%%
\begin{landscape}
%\centering
\begin{table}
\caption{Identified YSOs per region$^a$ }
\label{YSOs_class}
\begin{tabular}{lcccccccccccccc} 
\hline
ID & Region & R.A. & Dec. & $J$ & $H$ & $K$ & $[3.6]$ & $[4.5]$ & $[5.8]$ & $[8.0]$  & Class$^b$ & \ai & Method$^c$ \\
    &             & J(2000) & J(2000) & (mag)& (mag) & (mag) & (mag) & (mag) & (mag) & (mag)  &  &  &  \\
\hline
1 &W5-east &45.090149& 60.301571 &-                  &13.86(0.01) &13.64(0.01) &13.55(0.04) &13.52(0.04) &13.42(0.06) & 12.80(0.08) &II& -1.99& IRAC\\
2 &W5-east &45.097431 &60.530960 &16.46(0.07)  &15.36(0.04) &14.69(0.03) &13.56(0.04) &13.30(0.04) &12.94(0.05) &-& - &- &HK2\\
3 &W5-east &45.100380 &60.543320 &15.88(0.02)  &14.83(0.01) &14.23(0.01) &13.43(0.04) &13.02(0.03) &12.36(0.04) &- &- &- &HK2\\
4 &W5-east &45.106079 &60.469429 &13.73(0.01)  &-                 &-                  &11.57(0.02) &11.86(0.02) &10.87(0.02) &10.40(0.02) &II &-1.25 &IRAC\\
5 &W5-east &45.107029 &60.307461 &15.43(0.02)  &14.57(0.02) &14.04(0.01) &13.72(0.04) &13.30(0.04) &13.06(0.05) &12.22(0.05)& II& -1.32& IRAC\\
6 &W5-east &45.108589 &60.415009 &15.08(0.01) &14.27(0.01) &12.99(0.01) & - & -& -& - & - & -& NIR\\
7 &W5-east &45.113110 &60.461498 &17.16(0.07) &16.89(0.11) &15.55(0.04) &- & -& -& - & - & -& NIR\\
8 &W5-east &45.113350 &60.369598 &17.58(0.08) &16.85(0.09) &15.74(0.04) &- & -& -& - & - & -& NIR\\
9 &W5-east &45.114182 &60.460030 &12.87(0.01) &12.17(0.01) &11.55(0.01) &10.89(0.01)& 10.62(0.01)& 9.92(0.01)& 9.45(0.01)& II& -1.10& IRAC\\
10&W5-east&45.114819 &60.479900 &15.04(0.01) &14.07(0.01) &13.68(0.01) &13.49(0.04)& 13.16(0.04)&13.08(0.05)& 12.32(0.05)& II& -1.66 &IRAC\\
\hline
\end{tabular}
\end{table}
$^a$~Table~2 is published in its entirety in the electronic version of this paper. A portion is shown here as a guidance. Values in parentheses by photometry signify error in magnitudes.

$^b$~YSOs are classified in Class I or II following the description in \S~\ref{section_classification}.

$^c$~This column shows the method used to identify the YSO: \ai~slope (IRAC, see \S~\ref{irac_detections}), excess in the H, K, and [4.5] bands (HK2, see \S~\ref{section_HK2_detections}), excess in the NIR bands (NIR, see \S~\ref{nirexcess_section}) or detection in the MIPS 24 micronds band (MIPS, see \S~\ref{section_MIPS_detections}).
\end{landscape}
%%%%%%%%%%%%%%%%%%%%%%%%%%%%%%%%%%%%%%%%%%%%%%%%
%%%%%%%%%%%%%%%%%%%%%%%%%%%%%%%%%%%%%%%%%%%%%%%%

By adding the identified YSOs from the previously described methods, the total number of YSOs per region is 478, 690, 679, 512 and 662 for W5-east,  S235, S252, S254-S258 and NGC7538 respectively. Their coordinates, magnitudes and classification can be found in the electronic version of this paper (see Table~\ref{YSOs_class} for an example). The properties of YSOs per region are listed in Table~\ref{stars_detected}.Previous identification of YSOs using IRAC+NIR data have been done by \citet{koe08} and \citet{dew11} in regions W5-east and S235 respectively. A comparison between their findings and our work is presented in Appendix~B.

%%%%%%%%%%%%%%%%%%%%%%%%%%%%%%%%%%%%%%%%%%%%%%%%
%%%%%%%%%%%%%%%%%%%%%%%%%%%%%%%%%%%%%%%%%%%%%%%%
\begin{table*}
\centering
\begin{minipage}{100mm}
\caption{Properties of detected YSOs per region \protect\footnote{$\overline{A}_K$ values are corrected by interstellar extinction.} } 
\label{stars_detected}
\begin{tabular}{lccccc} 
\hline
 & W5-east & S235 & S252 & S254-S258 & NGC7538\\
\hline
$N_{\textrm{\tiny{YSO}}}$ & 478 & 690 & 679 & 512 & 662 \\
Parameter \Q$_{\textrm{\tiny{YSO}}}$ & 0.69 & 0.74 & 0.64 & 0.68 & 0.81 \\
$N_{\textrm{\tiny{I}}}$ & 62 & 113 & 169 & 87 & 108 \\
$N_{\textrm{\tiny{II}}}$ & 242 & 288 & 273 & 162 & 221 \\
$\overline{A}_{K\textrm{\tiny{, I}}}$ (mag)& 0.98 & 1.13 & 0.77 & 1.19 & 1.15 \\
$\overline{A}_{K\textrm{\tiny{, II}}}$ (mag)& 0.48 & 0.95 & 0.75 & 1.02 & 0.75 \\
$\overline{\Sigma}_{\textrm{\tiny{YSO, I}}}$ (pc$^{-2}$) & 58 & 82 & 35 & 52 & 26 \\
$\overline{\Sigma}_{\textrm{\tiny{YSO, II}}}$ (pc$^{-2}$) & 24 & 36 & 15 & 42 & 11 \\
$\overline{\sigma}_{\textrm{\tiny{CO, I}}}$ ($10^{16}$ cm$^{-2}$) & 1.50 & 2.38 & 2.79 & 2.13 & 2.60 \\
$\overline{\sigma}_{\textrm{\tiny{CO, II}}}$ & 0.74 & 1.53 & 1.90 & 1.46 & 1.72 \\
\% of YSOs in clusters& 0.52 & 0.61 & 0.63 & 0.73 & 0.69 \\
\% of I in clusters& 0.66 & 0.73 & 0.76 & 0.71 & 0.79 \\
\% of II in clusters& 0.33 & 0.46 & 0.42 & 0.62 & 0.51 \\
%
%M$_{\textrm{CO}}$ & 6200 & 7300 & 13800 & 5900 & 14000 \\
%SFE$_{\textrm{CO}}$ & 0.04 & 0.04 & 0.02 & 0.04 & 0.02 \\
%SFE$_{\textrm{CO(N*)}}$ & 0.07 & 0.08 & 0.04 & 0.08 & 0.04 \\
\hline
\end{tabular}
\end{minipage}
\end{table*}
%%%%%%%%%%%%%%%%%%%%%%%%%%%%%%%%%%%%%%%%%%%%%%%%
%%%%%%%%%%%%%%%%%%%%%%%%%%%%%%%%%%%%%%%%%%%%%%%%

\subsection{Spatial distribution of YSOs}\label{section_distribution}
Having identified the young population of stars, we investigate their spatial distribution by searching for concentrations of YSOs and by comparing the arrangement of different YSOs classes. 

\subsubsection{Minimum spanning tree (MST)}
The MST is defined as a network of branches connecting points such as the total length of the branches is minimized and there is no loops \citep{bat91}. This algorithm has lately become a popular tool to search for clusters of stars since it is independent from the stars density number. Nevertheless, it requires the definition of some maximum critical distance between cluster members ($d_c$). \citet{gut09} used the MST to study embedded clusters in low-mass star forming regions with $d_c$ given by the turnover of the MST branches length distribution (see Fig.~1 in their paper). In our case, since our observations cover larger fields/regions including several clusters, a single critical distance per region may not be representative for all clusters. Therefore, we divide each region in areas containing concentrations of YSOs either clearly separated (like AFGL4029, AFGL416 and G138.15+1.69 in W5-east, see Fig.~\ref{afgl4029irac_classes}) or associated with molecular material at different \vlsr~(see \S~\ref{section_cloud}). Then, we calculate the critical distance $d_c$ for each area using the method from \citet{gut09}. The MST for each region is shown in figures~\ref{afgl4029irac_classes} to~\ref{ngc7538irac_classes} for W5-east, S235, S252, S254-S258 and NGC7538 respectively. 

We use the MST to identify clusters as groups of $n=10$ or more YSOs connected by branches shorter than $d_c$ \citep[for region S254-S258 we used $n=7$ in order to identify all clusters from][]{cha08a}. We find in total 41 embedded clusters. Out of those, 15 have not been found in the literature until date. All clusters and their properties are listed in Table~\ref{clusters}: columns 1 to 7 list the cluster name, center position (given by the members coordinates mean), total number of members ($N_{\textrm{\tiny{YSO}}}$), number of Class I and Class II and ratio of Class I to Class II sources; column~8 list the cluster members average \ai~value ($\widetilde{\alpha}_{\mathrm{IRAC}}$); column~9 list the radius of the minimum area circle enclosing all cluster members ($R_{\textrm{\tiny{C}}}$); column~10 list the hull radius ($R_{\textrm{\tiny{H}}}$, see \S~\ref{section_Qparameter}); column~11 list the aspect ratio (AR or cluster roundness) $R_{\textrm{\tiny{C}}}^2/R_{\textrm{\tiny{H}}}^2$; columns~12 and 13 list, respectively, the mean and peak surface density of YSOs inside the cluster convex hull area; columns~14 and 15 list the critical distance ($d_c$) and the mean branches length between cluster members ($s_{\mathrm{YSO}}$); columns 16 and 17 list the structural \Q~parameter value and error (see \S~\ref{section_Qparameter}) and column 18 list the corrected number of cluster members ($N\textrm{*}_{\mathrm{YSO}}$, see Appendix~\ref{N_corrected}).

\subsubsection{IR-excess sources surface density map} \label{section_density3}
\begin{figure*}
\begin{center}
\includegraphics[width=17cm]{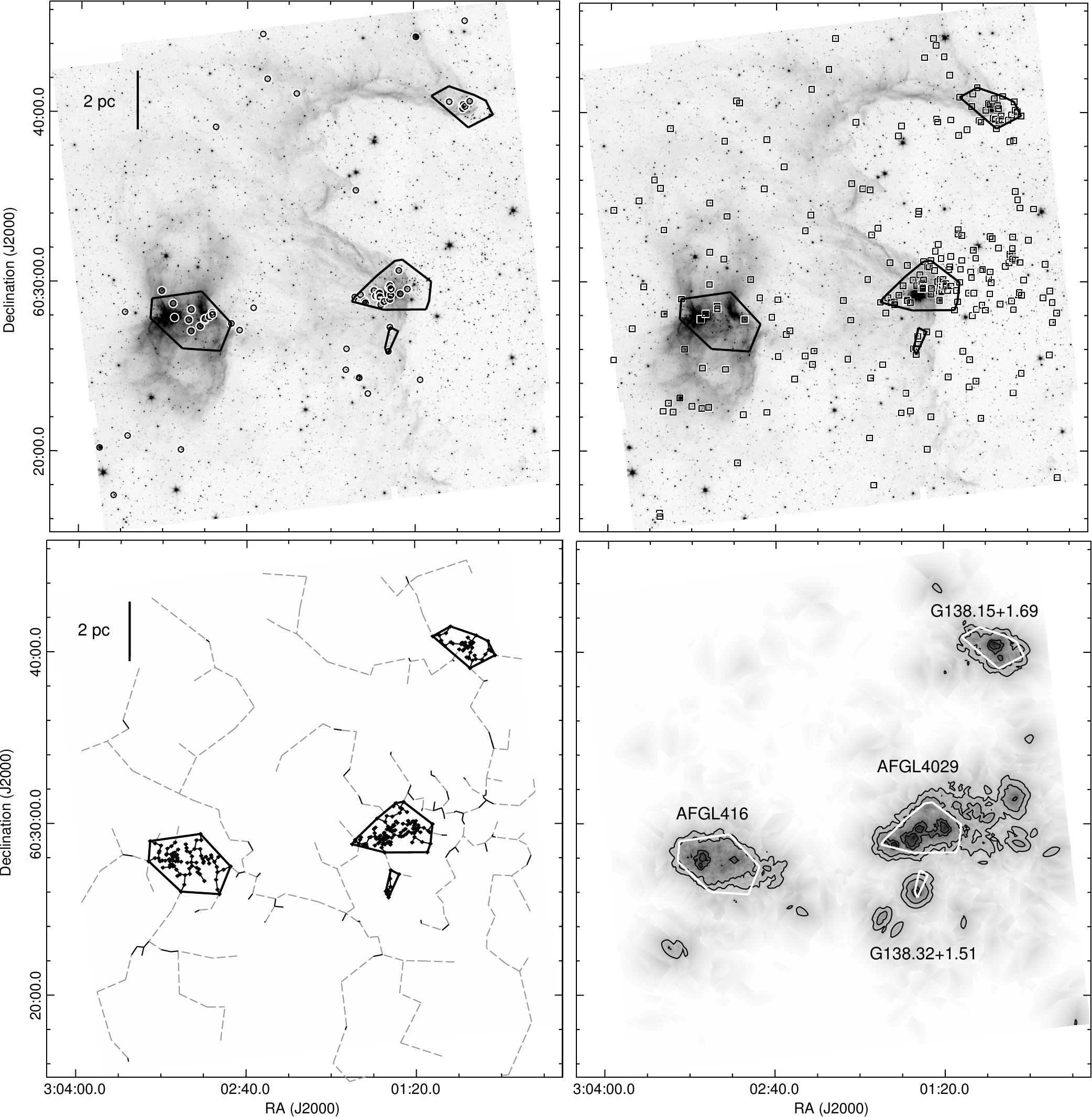}
\caption{Spatial distribution of YSOs in W5-east. Polygons enclosing cluster members (convex hull) are shown in all panels. \textbf{Upper:} Sources identified as Class I (\textbf{left}) and Class II (\textbf{right}) are shown over the IRAC 4.5 $\mu$m mosaic. \textbf{Lower-left:} Minimum spanning tree (MST) of YSOs positions (including Class I, Class II and IR-excess sources). Contiguous sources separated by a projected distance of less than the critical distance $d_c$ (see Table~\ref{clusters}) are joined by a continuous black line. Groups of YSOs that qualify as clusters are marked by black dots. \textbf{Lower-right:} Surface density of YSOs. Contour levels are at 5, 10, 50 and 100~stars~pc$^{-2}$. Cluster names are labeled in this panel.}
\label{afgl4029irac_classes}
\end{center}
\end{figure*}

\begin{figure*}
\begin{center}
\includegraphics[width=13cm]{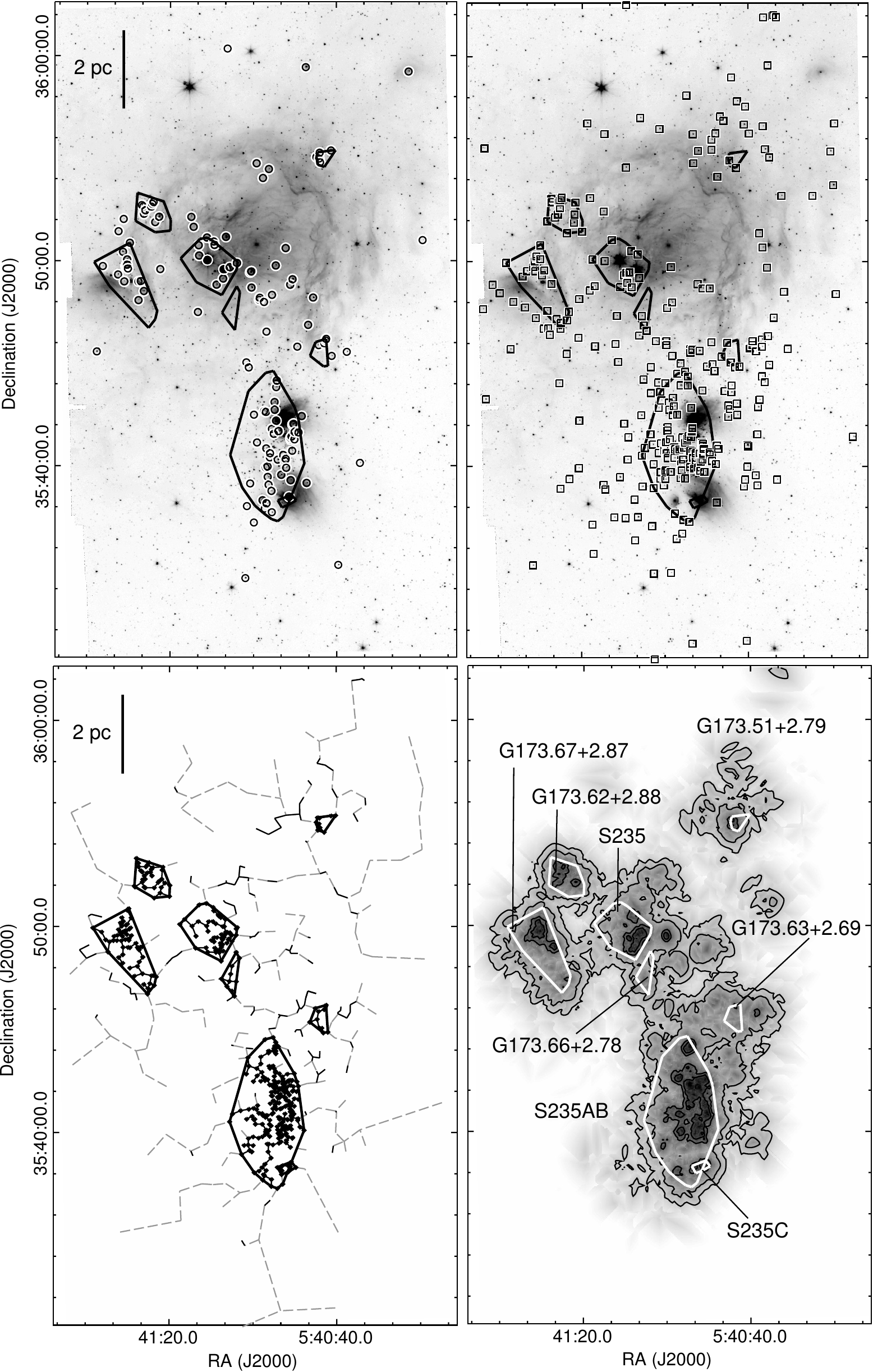}
\caption{Same as in Figure~\ref{afgl4029irac_classes} for region S235.}
\label{s235irac_classes}
\end{center}
\end{figure*}

We use YSOs surface density maps to complement the MST cluster finding algorithm. For each region, the FOV is divided into a 3 arc-second grid. Then, at each point of the grid, the YSOs surface density ($\Sigma_{\mathrm{YSO}}$) is calculated from:
\be
\Sigma_{\mathrm{YSO}}=\frac{N}{\pi r_{N}^2},
\ee
where $r_{N}$ is the distance to the $N=5$ nearest neighbor (NN). Surface density maps for the studied regions are shown in Figures~\ref{afgl4029irac_classes} to~\ref{ngc7538irac_classes}. Both MST and $\Sigma_{\mathrm{YSO}}$ methods identify basically the same YSOs groups as clusters. In general, MST clusters are enclosed by surface density contours of around $10-50$ stars per pc$^{-2}$. For a more detailed comparison between these and other cluster finding algorithms we recommend the work by \citet{sch11}. 

\subsubsection{The structural \Q~parameter and K-S tests}\label{section_Qparameter}
In this section we describe the tools used to compare in a more quantitatively way the distribution of different class YSOs in the observed regions and clusters: the \Q~parameter and the Kolmogorov-Smirnov test. The \Q~parameter \citep{car04,schm06} is used to measure the level of hierarchical vs. radial distribution of a set of $(x_N,y_N)$ points, and it is defined by:
\be
\mathcal{Q}=\frac{\bar{l}_{\textrm{\tiny{MST}}}}{\bar{s}},
\ee
where $\bar{l}_{\textrm{\tiny{MST}}}$ is the MST normalized mean branches length and $\bar{s}$ is the normalized mean separation between points. The MST mean branches length $l_{\textrm{\tiny{MST}}}$ is calculated directly from the MST total length divided by the number of branches ($N-1$). Then, it is normalized by $\sqrt{A/N}$, where the convex hull area $A$ is the area of a polygon enclosing all points, with internal angles between two contiguous sides of less than 180 degrees. In a similar way, the mean separation between points $s$ is normalized by $R_H$, the radius of a circle with area $A$. Using the normalized values $\bar{l}_{\textrm{\tiny{MST}}}$ and ${\bar{s}}$, the \Q~parameter becomes independent from the cluster size \citep{schm06}. According to \citet{car04}, both $\bar{l}_{\textrm{\tiny{MST}}}$ and $\bar{s}$ values decrease as the degree of radial concentration increases (or the level of hierarchy decreases). However, $\bar{s}$ decreases faster than $\bar{l}_{\textrm{\tiny{MST}}}$. This way, a group of $(x_N,y_N)$ points distributed radially will have a high \Q~value (\Q~$ > 0.8$) while clusters with a more fractal distribution will have a low \Q~value (\Q~$ < 0.8$) \citep{car04}. We also assign an error to \Q~given by the variation of its value over the range $d_c$ and $d_c +0.2\times d_c$.

% $R_H$ as the radius of a circle with area $A$ \citep[the same as $R_{\textrm{\tiny{hull}}}$ in][Table~7]{gut09}. 

The Kolmogorov-Smirnov (K-S) test is a statistical tool commonly used to investigate if two data sets are significantly different. In practice, the K-S test gives the probability that the hypothesis statement "both data sets are drawn from the same distribution", is true. We define that two data sets are significantly different if their K-S probability is of less than 1\% ($P < 0.01$).

%%%%%%%%%%%%%%%%%%%%%%%%%%%%%%%%%%%%%
%%%%%%%%%%%%%%%%%%%%%%%%%%%%%%%%%%%%%
%%%%%%%%%%%%%%%%%%%%%%%%%%%%%%%%%%%%%
%%%%%%%%%%%%%%%%%%%%%%%%%%%%%%%%%%%%%

%%%%%%%%%%%%%%%%%%%%%%%%%%%%%%%%%%%%%
%%%%%%%%%%%%%%%%%%%%%%%%%%%%%%%%%%%%%
%%%%%%%%%%%%%%%%%%%%%%%%%%%%%%%%%%%%%
%%%%%%%%%%%%%%%%%%%%%%%%%%%%%%%%%%%%%
\subsection{The molecular content around embedded clusters}\label{section_cloud}
We use extinction maps together with \coa~and \cob~maps to derive the physical properties of the molecular material associated with embedded clusters.

\subsubsection{K-band extinction maps}\label{section_extinctionmap}
The extinction maps were created from all sources detected in the NIR bands and located in the main sequence region in the [H--K] vs. [J--H] color-color diagram. Following the method from \citet{cha10}, we derive the reddening law $E_{J-H}/E_{H-K}$ towards each region (see \S~\ref{nirexcess_section}) and estimate the sources K-band extinction ($A_K$) by dereddening them to their main sequence colors. Then, the observed FOV is divided into a 3 arc-seconds grid and the average $A_K$ values between the 5 nearest neighbors to the grid center are assigned to each grid point. Since the studied regions are located at a few kpc from the Sun, foreground stars extinguished mostly by interstellar dust will "artificially" decrease the average $A_K$ value. This will lead to an underestimation of the extinction due to the molecular cloud. To correct this, we remove from each grid point the two sources with lower extinction if the grid $A_K$ standard deviation is more than 0.5 magnitudes (typical values are between 0.2 and 0.3 magnitudes). After applying this correction, the extinction in the most embedded areas increased by 0.3-0.5 magnitudes.

The clusters molecular mass were estimated from the extinction maps. We use the relations from \citet{dic78} and \citet*{car89} to estimate the H$_2$ column density in each grid from the average $A_K$ values using:
\be
N(H_2) = 1.25\times 10^{21}\left(\frac{A_K-A_{K,fg}}{0.114}\right)~cm^{-2}~mag^{-1}, \label{dust2.1}
\ee
where $A_{K,fg} = 0.15\times distance~[kpc]$ is the interstellar contribution to the extinction \citep{ind05}. Then, the cluster molecular mass is calculated by integrating the column density over each cluster convex hull area and multiplying by the H$_2$ molecule mass. Extinction maps are shown in Figures~\ref{w5_molecular} to~\ref{ngc7538_molecular}.

The properties of the molecular material associated with embedded clusters (meaning inside the convex hull area) are listed in Table~\ref{cloud}: the mean and peak K-band extinction are listed in columns 2 and 3 respectively; molecular mass derived from the cluster extinction and in regions where $A_K>0.8$ magnitudes (see \S~\ref{section_Ndense}) are listed in columns 4 and 5 respectively; mean and peak \cob~column densities are listed in columns 6 and 7 respectively (see \S~\ref{section_molecular_gas}); molecular mass derived from the cluster CO column density is listed in column 8; the \cob~\vlsr~and FWHM of the integrated \cob~emission over the hull area are listed in columns 9 and 10 respectively.
%%%%%%%%%%%%%%%%%%%%%%%%%%%%%%%%%%%%%
%%%%%%%%%%%%%%%%%%%%%%%%%%%%%%%%%%%%%
\begin{figure*}
\begin{center}
\includegraphics[width=17cm]{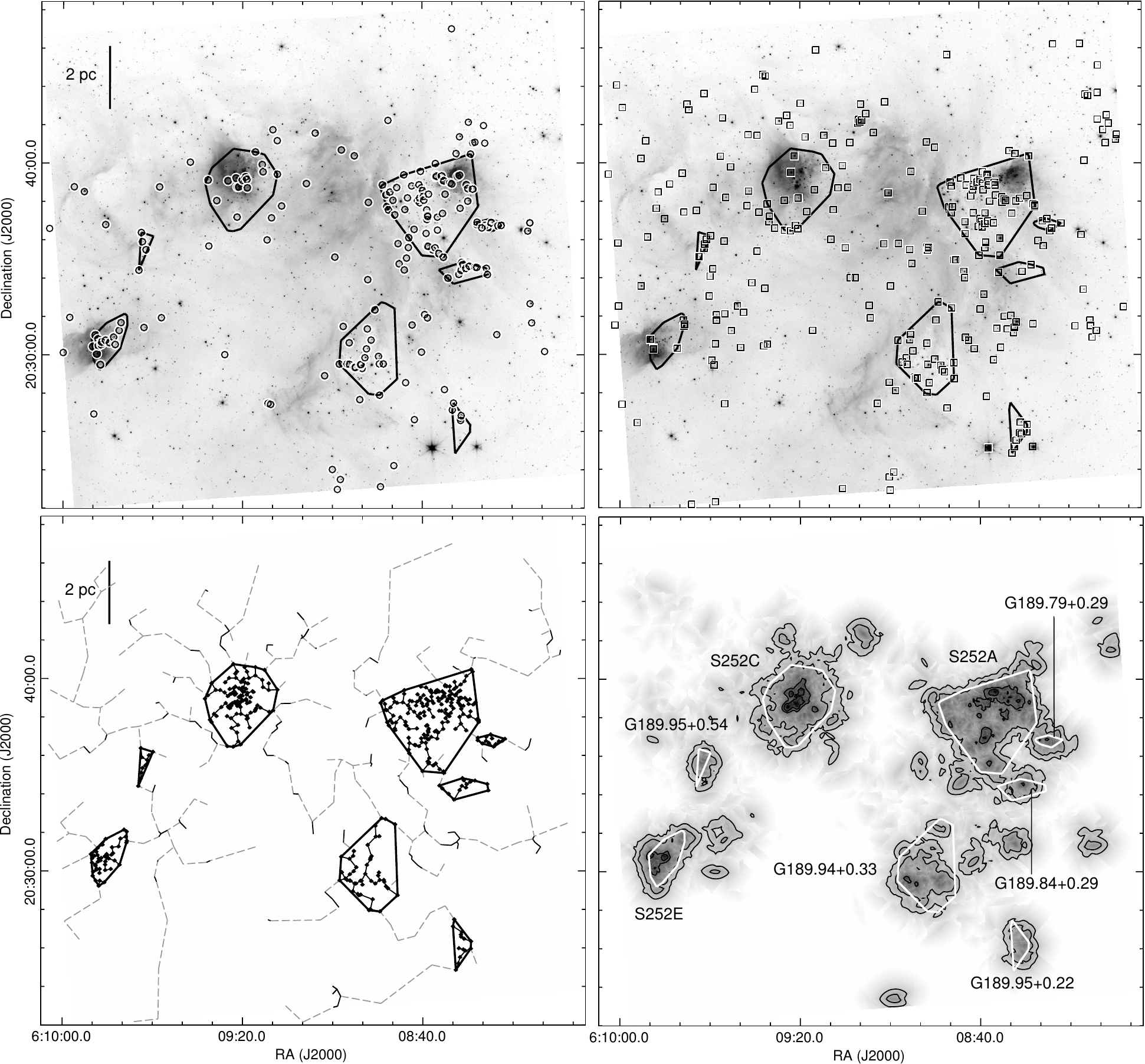}
\caption{Same as in Figure~\ref{afgl4029irac_classes} for region S252.}
\label{s252irac_classes}
\end{center}
\end{figure*}
%%%%%%%%%%%%%%%%%%%%%%%%%%%%%%%%%%%%%
%%%%%%%%%%%%%%%%%%%%%%%%%%%%%%%%%%%%%

\subsubsection{\coa~and \cob~emission}\label{section_molecular_gas}
The CO observations trace the less dense gas which is more difficult to detect using the extinction maps. In addition, CO maps contain velocity information that allows to disentangle different molecular components along the line of sight. 

We created \cob~column density maps for all regions by calculating the \cob~column density at each CO grid point (see \S~\ref{section_CO}) using \citep{hey89}:
\be
N(^{13}\mathrm{CO}) = 2.42\times10^{14}\left(\frac{\Delta v T_{ex}\tau}{1 - e^{-5.29/T_{ex}}}\right) cm^{-2},\label{13coeq1}
\ee
where
\be
T_{ex} = \frac{5.53}{ln\left(1 + \frac{5.53}{T_A + 0.82}\right)},\label{13coeq2}
\ee
\be
\tau = -ln\left(1 - \frac{T_{ex}}{\frac{5.29}{e^{5.29/T_{ex}}-1}-1}\right),\label{13coeq3}
\ee
$T_A$ is the observed peak temperature of the \coa, and $\Delta v$ correspond to the full width at half maximum (FWHM) of the \cob~line (see the maps in Figures~\ref{w5_molecular} to~\ref{ngc7538_molecular}). The FWHM was obtained by fitting a Gaussian profile to each grid velocity spectrum. From the \cob~column density maps, we estimate the gas mass by integrating the column density over the clusters convex hull area \citep*[assuming \coa$/$\cob~$= 50$ and \coa$/$H$_2=8.5\times 10^{-5}$ from][respectively]{isr03,fre82}. The derived \cob~column densities and gas masses are summarized in Table~\ref{cloud}.  

In order to distinguish different molecular components along the line of sight, we used the \cob~first moment maps to identify the \vlsr~of the peak emission at each grid point (Figures~\ref{w5_molecular} to~\ref{ngc7538_molecular}). In addition to this, we estimate the \vlsr~of the molecular material associated with each embedded cluster by integrating the \cob~spectra over the cluster convex hull area (assuming that \cob~is more optically thin than \coa) and then adjusting Gaussians to the integrated line profile. For simplicity, we assume that the \vlsr~of the molecular material associated with each cluster corresponds to the strongest Gaussian component.  As an example, Figure~\ref{gauss_afgl416} shows the two Gaussian components used to fit the \cob~integrated spectra for cluster AFGL416 (the integrated \cob~spectrum for the other clusters are shown in Figures~\ref{image_spectrum_w5} to \ref{image_spectrum_ngc7538}). The \vlsr~and FWHM of the Gaussian component associated to each cluster are also listed in Table~\ref{cloud}.

%%%%%%%%%%%%%%%%%%%%%%%%%%%%%%%%%%%%%
%%%%%%%%%%%%%%%%%%%%%%%%%%%%%%%%%%%%%
\begin{figure*}
\begin{center}
\includegraphics[width=17cm]{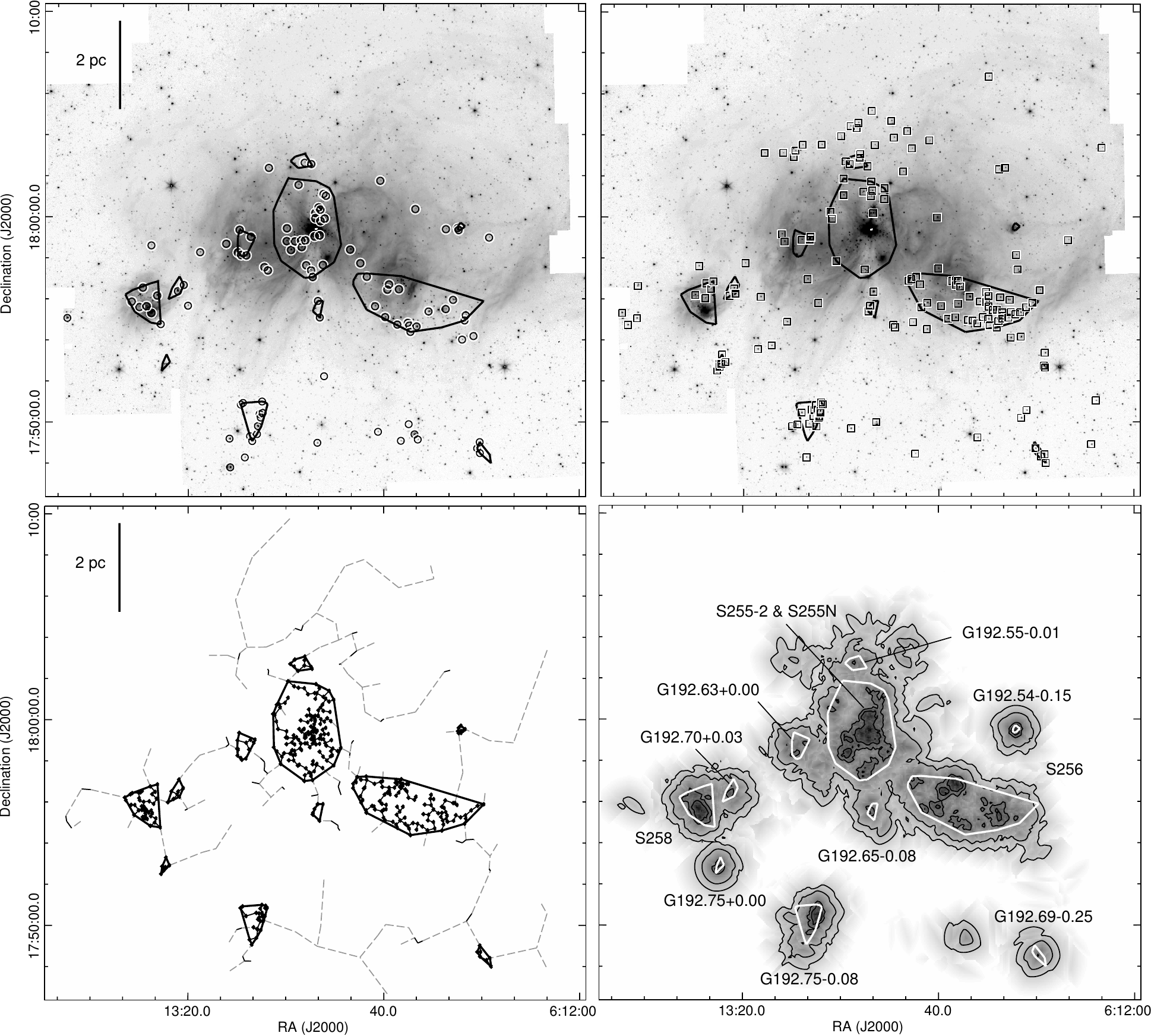}
\caption{Same as in Figure~\ref{afgl4029irac_classes} for region S254-S258.}
\label{s255irac_classes}
\end{center}
\end{figure*}

\begin{figure*}
\begin{center}
\includegraphics[width=17cm]{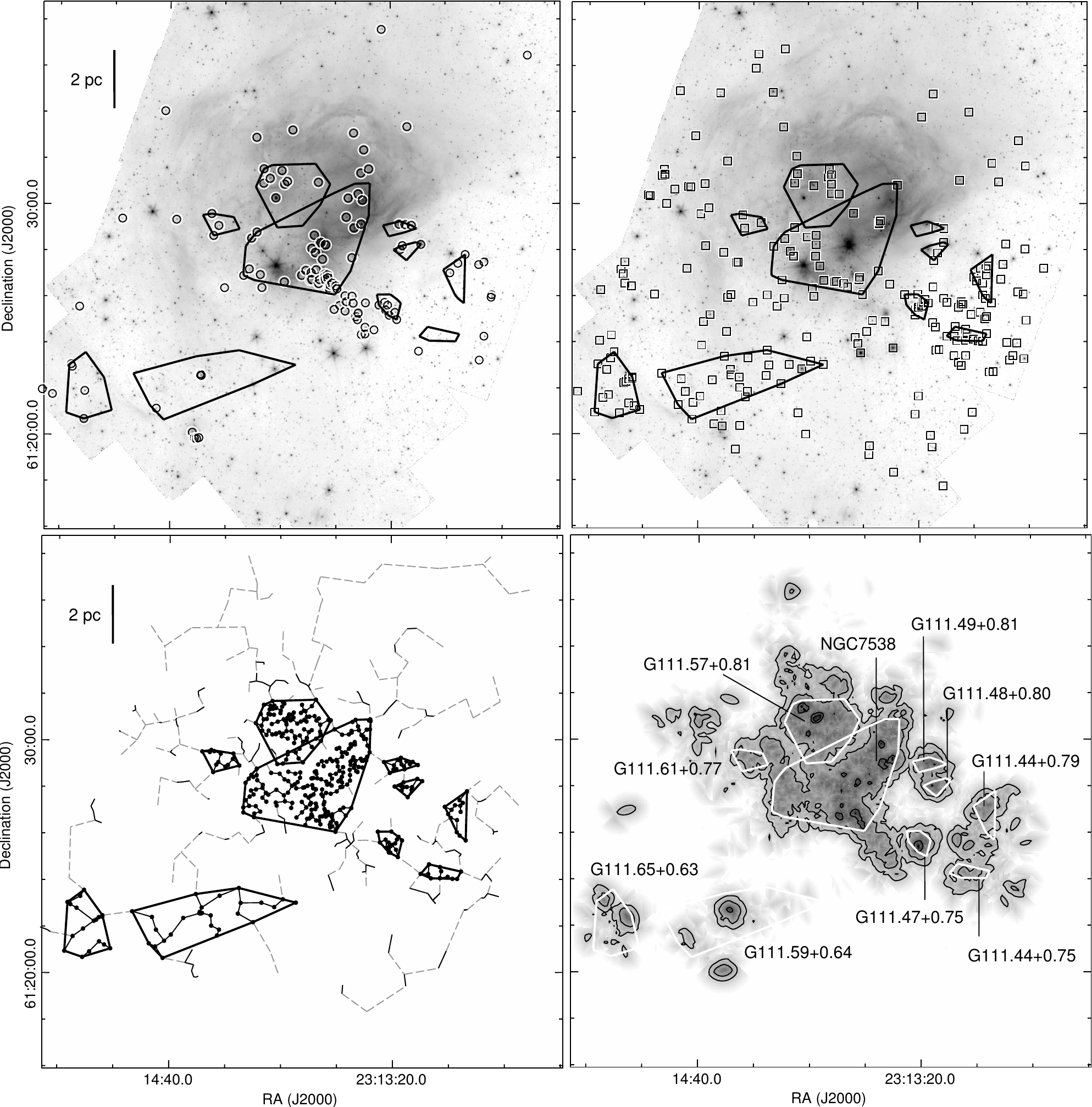}
\caption{Same as in Figure~\ref{afgl4029irac_classes} for region NGC7538.}
\label{ngc7538irac_classes}
\end{center}
\end{figure*}
%%%%%%%%%%%%%%%%%%%%%%%%%%%%%%%%%%%%%
%%%%%%%%%%%%%%%%%%%%%%%%%%%%%%%%%%%%%

\clearpage

%%%%%%%%%%%%%%%%%%%%%%%%%%%%%%%%%%%%%
%%%%%%%%%%%%%%%%%%%%%%%%%%%%%%%%%%%%%
\begin{figure*}
\begin{center}
\includegraphics[width=17cm]{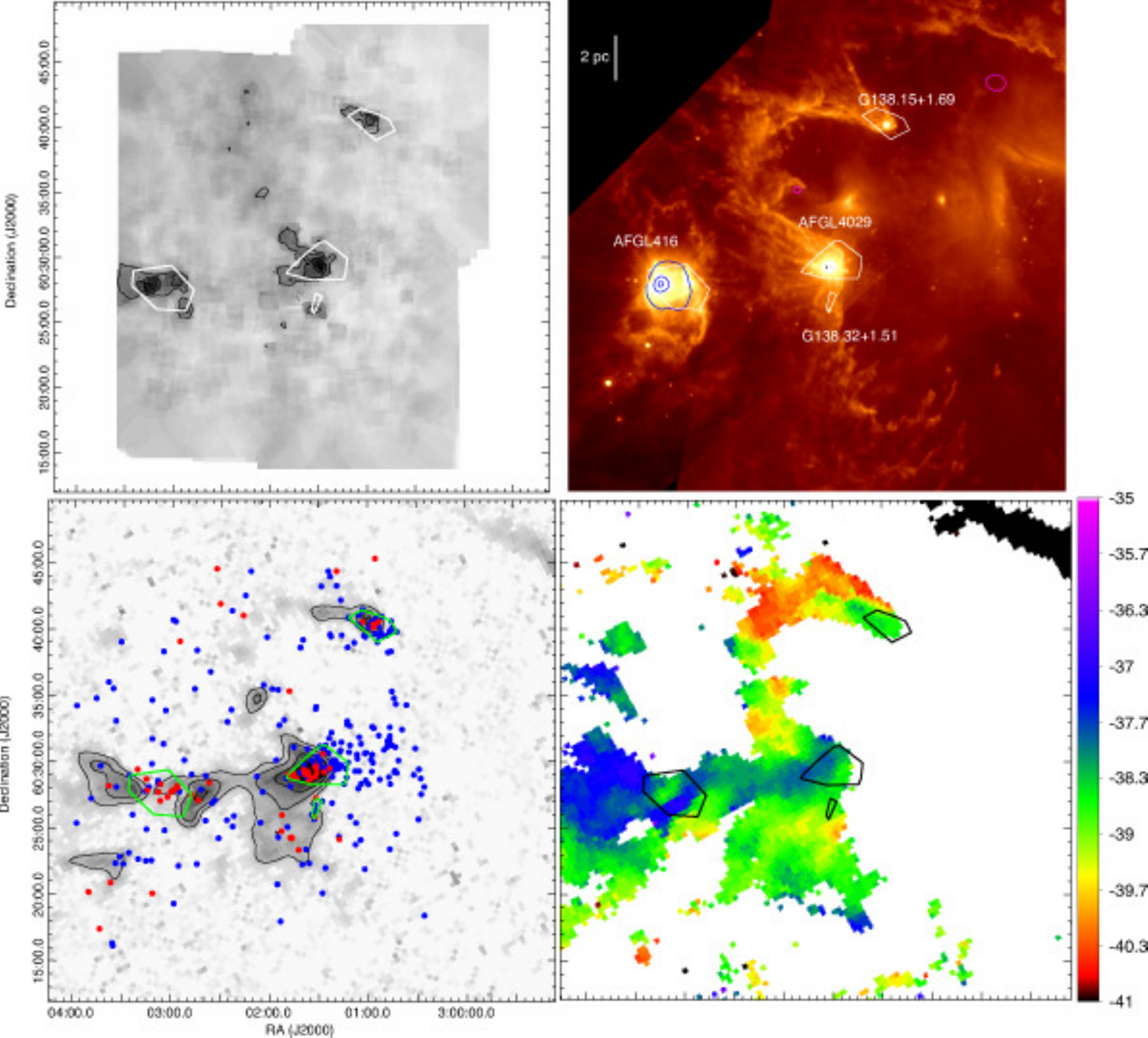}
\caption{Molecular material distribution in region W5-east. Clusters convex hull area are shown as white, green or black polygons. \textbf{Upper-left}: K-band extinction map. Black contours begin at $A_K=0.4$ (mag) and increase by 0.2. \textbf{Upper-right}: MIPS map at 24~$\mu$m. Blue and magenta contours correspond to the 1.4~GHz radio emission from VLA observations. Contours are at 10, 50 and 90\% of peak intensity. Blue contours correspond to emission associated with W5-east. Magenta contours show likely background emission. Cluster names are labeled in this panel. \textbf{Lower-left}: \cob~column density map. Black contours are spaced logarithmically between $0.6$ and $4\times 10^{16}$~cm$^{-2}$. Red dots correspond to Class I sources, blue dots correspond to Class II. \textbf{Lower-right}: \cob~first momentum map. The bar at the left shows the color-coded \vlsr~in \kms.}
\label{w5_molecular}
\end{center}
\end{figure*}

\begin{figure*}
\begin{center}
\includegraphics[width=14cm]{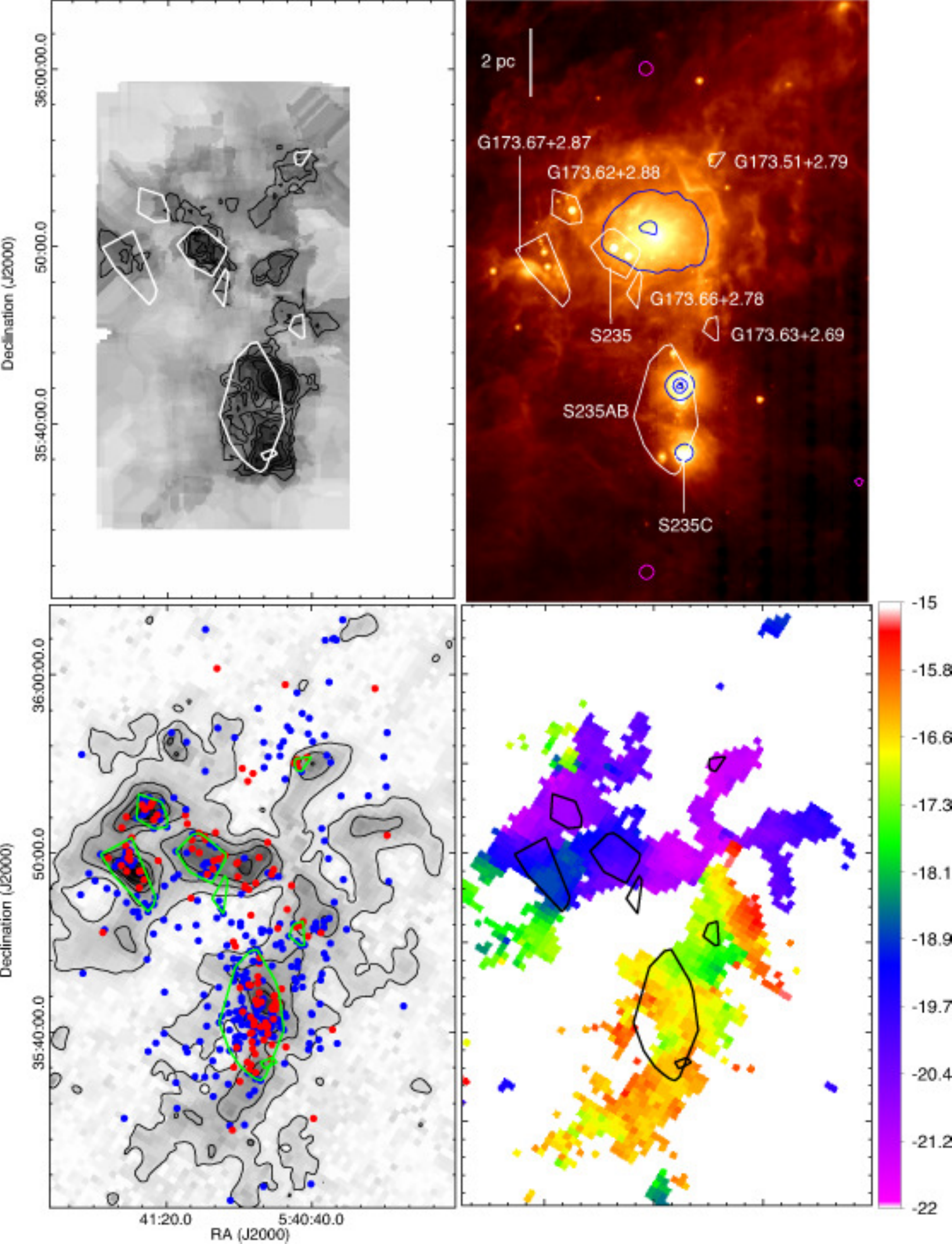}
\caption{Same as Figure~\ref{w5_molecular} for region S235. Contours in the K-band extinction map (upper-right) begin at $A_K=0.8$ (mag) and increase by 0.2. \cob~column density contours (lower-left) are logarithmically spaced between 0.6 and $4.2\times 10^{16}$~cm$^{-2}$.}
\label{s235_molecular}
\end{center}
\end{figure*}

\begin{figure*}
\begin{center}
\includegraphics[width=17cm]{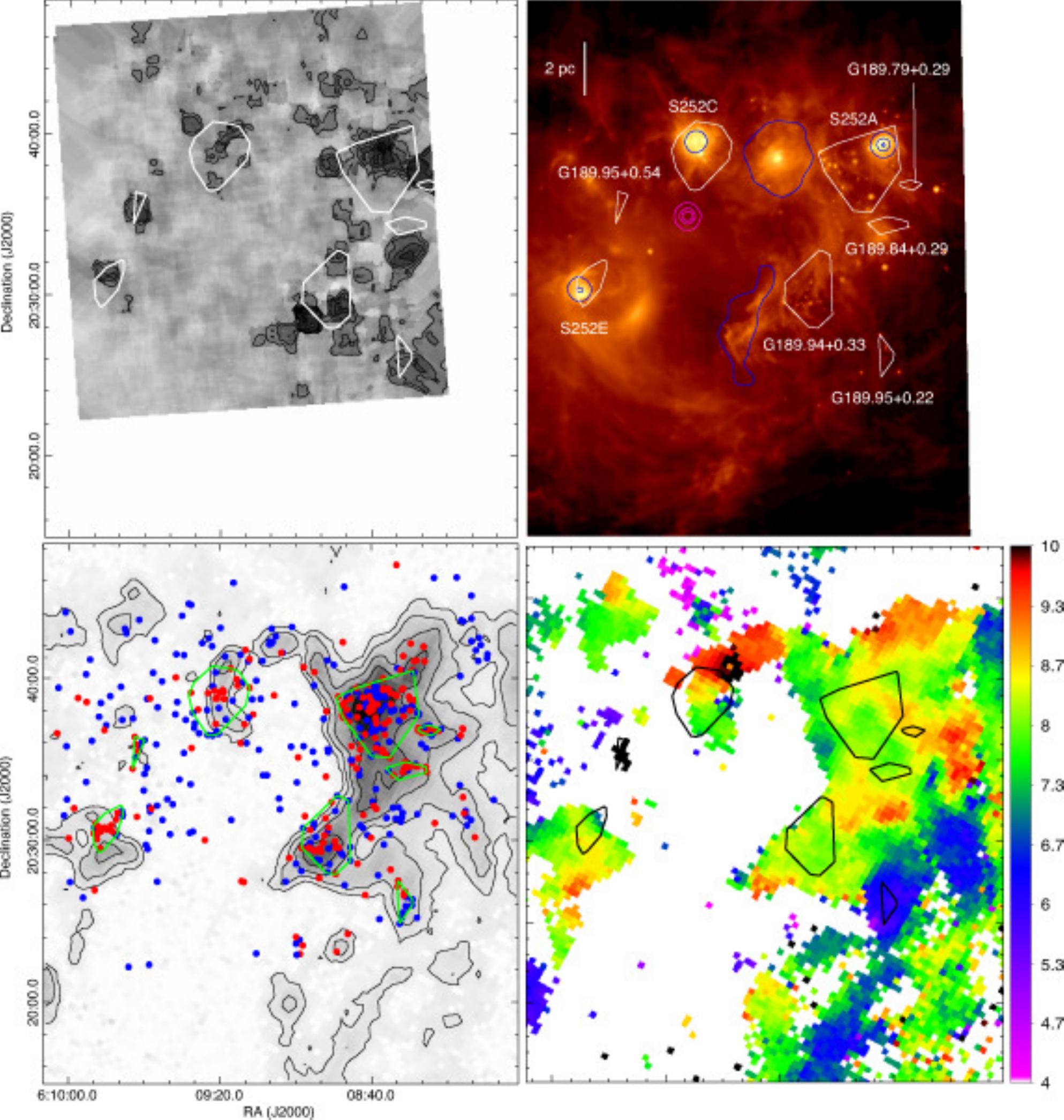}
\caption{Same as Figure~\ref{w5_molecular} for region S252. Contours in the K-band extinction map (upper-right) begin at $A_K=0.6$ (mag) and increase by 0.2. \cob~column density contours (lower-left) are logarithmically spaced between 0.6 and $6\times 10^{16}$~cm$^{-2}$.}
\label{s252_molecular}
\end{center}
\end{figure*}

\begin{figure*}
\begin{center}
\includegraphics[width=17cm]{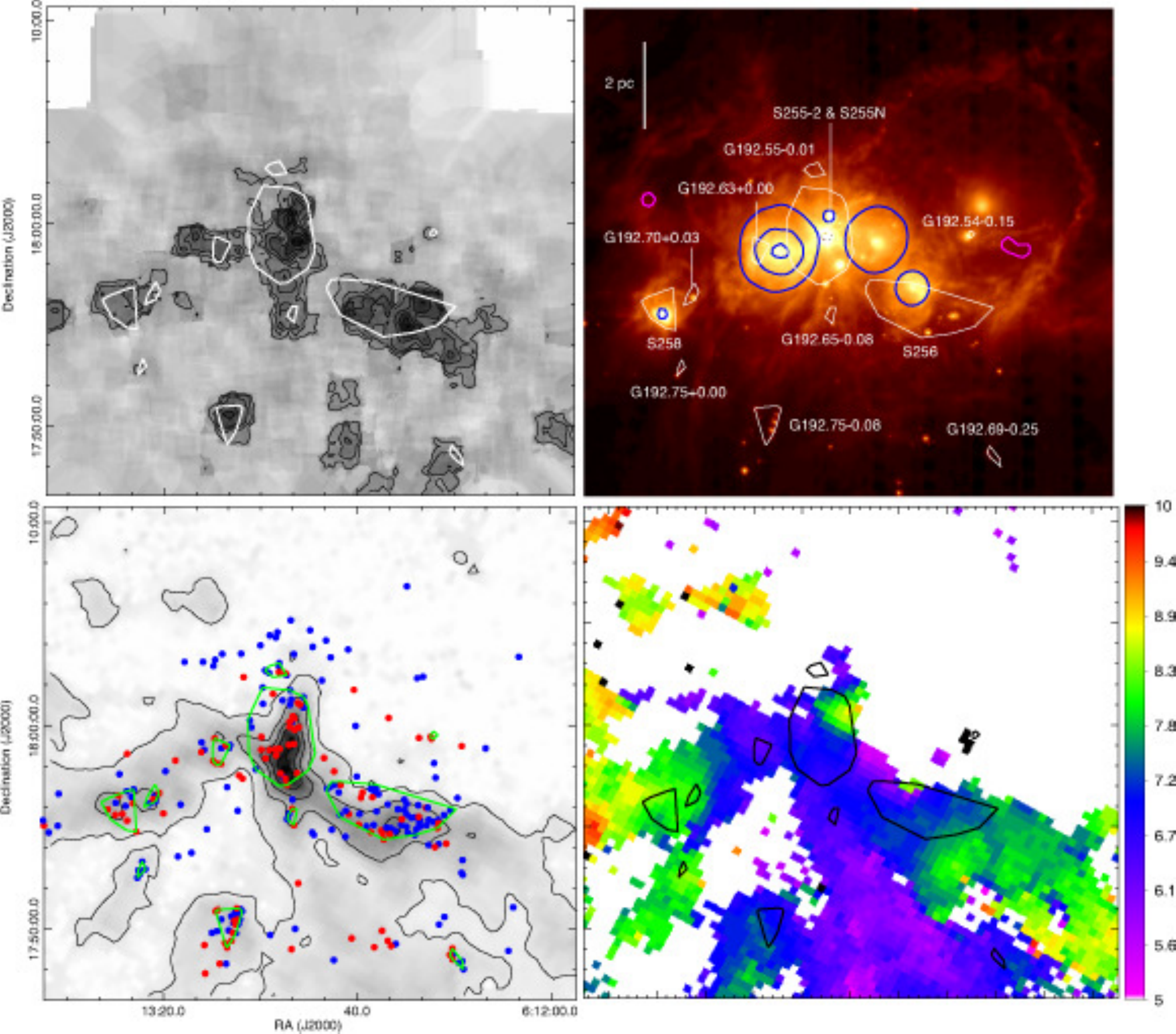}
\caption{Same as Figure~\ref{w5_molecular} for region S254-S258. Contours in the K-band extinction map (upper-right) begin at $A_K=1.0$ (mag) and increase by 0.2. \cob~column density contours (lower-left) are logarithmically spaced between 0.6 and $6\times 10^{16}$~cm$^{-2}$.}
\label{s255_molecular}
\end{center}
\end{figure*}

\begin{figure*}
\begin{center}
\includegraphics[width=17cm]{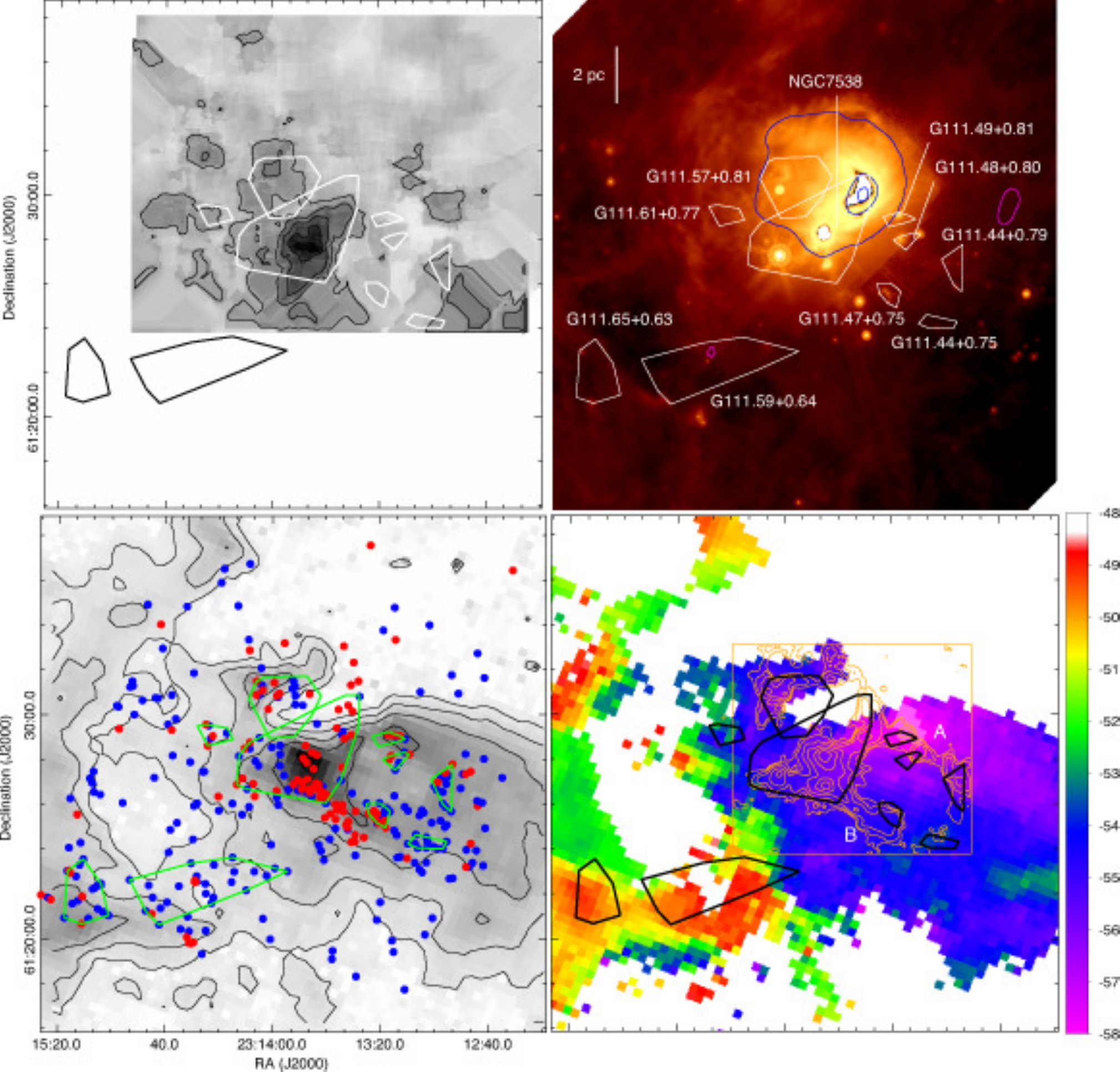}
\caption{Same as Figure~\ref{w5_molecular} for region NGC7538. The contours in the K-band extinction map (upper-right) begin at $A_K=0.6$ (mag) and increase by 0.2.  \cob~column density contours (lower-left) are logarithmically spaced between 0.6 and $6\times 10^{16}$~cm$^{-2}$. The \cob~first momentum map (lower-right) includes the SCUBA 850 microns emission in orange contours (spaced logarithmically between 0.063 and 18 Jy). The orange box shows the SCUBA observed FOV. Filaments A and B are also labeled (see NGC7538 in \S~\ref{section_NGC7538}).}
\label{ngc7538_molecular}
\end{center}
\end{figure*}
%%%%%%%%%%%%%%%%%%%%%%%%%%%%%%%%%%%%%
%%%%%%%%%%%%%%%%%%%%%%%%%%%%%%%%%%%%%
\clearpage

%%%%%%%%%%%%%%%%%%%%%%%%%%%%%%%%%%%%%%%%%%%%%%%%
%%%%%%%%%%%%%%%%%%%%%%%%%%%%%%%%%%%%%%%%%%%%%%%%
\begin{landscape}
\begin{table}
\begin{minipage}{150mm}
\caption{Properties of embedded clusters \protect\footnote{See a detailed description of the columns at the end of \S~4.2.1.} }
\label{clusters}
\begin{tabular}{lccccccrcccccccccc} 
\hline
Name & RA & Dec & $N_{\mathrm{YSO}}$ & I & II & I/II & $\widetilde{\alpha}_{\mathrm{IRAC}}$ & $R_{\mathrm{C}}$ & $R_{\mathrm{H}}$ & AR & $\overline{\Sigma}_{\mathrm{YSO}}$ & $\Sigma_{\mathrm{YSO}}$ & $d_c$ & $s_{\mathrm{YSO}}$ & \Q$_{\mathrm{YSO}}$ & $\delta$\Q & $N\textrm{*}_{\mathrm{YSO}}$\\
& (J2000)& (J2000)& & & & & & (pc) & (pc) & & (pc$^{-2}$) & (pc$^{-2}$) & (pc) & (pc) & & &\\
\hline
                  %       RA                  DEC          N       I     II      II/I       alpha   Rc     Rh      AR  meanS  peakS  Dcritic  m      Q       Qerr  N_corrected                                                            
G138.15+1.69 &  03:00:55.5 &  60:40:18.7 & 	42 & 6 & 23 & 0.26 & -0.98 & 1.10 & 0.76 & 2.12 & 28 & 591 & 0.30 & 0.140 & 0.82 & 0.01 & 71 \\ 
AFGL4029 &  03:01:30.6 &  60:29:31.4 & 	114 & 23 & 43 & 0.53 & -0.67 & 1.40 & 1.03 & 1.86 & 42 & 708 & 0.32 & 0.104 & 0.73 & 0.05 & 206 \\ 
G138.32+1.51 &  03:01:32.3 &  60:26:20.2 & 	11 & 2 & 6 & 0.33 & -1.04 & 0.46 & 0.33 & 1.97 & 33 & 335 & 0.32 & 0.142 & 0.86 & 0.01 & 17 \\ 
AFGL416 &  03:03:08.8 &  60:27:52.6 & 	83 & 12 & 6 & 2.00 & 0.05 & 1.48 & 1.17 & 1.60 & 22 & 220 & 0.35 & 0.146 & 0.82 & 0.02 & 143 \\ 
\\
G173.51+2.79 &  05:40:43.5 &  35:55:02.3 & 	10 & 5 & 2 & 2.50 & 0.57 & 0.29 & 0.31 & 0.88 & 45 & 150 & 0.37 & 0.116 & 0.91 & 0.13 & 24 \\ 
G173.63+2.69 &  05:40:43.7 &  35:45:32.7 & 	11 & 4 & 5 & 0.80 & -0.35 & 0.35 & 0.35 & 1.01 & 25 & 72 & 0.23 & 0.133 & 0.77 & 0.03 & 11 \\ 
S235C &  05:40:52.0 &  35:38:18.1 & 	14 & 1 & 0 & 0.00 & 0.00 & 0.23 & 0.20 & 1.41 & 114 & 327 & 0.23 & 0.077 & 0.91 & 0.20 & 44 \\ 
S235AB &  05:40:54.6 &  35:41:16.6 & 	226 & 41 & 79 & 0.52 & -0.44 & 1.92 & 1.27 & 2.28 & 58 & 950 & 0.23 & 0.086 & 0.70 & 0.02 & 373 \\ 
G173.66+2.78 &  05:41:05.0 &  35:47:34.8 & 	10 & 0 & 3 & 0.00 & -1.15 & 0.56 & 0.39 & 2.02 & 18 & 40 & 0.28 & 0.187 & 0.86 & 0.07 & 12 \\ 
S235 &  05:41:08.6 &  35:49:40.2 & 	58 & 10 & 12 & 0.83 & -0.62 & 0.77 & 0.69 & 1.24 & 45 & 330 & 0.28 & 0.108 & 0.85 & 0.06 & 104 \\ 
G173.62+2.88 (E2)&  05:41:24.5 &  35:52:23.1 & 	33 & 9 & 14 & 0.64 & -0.52 & 0.64 & 0.51 & 1.56 & 44 & 150 & 0.28 & 0.115 & 0.85 & 0.02 & 57 \\ 
G173.67+2.87 (E1)&  05:41:29.8 &  35:49:06.2 & 	61 & 14 & 18 & 0.78 & -0.45 & 1.11 & 0.74 & 2.24 & 40 & 444 & 0.28 & 0.116 & 0.71 & 0.05 & 83 \\ 
\\
G189.79+0.29 &  06:08:24.9 &  20:36:46.5 & 	13 &  8 &  5 &  1.60 &  0.03 &  0.43 &  0.37 &  1.37 &  33 &  70 &  0.31 &  0.132 &  0.85 &  0.01 & -\\
G189.84+0.29 &  06:08:31.0 &  20:34:23.4 & 	14 & 9 & 3 & 3.00 & 0.18 & 0.78 & 0.63 & 1.51 & 15 & 72 & 0.41 & 0.195 & 0.78 & 0.01 & 21 \\ 
G189.95+0.22 &  06:08:31.4 &  20:26:12.4 & 	14 & 5 & 8 & 0.63 & -0.83 & 0.80 & 0.50 & 2.61 & 16 & 37 & 0.41 & 0.184 & 0.76 & 0.01 & 24 \\ 
S252A &  06:08:37.5 &  20:38:13.0 & 	174 & 54 & 52 & 1.04 & -0.10 & 1.79 & 1.50 & 1.43 & 29 & 535 & 0.31 & 0.122 & 0.72 & 0.02 & 700 \\ 
G189.94+0.33 &  06:08:52.5 &  20:29:56.3 & 	48 & 18 & 22 & 0.82 & -0.23 & 1.48 & 1.21 & 1.50 & 12 & 113 & 0.41 & 0.186 & 0.71 & 0.05 & 49 \\ 
S252C &  06:09:20.6 &  20:38:55.2 & 	109 & 14 & 12 & 1.17 & 0.04 & 1.35 & 1.20 & 1.25 & 30 & 508 & 0.32 & 0.128 & 0.91 & 0.01 & 173 \\ 
G189.95+0.54 &  06:09:41.8 &  20:35:32.5 & 	11 & 4 & 5 & 0.80 & -0.05 & 0.60 & 0.32 & 3.53 & 21 & 51 & 0.32 & 0.171 & 0.80 & 0.01 & 16 \\ 
S252E &  06:09:50.6 &  20:30:40.0 & 	50 & 14 & 7 & 2.00 & 0.36 & 0.98 & 0.69 & 2.03 & 37 & 278 & 0.32 & 0.118 & 0.77 & 0.02 & 78 \\ 
\\
G192.69--0.25 &  06:12:19.6 &  17:48:31.5 & 	7 & 3 & 4 & 0.75 & -0.30 & 0.27 & 0.25 & 1.17 & 30 & 61 & 0.25 & 0.127 & 1.01 & 0.01 & 6 \\ 
G192.54--0.15 &  06:12:24.3 &  17:59:29.4 & 	8 & 2 & 0 & 0.00 & 1.61 & 0.49 & 0.80 & 0.37 & 126 & 334 & 1.18 & 0.056 & 1.10 & 0.01 & 19 \\ 
S256 &  06:12:35.0 &  17:56:54.8 & 	111 & 16 & 42 & 0.38 & -0.88 & 1.49 & 0.92 & 2.64 & 46 & 530 & 0.21 & 0.103 & 0.64 & 0.02 & 154 \\ 
G192.65--0.08 &  06:12:53.8 &  17:55:33.0 & 	7 & 2 & 3 & 0.67 & -0.37 & 0.19 & 0.34 & 0.33 & 44 & 87 & 0.21 & 0.099 & 0.95 & 0.02 & 11 \\ 
S255-2 \& S255N &  06:12:55.3 &  17:59:20.1 & 	142 & 21 & 13 & 1.62 & 0.26 & 1.15 & 0.97 & 1.42 & 57 & 664 & 0.21 & 0.093 & 0.81 & 0.02 & 251 \\ 
G192.55--0.01 &  06:12:56.8 &  18:02:41.8 & 	9 & 2 & 5 & 0.40 & -0.77 & 0.25 & 0.30 & 0.69 & 35 & 61 & 0.21 & 0.122 & 0.97 & 0.01 & 16 \\ 
G192.75--0.08 &  06:13:05.7 &  17:50:20.0 & 	27 & 10 & 15 & 0.67 & -0.51 & 0.47 & 0.40 & 1.35 & 66 & 538 & 0.25 & 0.089 & 0.80 & 0.03 & 50 \\ 
G192.63+0.00 &  06:13:08.6 &  17:58:39.1 & 	13 & 5 & 3 & 1.67 & 0.56 & 0.31 & 0.36 & 0.72 & 38 & 88 & 0.21 & 0.119 & 0.90 & 0.01 & 16 \\ 
G192.70+0.03 &  06:13:22.4 &  17:56:26.9 & 	8 & 2 & 3 & 0.67 & -0.10 & 0.29 & 0.26 & 1.23 & 44 & 129 & 0.22 & 0.133 & 1.10 & 0.01 & 8 \\ 
G192.75+0.00 &  06:13:24.8 &  17:52:51.1 & 	7 & 0 & 7 & 0.00 & -0.91 & 0.19 & 0.15 & 1.71 & 93 & 177 & 0.22 & 0.090 & 1.11 & 0.01 & 7 \\ 
S258 &  06:13:28.9 &  17:55:47.2 & 	43 & 7 & 5 & 1.40 & -0.13 & 0.52 & 0.44 & 1.42 & 80 & 646 & 0.22 & 0.088 & 0.88 & 0.01 & 83 \\ 
\\
G111.44+0.79 &  23:12:56.5 &  61:26:56.1 & 	14 & 3 & 10 & 0.30 & -0.53 & 0.83 & 0.62 & 1.80 & 12 & 41 & 0.43 & 0.221 & 0.83 & 0.17 & 28 \\ 
G111.44+0.75 &  23:13:02.5 &  61:24:15.4 & 	11 & 0 & 11 & 0.00 & -1.05 & 0.68 & 0.50 & 1.83 & 10 & 24 & 0.43 & 0.232 & 0.89 & 0.23 & 28 \\ 
G111.48+0.80 &  23:13:14.4 &  61:28:00.4 & 	11 & 3 & 3 & 1.00 & -0.59 & 0.47 & 0.40 & 1.41 & 23 & 71 & 0.43 & 0.177 & 1.01 & 0.16 & 73 \\ 
G111.49+0.81 &  23:13:16.7 &  61:28:57.4 & 	15 & 3 & 2 & 1.50 & 0.76 & 0.57 & 0.42 & 1.88 & 27 & 64 & 0.43 & 0.157 & 0.94 & 0.09 & 100 \\ 
G111.47+0.75 &  23:13:20.1 &  61:25:30.6 & 	18 & 7 & 11 & 0.64 & -0.17 & 0.54 & 0.48 & 1.29 & 32 & 131 & 0.43 & 0.153 & 1.03 & 0.37 & 53 \\ 
NGC7538 &  23:13:45.9 &  61:28:12.9 & 	192 & 36 & 22 & 1.64 & 0.57 & 2.65 & 1.88 & 1.99 & 21 & 166 & 0.32 & 0.146 & 0.64 & 0.08 & 1227 \\ 
G111.57+0.81 &  23:13:56.3 &  61:30:39.1 & 	90 & 8 & 9 & 0.89 & -0.23 & 1.27 & 1.18 & 1.16 & 24 & 192 & 0.32 & 0.149 & 0.82 & 0.10 & 321 \\ 
G111.61+0.77 &  23:14:21.7 &  61:29:12.2 & 	13 & 3 & 3 & 1.00 & -0.06 & 0.67 & 0.55 & 1.47 & 12 & 28 & 0.32 & 0.209 & 0.87 & 0.15 & 27 \\ 
G111.59+0.64 &  23:14:26.5 &  61:22:23.2 & 	28 &  4 &  24 &  0.17 &  -1.11 &  2.77 &  1.67 &  2.73 &  4 &  135 &  0.76 &  0.411 &  0.70 &  0.03 & -\\
G111.65+0.63 &  23:15:09.6 &  61:22:06.7 & 	19 &  4 &  15 &  0.27 &  -0.77 &  1.13 &  1.06 &  1.14 &  6 &  49 &  0.76 &  0.304 &  0.77 &  0.01 & -\\
%\tablenotetext{a}{See a detailed description of the columns at the end of \S~\ref{section_distribution}.}
\hline
\end{tabular}
\end{minipage}
\end{table}
\end{landscape}
%%%%%%%%%%%%%%%%%%%%%%%%%%%%%%%%%%%%%%%%%%%%%%%%
%%%%%%%%%%%%%%%%%%%%%%%%%%%%%%%%%%%%%%%%%%%%%%%%

%%%%%%%%%%%%%%%%%%%%%%%%%%%%%%%%%%%%%%%%%%%%%%%%
%%%%%%%%%%%%%%%%%%%%%%%%%%%%%%%%%%%%%%%%%%%%%%%%
%\begin{table*}
%\caption{}
%\label{clusters}
%\vbox to220mm{\vfil Landscape table 3 goes here
%\vfil}
%\end{table*}
%%%%%%%%%%%%%%%%%%%%%%%%%%%%%%%%%%%%%%%%%%%%%%%%
%%%%%%%%%%%%%%%%%%%%%%%%%%%%%%%%%%%%%%%%%%%%%%%%

%%%%%%%%%%%%%%%%%%%%%%%%%%%%%%%%%%%%%
%%%%%%%%%%%%%%%%%%%%%%%%%%%%%%%%%%%%%
\begin{table*}
\caption{Properties of molecular material associated with embedded clusters  }
\label{cloud}
\centering
\begin{tabular}{lccccccccc} 
\hline
Name & $\overline{A}_K$ & $A_K$ & $M_{A_K}$ & $M_{0.8}$ & $\overline{\sigma}_{\mathrm{CO}}$ & $\sigma_{\mathrm{CO}}$ & $M_{\sigma_{\mathrm{CO}}}$ & \vlsr & \cob$_\mathrm{FWHM}$ \\
& (mag) & (mag) & (\msun) & (\msun) & ($10^{16}$ cm$^{-2}$) & ($10^{16}$ cm$^{-2}$) & (\msun) & (\kms) & (\kms)\\
\hline
                               %      meanA peakA massAk meanCO peakCO massCO eff     vlsr                                                                
G138.15+1.69 & 0.71 & 2.33 & 209 & 46 & 0.95 & 2.24 & 131 & -38.5 & 1.65 \\ 
AFGL4029 & 0.83 & 2.61 & 477 & 143 & 1.60 & 3.75 & 435 & -38.3 & 1.81 \\ 
G138.32+1.51 & 0.52 & 1.38 & 19 & 2 & 0.62 & 1.06 & 11 & -38.7 & 1.74 \\ 
AFGL416 & 0.88 & 2.24 & 665 & 155 & 0.68 & 1.78 & 242 & -37.7 & 2.26 \\ 
\\
G173.51+2.79 & 1.33 & 1.51 & 32 & 13 & 1.68 & 2.29 & 19 & -21.4 & 1.39 \\ 
G173.63+2.69 & 1.01 & 1.69 & 42 & 13 & 1.53 & 1.97 & 30 & -17.4 & 1.06 \\ 
S235C & 1.99 & 2.16 & 31 & 18 & 2.13 & 2.53 & 15 & -16.2 & 1.72 \\ 
S235AB & 1.43 & 2.37 & 1305 & 580 & 1.79 & 4.03 & 770 & -16.7 & 1.76 \\ 
G173.66+2.78 & 1.15 & 1.67 & 54 & 17 & 0.72 & 2.07 & 16 & -19.7 & 4.14 \\ 
S235 & 1.54 & 3.40 & 375 & 192 & 2.60 & 3.46 & 297 & -19.8 & 1.74 \\ 
G173.62+2.88 (E2)& 0.75 & 1.38 & 94 & 17 & 3.29 & 4.89 & 194 & -20.7 & 1.36 \\ 
G173.67+2.87 (E1)& 0.98 & 1.63 & 294 & 85 & 2.77 & 5.12 & 391 & -18.8 & 1.69 \\ 
\\
G189.79+0.29 & - & - & - & - & 2.80 & 3.26 & 53 & 10.0 & 1.93 \\ 
G189.84+0.29 & 0.51 & 1.22 & 63 & 10 & 3.29 & 4.03 & 191 & 8.8 & 2.82 \\ 
G189.95+0.22 & 1.25 & 1.79 & 122 & 46 & 1.56 & 1.84 & 71 & 5.8 & 1.88 \\ 
S252A & 0.87 & 2.15 & 1121 & 318 & 4.60 & 7.34 & 2788 & 8.3 & 3.57 \\ 
G189.94+0.33 & 0.94 & 2.37 & 623 & 222 & 2.49 & 4.82 & 778 & 8.7 & 2.16 \\ 
S252C & 0.66 & 2.39 & 540 & 83 & 0.89 & 2.37 & 338 & 8.7 & 2.94 \\ 
G189.95+0.54 & 1.00 & 1.64 & 47 & 17 & 0.43 & 0.61 & 10 & 9.5 & 1.27 \\ 
S252E & 0.78 & 1.86 & 183 & 36 & 1.01 & 2.45 & 112 & 8.5 & 1.72 \\ 
\\
G192.69--0.25 & 1.16 & 1.62 & 13 & 4 & 1.03 & 1.12 & 5 & 6.7 & 2.44 \\ 
G192.54--0.15 & 0.88 & 0.97 & 3 & 1 & 0.34 & 0.34 & 1 & 24.1 & 1.67 \\ 
S256 & 1.35 & 2.48 & 626 & 269 & 2.00 & 3.56 & 435 & 7.3 & 2.16 \\ 
G192.65--0.08 & 1.59 & 1.74 & 16 & 8 & 2.28 & 2.39 & 11 & 6.5 & 1.90 \\ 
S255-2 \& S255N & 1.24 & 3.08 & 640 & 264 & 2.79 & 7.02 & 677 & 6.6 & 2.14 \\ 
G192.55--0.01 & 0.68 & 0.85 & 13 & 1 & 0.19 & 0.40 & 2 & 5.9 & 2.19 \\ 
G192.75--0.08 & 1.57 & 2.10 & 97 & 49 & 1.47 & 1.87 & 42 & 7.1 & 1.97 \\ 
G192.63+0.00 & 1.43 & 1.61 & 44 & 19 & 1.73 & 1.95 & 25 & 6.5 & 1.57 \\ 
G192.70+0.03 & 0.83 & 1.71 & 13 & 1 & 1.96 & 2.11 & 14 & 8.0 & 1.39 \\ 
G192.75+0.00 & 0.91 & 1.92 & 5 & 1 & 0.81 & 0.92 & 2 & 8.2 & 1.81 \\ 
S258 & 1.14 & 1.92 & 108 & 35 & 1.82 & 2.50 & 81 & 8.2 & 1.72 \\ 
\\
G111.44+0.79 & 1.12 & 1.72 & 150 & 46 & 2.83 & 3.16 & 177 & -57.8 & 5.05 \\ 
G111.44+0.75 & 0.87 & 1.29 & 73 & 27 & 2.34 & 2.77 & 92 & -54.5 & 6.53 \\ 
G111.48+0.80 & 0.83 & 1.06 & 43 & 4 & 2.59 & 2.98 & 63 & -57.6 & 5.03 \\ 
G111.49+0.81 & 0.18 & 1.06 & 12 & 1 & 3.42 & 3.96 & 102 & -58.9 & 3.92 \\ 
G111.47+0.75 & 0.83 & 0.89 & 71 & 6 & 2.87 & 3.08 & 115 & -53.2 & 2.14 \\ 
NGC7538 & 1.27 & 2.97 & 2512 & 1211 & 2.82 & 6.82 & 2616 & -56.4 & 4.00 \\ 
G111.57+0.81 & 0.76 & 1.45 & 577 & 90 & 1.20 & 3.10 & 430 & -55.5 & 4.16 \\ 
G111.61+0.77 & 0.84 & 1.49 & 94 & 25 & 0.91 & 1.17 & 48 & -52.8 & 2.23 \\ 
G111.59+0.64 & - & - & - & - & 1.61 & 0.89 & 615 & -48.8 & 2.02 \\ 
G111.65+0.63 & - & - & - & - & 1.90 & 2.77 & 535 & -50.5 & 3.60 \\ 
\hline
\end{tabular}
\end{table*}
\clearpage

\section{Discussion}\label{section_discussion3}
%We complement the physical parameters of detected YSOs with the properties of the molecular material they are embedded in. This information is used to build a more comprehensive plot of the evolution of such star forming regions. 

%The studied star forming regions harbor embedded clusters spanning a wide range of evolutionary stages and masses. They are still part of their natal molecular cloud, hence they are an excellent laboratory to study clusters dynamic, to test star formation theories and to compare with synthetic cluster models.

%We compare our results with the synthetic cluster analysis from \citet{bon03,bon08,mas10}, and \citet{kru12}. Those hydrodynamic models are based on the turbulent fragmentation of a $10^3$ and $10^4$~\msun~sphere and cylinder respectively, forming clusters of stars in timescales of less than 1 Myr. The election of such models was based on their similarity in both spatial scale and star numbers with our observed regions. Similarly, the analysis performed by \citet{mas10} and \citet{kru12} are also based on MST cluster finding algorithm and the \Q~parameter. This allows a much better comparison with our observations.

\subsection{Spatial distribution of identified YSOs}\label{section_general}
In general, YSOs are distributed in groups surrounded by a more scattered population (see Figures~\ref{afgl4029irac_classes} to~\ref{ngc7538irac_classes}). The distribution parameter \Q~(see \S~\ref{section_Qparameter}) for all YSOs in each region has values in the hierarchical range between 0.64 and 0.81 (see Table~\ref{stars_detected}). This is not surprising, since each region contains several clusters resembling a fractal distribution. Synthetic cluster analysis from \citet{mas10} give also hierarchical \Q~values for a globally unbound $10^4$~\msun~cloud, which is comparable in mass and size to our regions.

%%%%%%%%%%%%%%%%%%%%%%%%%%%%%%%%%%%%%
%%%%%%%%%%%%%%%%%%%%%%%%%%%%%%%%%%%%%
\begin{figure}
\begin{center}
\includegraphics[width=7cm]{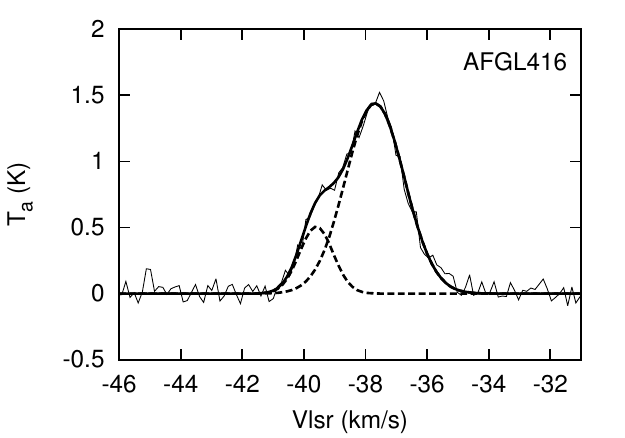}
\caption{Integrated \cob~spectra over the convex hull area for cluster AFGL416. The spectra is fitted (continuous black line) by combining Gaussian components (dashed lines). We assume that AFGL416 is associated with the component at \vlsr~$=-37.7$~\kms.}
\label{gauss_afgl416}
\end{center}
\end{figure}
%%%%%%%%%%%%%%%%%%%%%%%%%%%%%%%%%%%%%
%%%%%%%%%%%%%%%%%%%%%%%%%%%%%%%%%%%%%

%Simple visual inspection of Figures~\ref{afgl4029irac_classes} to~\ref{ngc7538irac_classes} shows that YSOs are distributed in clustered groups creating surface density enhancements over a more spread and low density layer. The location of those groups correlates well with the most dense molecular material on each region (see Fig.~\ref{w5_molecular} to \ref{ngc7538_molecular}). Indicating that those YSOs are still embedded in their natal molecular cloud. In addition, the presence of star formation signatures in those locations (see \S ~\ref{section_description3}), and a higher concentration of Class I sources than Class II, suggests that those groups are also young (with ages between 1-5 Myr) and that is in those groups where stars are mainly being formed. These YSO groups are efficiently selected by the MST algorithm as embedded clusters. However, an important percentage of the total YSOs is not consider part of a cluster, they are the so called isolated or scattered population.

\subsubsection{Scattered YSOs population}
%Those percentages change into 10 and 30\% when we search for clusters using a critical distance of $2\times d_c$.
The scattered or isolated population are YSOs not included in clusters. In our case, they range between 30 and 50\% of the detected YSOs per region (see Table~\ref{stars_detected}). This is in agreement with the 10 to 50\% expected depending on the adopted cluster definition \citep[][suggests that there is at least 10\% of scattered YSOs in massive star forming regions]{koe08}. 

The presence of a scattered population can be due to several reasons: dynamical interaction between cluster members, small groups merging, cluster definition and isolated star formation. In the first case, close encounters between cluster members may expel some at high velocities and place them far from its former companions. In the second case, the merging between small YSOs groups (with less members than the required number to be considered as a cluster) may dissipate cluster members forming a scattered halo whose members have a longer critical distance and will not be considered as part of the cluster \citep[numerical simulations on hierarchically formed star clusters gives sub-clusters merging time-scales of a few $10^5$ years, which is comparable to the observed clusters age,][]{mas10}. Third, the number of cluster members depends on the critical distance $d_c$. However, cluster dynamics and cluster merging creates a mixture of YSOs with different properties that may not be represented by a single $d_c$ value \citep[e.g.][showed that the percentage of cluster members for a $10^3$~\msun~simulation slowly decreases due to the merging of several small groups]{mas10}. In such a case, the used cluster finding algorithm may not identify properly all cluster members. Finally, the formation of isolated stars can not be ruled out as another alternative. \citet{bre12} and \citet{oey13} claim to have found several isolated young OB stars in 30 Doradus and the Small Magellanic Cloud respectively.

When we compare the scattered population in the studied regions per class, we find that it is mainly composed by Class II sources (67\%, see Table~\ref{stars_detected}). However, this difference decreases if we use a critical distance of $2\times d_c$ (53\% of Class II). If we assume that Class II sources are more evolved than Class I, the former are expected to be more dynamically relaxed and therefore more spread than the Class~I. As a consequence, some Class II sources located at the outer parts of the cluster are likely to have a projected distance larger than $d_c$ to the nearest cluster companion and will not be considered as cluster members. Following this argument, we estimate that between 10 and 20\% of the scattered population in the observed regions may correspond to dynamically evolved cluster members.

%To study the impact of $d_c$ over the member class percentage, we calculated the number of Class I and Class II sources in clusters using a critical distance of $2\times d_c$. We found than in average the percentage of Class II sources in clusters raised by 30\% while the percentage of Class I raised by 17\%. This suggest that at least part of the isolated population may correspond to Class II members located in the outer parts of clusters.

%The YSOs local surface density ($\Sigma_{\textrm{\tiny{YSO}}}$) is in average of the order of a few tens of stars per pc$^{-2}$ (see Table~\ref{stars_detected}), with a maximum of several hundreds of stars per pc$^-2$ (see Table~\ref{clusters}). The $\Sigma_{\textrm{\tiny{YSO}}}$ cummulative distribution function (see Fig.~\ref{CDF_regions}) shows that the percentage of sources with local surface density of more than 100 stars per pc$^{-2}$ is between 5 and 20\%. Those numbers are somehow lower than what is found in Orion by \citet{meg2005}

%%%%%%%%%%%%%%%%%%%%%%%%%%%%%%%%%%%%%
%%%%%%%%%%%%%%%%%%%%%%%%%%%%%%%%%%%%%
\begin{figure*}
\begin{center}
\includegraphics[width=12cm]{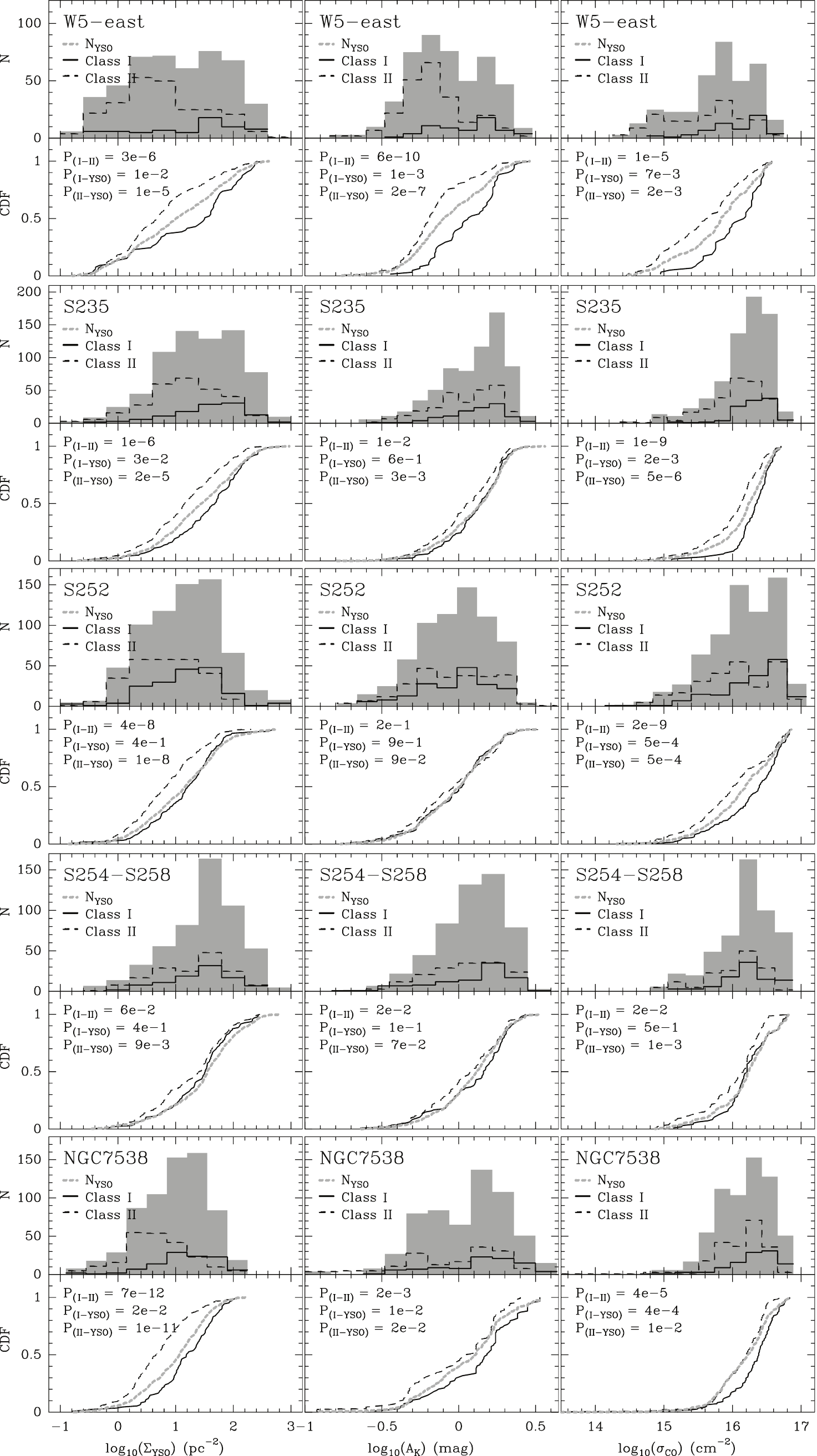}
\caption{YSOs distribution over molecular material per region. Each panel shows the distribution histogram and cumulative distribution function (CDF) of all detected YSOs ($N_{\textrm{\tiny{YSO}}}$, grey histogram and grey dashed line), Class I (continuous black line) and Class II (dashed black line) according to the surface density of YSOs (left), K-band extinction (center) and \cob~surface density (right). Kolmogorov-Sminrov similarity probabilities are shown for two data sets comparison.}
\label{CDF_regions}
\end{center}
\end{figure*}
%%%%%%%%%%%%%%%%%%%%%%%%%%%%%%%%%%%%%
%%%%%%%%%%%%%%%%%%%%%%%%%%%%%%%%%%%%%

\subsubsection{Class I vs. Class II sources}
In average, Class I sources are located towards places with higher extinction ($A_K$) and column density ($\sigma_{\textrm{\tiny{CO}}}$) than Class II sources (see Table~\ref{stars_detected} and Fig.~\ref{CDF_regions}). The histograms and cumulative distribution functions in Figure~\ref{CDF_regions} show that both Class I and Class II are distributed over a similar range of $A_K$, $\sigma_{\textrm{\tiny{CO}}}$ and $\Sigma_{\textrm{\tiny{YSO}}}$ values. However, Class I histograms are skewed towards higher values. This difference is better quantified by the K-S tests, which show that the probability that both classes have the same distribution is in general low (P$_{\mathrm{(I-II)}}\la 10^{-2}$, see Fig~\ref{CDF_regions}). In addition, we find that Class I sources are more close to each other than Class II (see $\overline{\Sigma}_{\textrm{\tiny{YSO, I}}}$ vs. $\overline{\Sigma}_{\textrm{\tiny{YSO, II}}}$ in Table~\ref{stars_detected}). These properties agree well with previous findings in region W5 \citep{koe08,deh12} and with the assumed evolutionary stage of both classes: the youngest Class I sources are more clustered and associated with the most dense molecular material in which they were born while the Class II sources are moving away from their birthplace due to dynamic interactions. Also, as shown in \S~\ref{section_molecular_material}, the embedded clusters in these regions are likely gravitationally unbound. This way, the more evolved cluster members will move further away due to the weak pulling from the cluster gravitational well.

\subsection{Physical properties of embedded clusters}\label{section_physical}
In this section we investigate the physical properties of identified embedded clusters and look for trends that will help us to better understand their formation process. Clusters G192.54$-$0.15 in region S254-S258, G111.59$+$0.64 and G111.65$+$0.63 in region NGC7538 and G189.79+0.29 in S252 will not be included in our analyzes. The former is likely located at a much further distance than the rest of clusters in the region \citep{cha08a}. The other three were not observed completely in the K-band, which reduces number of identified cluster members. We also assume that cluster S235C is part of cluster S235AB due to its location and similar \vlsr.

The 36 remaining clusters are divided in two sets: massive embedded clusters (MEC) and low-mass embedded clusters (LEC). MEC are clusters known to harbor massive stars while LEC show no evidence of containing massive stars, even though some of them may actually form massive stars in the future. Following this definition, 12 embedded clusters are MEC: AFGL4029, AFGL416, S235AB, S235, S252A, S252C, S252E, S256, S255-2, G192.63-0.00, S258, and NGC7538. The other 24 clusters are LEC.

%%%%%%%%%%%%%%%%%%%%%%%%%%%%%%%%%%%%%
%%%%%%%%%%%%%%%%%%%%%%%%%%%%%%%%%%%%%
\begin{figure}
\begin{center}
\includegraphics[width=8cm]{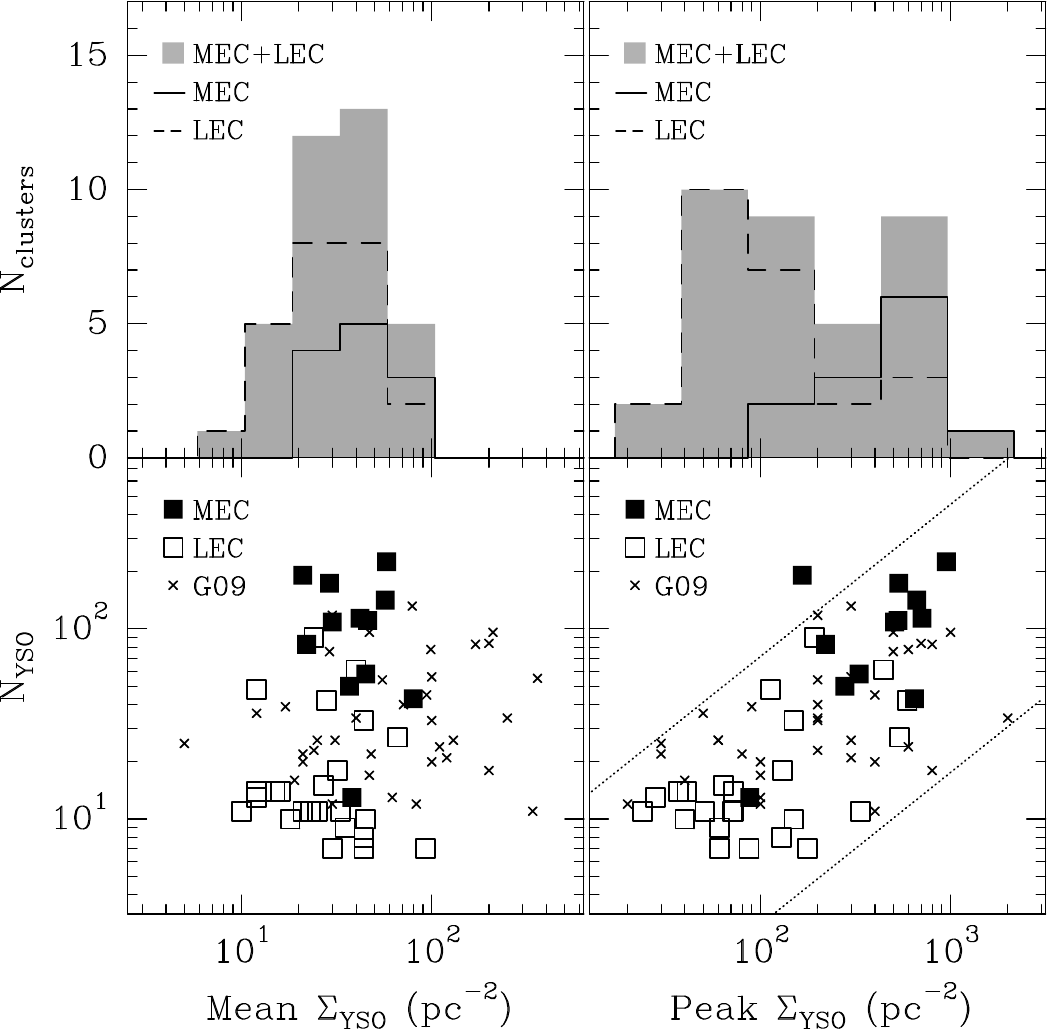}
\caption{Clusters mean (left) and peak (right) YSOs surface density. The histograms (upper charts) show the $\Sigma_{\mathrm{YSO}}$ distribution for all clusters (gray), MEC (continuous line) and LEC (segmented line). The YSOs surface densities vs. the number of cluster members are shown in the lower charts. The crosses show the values obtained by \citet{gut09} in low-mass embedded clusters. Dotted lines in the lower-right chart enclose most clusters (except cluster NGC7538) and show a weak correlation with index $0.8\pm 0.2$ and correlation index $r = 0.75$.}
%The segmented line shows the same correlation but for EC identified using a critical distance of $2\times d_c$ ($N_{\textrm{\tiny{I}}} \propto {\Sigma_{\textrm{\tiny{YSO}}}}^{0.83\pm 0.15}$, and $r = 0.75$)
\label{low-high_SD}
\end{center}
\end{figure}
%%%%%%%%%%%%%%%%%%%%%%%%%%%%%%%%%%%%%
%%%%%%%%%%%%%%%%%%%%%%%%%%%%%%%%%%%%%

\subsubsection{Spatial distribution of cluster members}
We characterize the distribution of members in embedded clusters by their surface density ($\Sigma_{\mathrm{YSO}}$). The YSOs in our cluster sample have mean surface densities between 10 and 100 pc$^{-2}$ with not significant difference between MEC and LEC (see Fig.~\ref{low-high_SD}). However, MEC show significantly higher peak surface densities than LEC. In addition, we find a weak trend between the peak surface density and the number of cluster members, with a correlation index $r=0.75$ and exponent of $0.8\pm 0.2$. Since the embedded clusters have their most agglomerated population concentrated in small areas compared to the cluster extent (which can be also seen in figures \ref{afgl4029irac_classes} to \ref{ngc7538irac_classes}, where the surface density contours of more than 50 stars pc$^{-2}$ enclose small areas compared to the cluster size), this correlation suggest that clusters are better characterized by their peak YSOs surface density.

%%%%%%%%%%%%%%%%%%%%%%%%%%%%%%%%%%%%%
%%%%%%%%%%%%%%%%%%%%%%%%%%%%%%%%%%%%%
\begin{figure}
\begin{center}
\includegraphics[width=7cm]{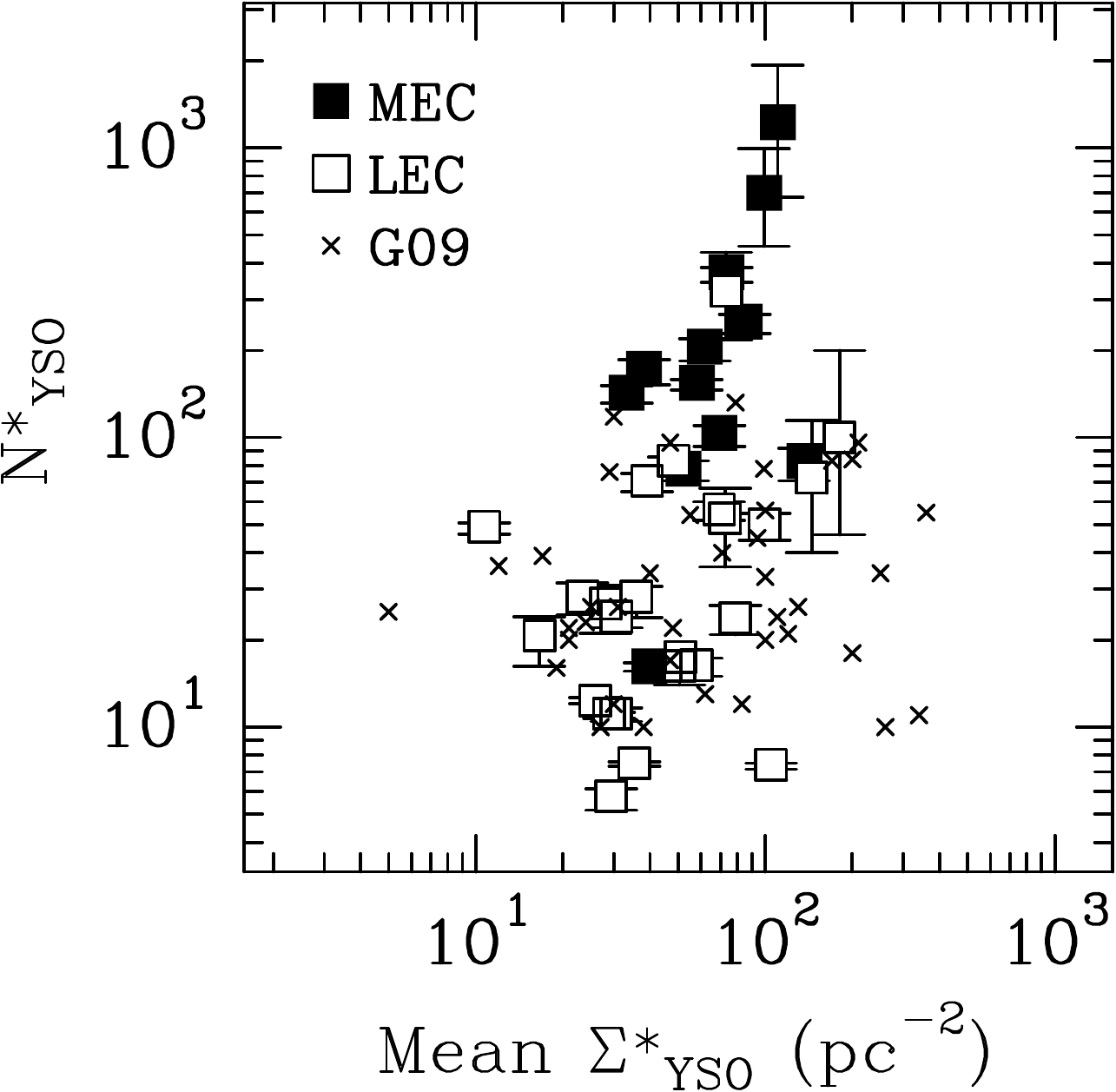}
\caption{Mean YSOs surface density per number of members derived from the corrected number of cluster members ${N^*}_{\mathrm{YSO}}$. Crosses show the mean surface densities from \citet{gut09}.}
\label{low_high_SD_corrected}
\end{center}
\end{figure}
%%%%%%%%%%%%%%%%%%%%%%%%%%%%%%%%%%%%%
%%%%%%%%%%%%%%%%%%%%%%%%%%%%%%%%%%%%%

Figure~\ref{low-high_SD} also shows the surface densities derived by \citet{gut09} in their low-mass EC sample. Their mean surface densities are somehow higher than our measurements. This may be a consequence of the shorter distance to their clusters, which allows the detection of fainter cluster members. We tested this hypothesis by estimating the mean $\Sigma_{\mathrm{YSO}}$ using the corrected number of cluster members ${N^*}_{\mathrm{YSO}}$ (see \S~\ref{N_corrected} and Table~\ref{clusters}). Figure~\ref{low_high_SD_corrected} shows that the surface densities from \citet{gut09} are in agreement with our corrected mean $\Sigma_{\mathrm{YSO}}$ values.

\begin{figure}
\begin{center}
\includegraphics[width=7cm]{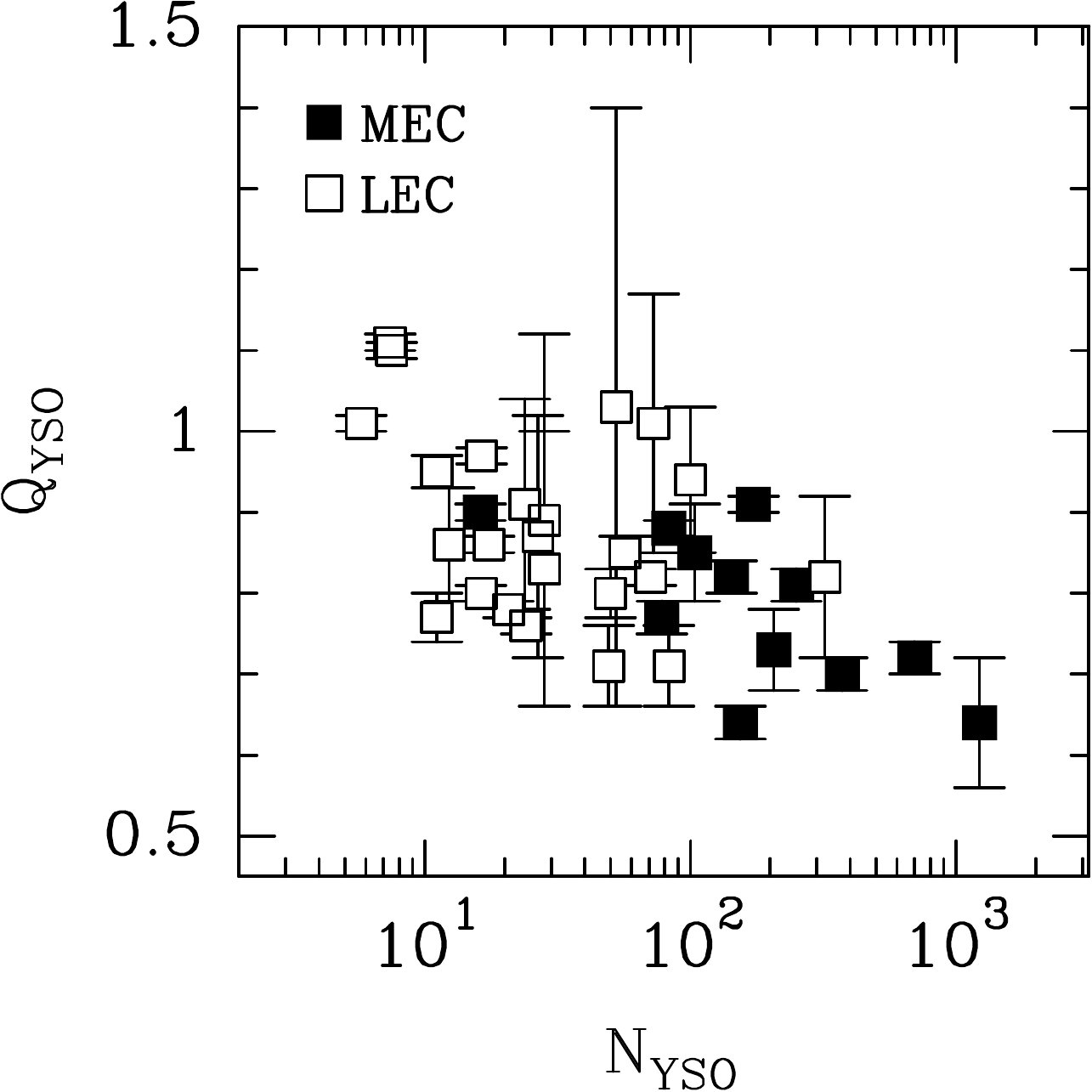}
\caption{Structural \Q~parameter (\Q$_{\mathrm{YSO}}$) for cluster members. Black squares correspond to MEC while empty squares correspond to LEC.}
\label{low-high_Q}
\end{center}
\end{figure}

The spatial distribution of cluster members is also investigated by their structural \Q~parameter value. We find that embedded clusters have an average \Q~value of $0.85\pm0.1$, ranging from 0.64 to 1.11. If we compare the \Q~values between MEC and LEC, we find that MEC have slightly lower values than LEC (with average of $0.78\pm0.1$ and $0.88\pm0.1$ respectively). This is also seen as a weak trend in the distribution of \Q~values per members number (see Fig.~\ref{low-high_Q}, where $N_{\textrm{\tiny{YSO}}} \propto {Q_{\textrm{\tiny{YSO}}}}^{-0.17\pm 0.05}$, and $r = -0.68$). Such a correlation may be due to a higher occurrence of sub-clusters merging in the most massive clusters caused by their higher potential well, which decreases the value of the \Q~parameter \citep[see models from][]{mas10}.

%%%%%%%%%%%%%%%%%%%%%%%%%%%%%%%%%%%%%
%%%%%%%%%%%%%%%%%%%%%%%%%%%%%%%%%%%%%
%%%%%%%%%%%%%%%%%%%%%%%%%%%%%%%%%%%%%
%%%%%%%%%%%%%%%%%%%%%%%%%%%%%%%%%%%%%

%%%%%%%%%%%%%%%%%%%%%%%%%%%%%%%%%%%%%
%%%%%%%%%%%%%%%%%%%%%%%%%%%%%%%%%%%%%
\begin{table}
\caption{Structural \Q~parameter for different YSOs classes in selected clusters.}
\label{Qparameter}
\centering
\begin{tabular}{lcc} 
\hline
Cluster name & \Q$_{\textrm{\tiny{I}}}$ & \Q$_{\textrm{\tiny{II}}}$ \\
\hline
AFGL4029                & 0.79 & 0.74  \\  
S235AB                    & 0.67 &  0.74  \\
S252A                      & 0.75 & 0.79  \\
S255-2 \& S255N    & 0.76 & 0.94  \\
NGC7538                  & 0.68 & 0.76  \\ 
\hline
\end{tabular}
\end{table}
%%%%%%%%%%%%%%%%%%%%%%%%%%%%%%%%%%%%%
%%%%%%%%%%%%%%%%%%%%%%%%%%%%%%%%%%%%%

We compare the \Q~parameter for Class I and Class II sources in the most numerous clusters in each region (see Table~\ref{Qparameter}). With the exception of cluster AFGL4029, all clusters have the Class I sources distributed more hierarchically than the Class II (\Q$_{\textrm{\tiny{I}}}<$ \Q$_{\textrm{\tiny{II}}}$). However, cluster AFGL4029 is a special case since it is located near a group of Class II sources that may be cluster members as well (see \S~\ref{section_w5east_clusters}). If we re-calculate the \Q~parameter for those clusters but increasing $d_c$ by 25\% (hence including the western Class II sources in AFGL4029), we find that the Class I sources are distributed more hierarchically than the Class II in the five clusters. Similar results have been also found in low-mass star forming regions \citep{sch08} and is likely a consequence of the cluster dynamic relaxation.

\subsubsection{Clusters morphology}
\begin{figure}
\begin{center}
\includegraphics[width=8cm]{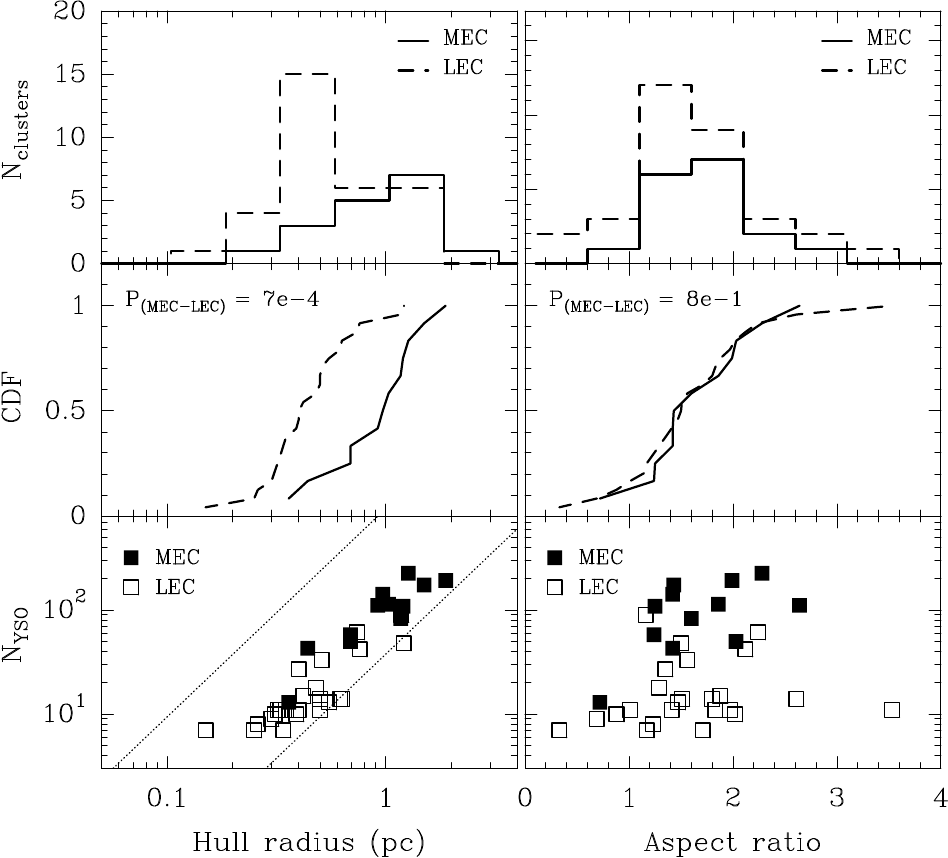}
\caption{Morphology of embedded clusters. \textbf{Left}: hull radius distribution. The upper chart shows the $R_H$ distribution for MEC and LEC. The middle chart shows the $R_H$ cumulative distribution function for MEC (continuous line) and LEC (dashed line). The K-S similarity test value is also shown. The lower chart shows the $R_H$ vs. the number of cluster members. The dot-dashed lines shows constant surface densities at 12 and 300 pc$^{-2}$. Those correspond to the range spanned by the embedded clusters from \citet{gut09}. \textbf{Right}: the same for the aspect ratio AR distribution.}
\label{low-high_structure}
\end{center}
\end{figure}

We use the clusters convex hull radius ($R_H$) and aspect ratio (AR) to investigate their morphology (see Fig.~\ref{low-high_structure}). Clusters $R_H$ values range between 0.1 and 2.0 parsecs with a mean value of $0.67\pm0.4$~pc. We find that MEC are more extended than LEC, with mean $R_H$ of $1.01\pm0.4$~pc and $0.50\pm0.3$~pc respectively. This difference agrees with a low K-S similarity test value of $\sim10^{-4}$. On the other hand, the clusters elongation, given by their aspect ratio (AR), range between 0.2 and 3.5 with a mean of $1.62\pm0.6$. This means that clusters are in general elongated (a circular cluster will have AR of unity). We find no significant differences between the LEC and MEC aspect ratios.

%AR averages of $1.59\pm0.7$ for LEC and $1.66\pm0.5$ for MEC.

%We compared our $R_H$ measurements with the values from \citep{gut09} and find that our results are spanned in a similar range as their "cores". The K-S test shows that their $R_H$ distribution is more similar to the LEC (see Fig.~\ref{low-high_structure}). Figure~\ref{low-high_structure} also shows a lack of members in our clusters (seen as a down shift of the LEC with respect to the G09 clusters in the lower-left chart from Fig.~\ref{low-high_structure}). This is likely due to the shorter distance to their clusters ($\lesssim$ 1~kpc), which allows them to detect fainter sources than in our case.

%The AR values from \citet{gut09} are in average slightly higher (AR~$=2.0$) than what we obtain. This may be also due to their higher detection limit. For clusters presenting some level of mass segregation \citep[previous studies suggest that mass segregation is present in clusters at very early formation stages, eg;][]{cha10,mas10}, faint and less massive members will be likely located in the outer parts of the cluster. Those may be not detected in our data making our clusters somehow less elongated.

%%%%%%%%%%%%%%%%%%%%%%%%%%%%%%%%%%%%%
%%%%%%%%%%%%%%%%%%%%%%%%%%%%%%%%%%%%%

\subsubsection{Associated molecular material}\label{section_molecular_material}
%\coa~and \cob~data have the advantage of showing material which is close in velocity to the cluster \vlsr, and hence unaffected by foreground emission. However, its spatial resolution ($\sim 22$ arc-seconds) allows to study structures with scale-size only bigger than 0.5~pc. Conversely, extinction maps have a much higher spatial resolution (only a few arc-seconds in the most crowded areas), but are affected by foreground interstellar extinction.

%%%%%%%%%%%%%%%%%%%%%%%%%%%%%%%%%%%%%
%%%%%%%%%%%%%%%%%%%%%%%%%%%%%%%%%%%%%
\begin{figure}
\begin{center}
\includegraphics[width=8cm]{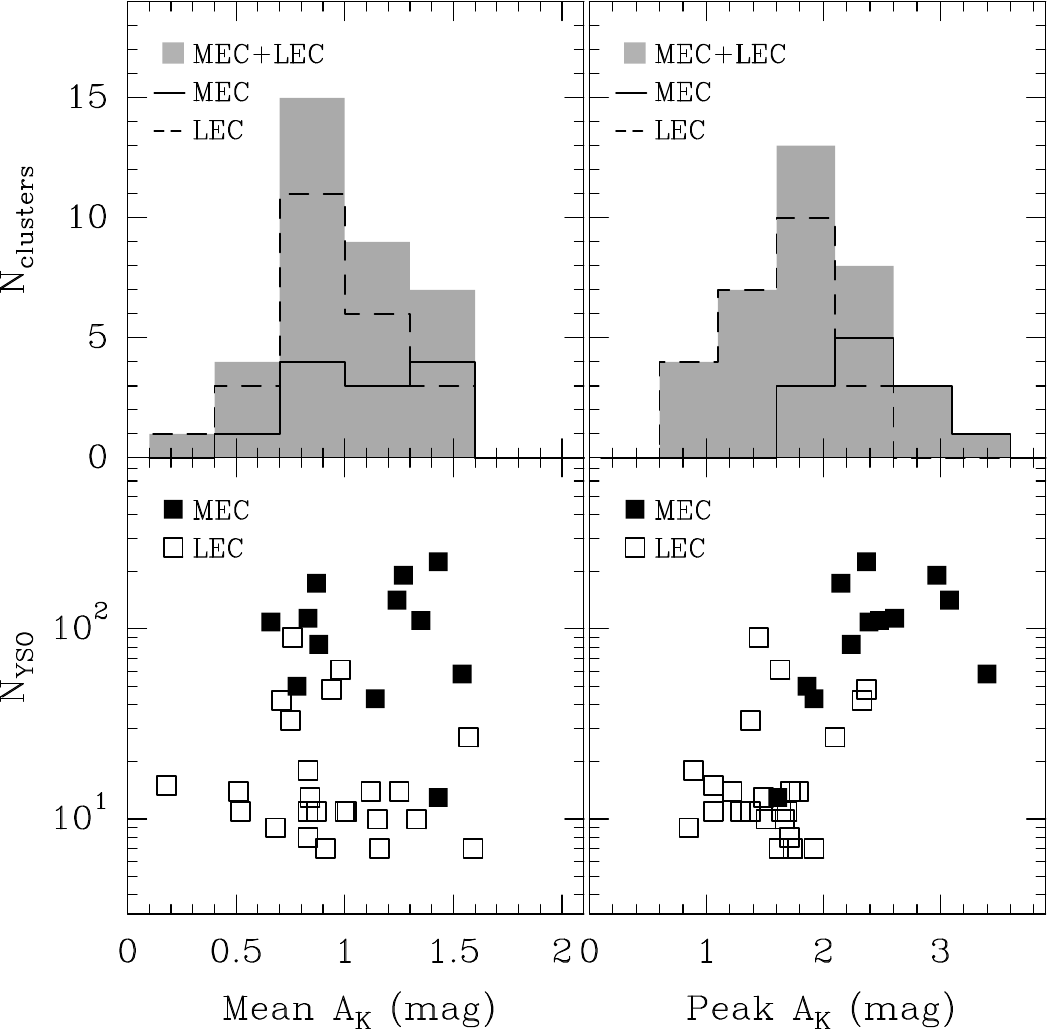}
\caption{Embedded clusters K-band extinction. \textbf{Left}: the upper charts shows the mean $A_K$ for LEC and MEC. The lower chart show the mean $A_K$ vs. the number of cluster members. \textbf{Right}: the same for the peak $A_K$ values.}
%The segmented line shows the same correlation but for EC identified using a critical distance of $2\times d_c$ ($A_{\textrm{\tiny{K}}} \propto {N_{\textrm{\tiny{YSO}}}}^{0.28\pm 0.05}$, and $r = 0.75$)
\label{low-high_Ak}
\end{center}
\end{figure}
%%%%%%%%%%%%%%%%%%%%%%%%%%%%%%%%%%%%%
%%%%%%%%%%%%%%%%%%%%%%%%%%%%%%%%%%%%%

We find that clusters have mean $A_K$ values (after subtracting the interstellar contribution) between 0.2 and 1.6 magnitudes with an average of $1.1\pm0.3$ (see Fig.~\ref{low-high_Ak}). We find no differences between the mean $A_K$ for MEC and LEC. However, the peak $A_K$ values are higher for MEC, with an average of $\sim2.5$ magnitudes versus $1.5$ magnitudes for LEC. As for the peak YSOs surface density, there is a weak correlation between the peak $A_K$ and the number of cluster members. %This correlation may be related with the higher potential well of the more numerous/massive clusters.

\begin{figure}
\begin{center}
\includegraphics[width=8cm]{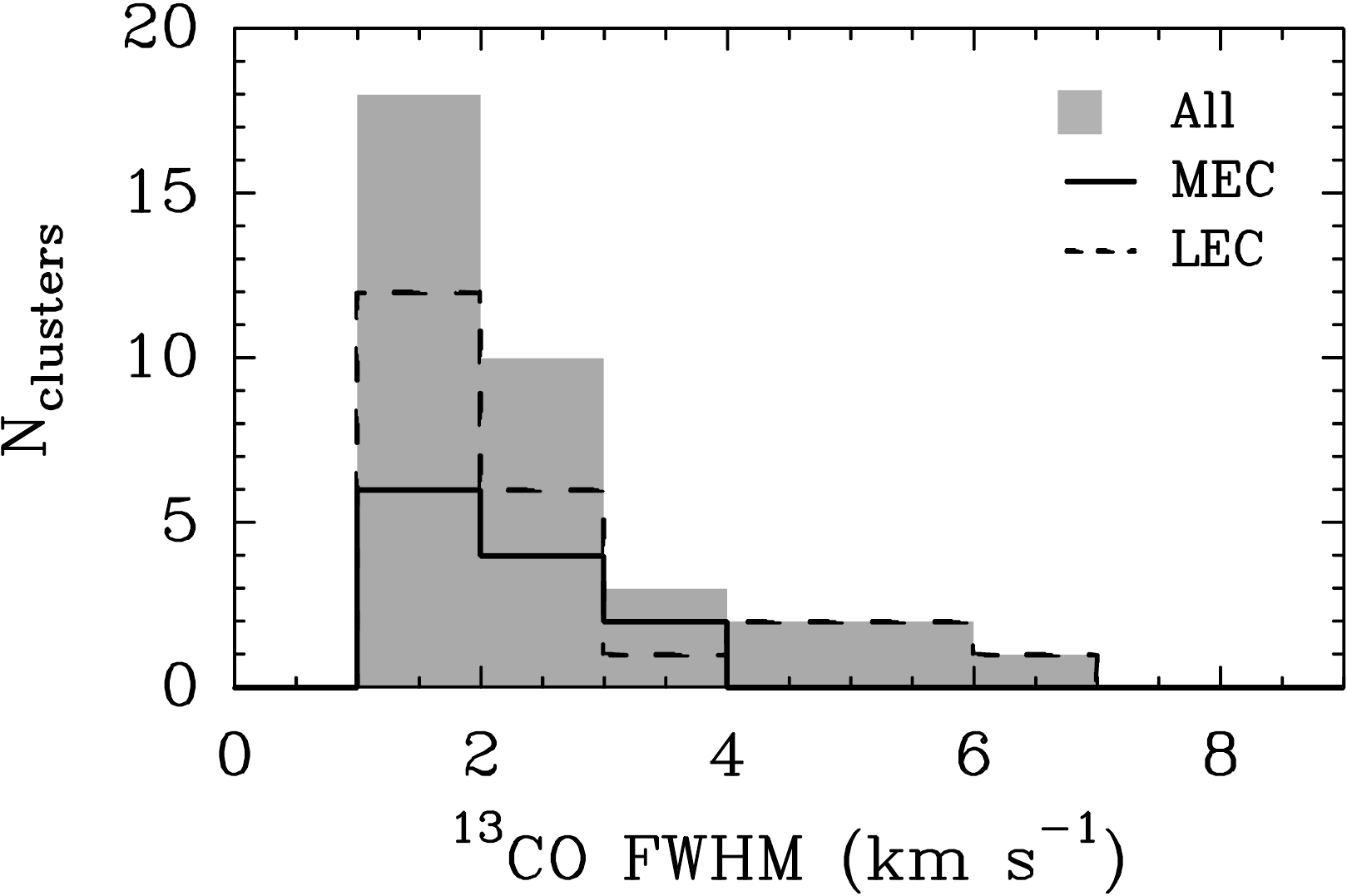}
\caption{Distribution of clusters \cob~line-width (FWHM). The thermal broadening for a 30 K cloud is $\sim 0.3~\mathrm{km~s^{-1}}$.}
\label{low-high_fwhm}
\end{center}
\end{figure}

We integrate the \cob~spectra over the clusters convex hull area to investigate the molecular cloud dynamics at a cluster-size scale. We use the \cob~line since is more optically thin that the \coa~line and traces more dense material as well as it may be less affected by outflow emission. We find no correlations between the \cob~full width at half maximum (FWHM) and cluster physical properties like YSOs surface density, size, extinction or cluster mass. However, all clusters have integrated \cob~line widths larger than the thermal line broadening for an assumed cloud temperature of 30~K (see Fig.~\ref{low-high_fwhm}):
\be
\Delta v_{\mathrm{th}} =  \left(\frac{8kT\ln(2)}{\mu_{\mathrm{CO}} m_{\mathrm{H}}}\right)^{1/2}\sim 0.3~\mathrm{km~s^{-1}},
\ee
where $k$ is the Boltzmann constant, $m_\mathrm{H}$ is the hydrogen mass and $\mu_\mathrm{CO}$ the \cob~atomic weight. This implies that the main support against gravity in all clusters is mainly due to non-thermal processes like turbulence. We find no significant differences between the MEC and LEC \cob~line broadenings. 

%%%%%%%%%%%%%%%%%%%%%%%%%%%%%%%%%%%%%
%%%%%%%%%%%%%%%%%%%%%%%%%%%%%%%%%%%%%
%%%%%%%%%%%%%%%%%%%%%%%%%%%%%%%%%%%%%
%%%%%%%%%%%%%%%%%%%%%%%%%%%%%%%%%%%%%

The Virial parameter ($\alpha_{\mathrm{vir}}$) measures the ratio between the kinetic and gravitational energy and is defined by:
\be
\alpha_{\mathrm{vir}} = \frac{5{\Delta v_{\mathrm{CO}}}^2 R_{\mathrm{H}}}{GM_{\mathrm{T}}},\label{equation_virial}
\ee
where $\Delta v_{\mathrm{CO}} = \mathrm{FWHM}/(8\ln2)^{1/2}$ is the dispersion velocity (of \cob), $G$ is the gravitational constant, $M_{\mathrm{T}}$ is the total cluster mass (gas and stars) and $R_{\mathrm{H}}$ is the convex hull radius. The total mass is given by the maximum between $M_{A_K}$ and $M_{\sigma_{\mathrm{CO}}}$ plus the corrected cluster members mass (assuming a mean mass of 0.5~\msun~per source, see \S~\ref{N_corrected}). We calculated $\alpha_{\mathrm{vir}}$ for all clusters and find values $>1$ in all cases. This suggest that clusters are mostly gravitationally unbound. However, the derivation of equation~(\ref{equation_virial}) has several assumption which may not apply for the observed clusters and this result must be treated carefully.

\begin{figure}
\begin{center}
\includegraphics[width=8cm]{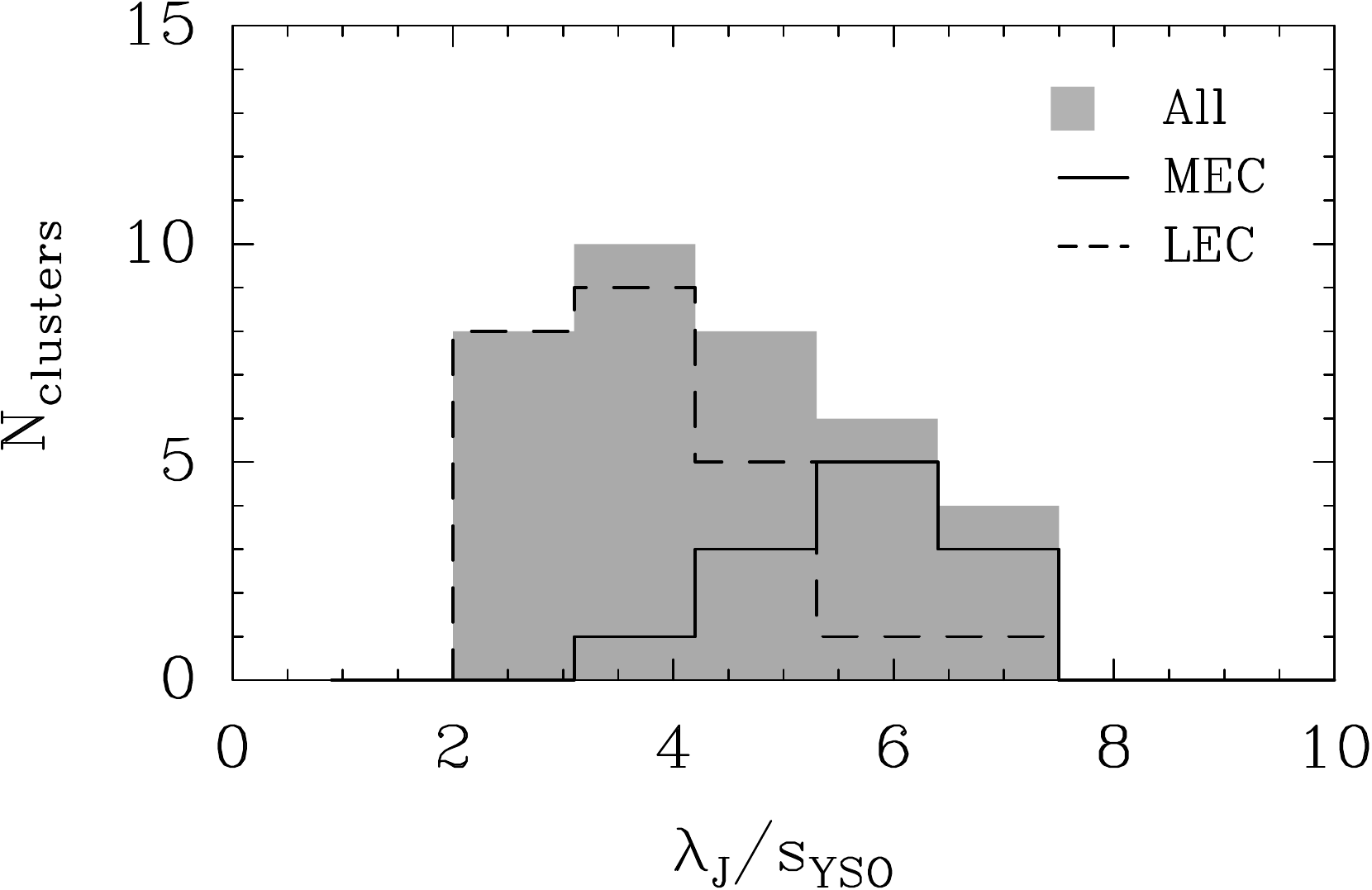}
\caption{Ratio between the cluster Jeans length ($\lambda_J$) and the mean projected distance between cluster members ($s_{\mathrm{YSO}}$). Massive embedded clusters are shown by the continuous line. Low-mass EC are shown by the segmented line.}
\label{low-high_jeans}
\end{center}
\end{figure}

We also investigate the clusters fragmentation scale by calculating their Jeans length ($\lambda_J$). This is the minimum radius for the gravitational collapse of an homogeneous isothermal sphere, and is given by: 
\be
\lambda_J = \left(\frac{15 k T}{4\pi G m_{\mathrm{H} } \rho_o}\right)^{1/2},
\ee
where $m_{\mathrm{H}}$ is the hydrogen atom mass, $T$ is the cloud temperature (30~K) and the mean density $\rho_o$ is defined by:
\be
\rho_o = \frac{3}{4\pi}\frac{M_{\mathrm{T}}}{R_{\mathrm{H}}^3}.
\ee
We compare $\lambda_J$ and the mean separation between cluster members ($s_{\mathrm{YSO}}$, Table~\ref{clusters}) and find that the ratio $\lambda_J/s_{\mathrm{YSO}}$ has values higher than 1 for all clusters, with an average of $4.3\pm1.5$ (see Fig.~\ref{low-high_jeans}). We also find that MEC have larger $\lambda_J/s_{\mathrm{YSO}}$ ratios than LEC. Those results agree with a non-thermal driven fragmentation since it occurs at scales smaller than the Jeans length. They also suggest that non-thermal processes may have a more pronounced effect in clusters harboring massive stars. Alternatively, a higher potential well in MEC may keep the cluster members closer to each other and for a longer time. This can also explain the $\lambda_J/s_{\mathrm{YSO}}$ ratio differences between LEC and MEC.

%%%%%%%%%%%%%%%%%%%%%%%%%%%%%%%%%%%%%
%%%%%%%%%%%%%%%%%%%%%%%%%%%%%%%%%%%%%
%%%%%%%%%%%%%%%%%%%%%%%%%%%%%%%%%%%%%
%%%%%%%%%%%%%%%%%%%%%%%%%%%%%%%%%%%%%

We note that the $s_{\mathrm{YSO}}$ values are calculated from the projected distance between cluster members. This means that the real distance is in average longer by a factor 2-3. However, this effect could be counteracted by including the non-detected cluster members, which decreases the $s_{\mathrm{YSO}}$ values by a similar factor.  %If we repeat those calculations using the corrected number of cluster members (see Appendix~\ref{N_corrected}) and assuming a mean mass of 0.5~\msun~per source, we find that those ratios increase, with a mean of $4.9\pm2$, for all clusters, $4.2\pm2$ for LEC and $6.5\pm2$ for MEC.

Magnetic fields may also play an important role on the clusters fragmentation level. Observations show that magnetic fields become stronger at higher densities in star-forming regions, which means that they stay coupled to the gas for a wide range of densities \citep{vle08}. \citet{tro08} showed that the ratio between turbulent and magnetic energy in dark clouds cores is roughly a factor of 2. \citet{vle08} showed that the energy due to turbulence and magnetic fields may be also of the same order for more evolved and massive cores. More detailed studies on the density structure and magnetic field measurements at similar scales are needed to disentangle which force dominates the support against gravitation.

%The line-width is given by the thermal broadening plus other non-thermal processes like for example turbulence and outflow-driven shocks. If we calculate the thermal width of the \cob~line from:
%\be
%\Delta v_{\mathrm{thermal}} =  \left(\frac{8\ln(2)~k T}{\mu_{\mathrm{CO}} m_{\mathrm{H}}},\right)^{1/2}
%\ee
%(where $k$ is the Boltzmann constant, $m_\mathrm{H}$ is the hydrogen mass and $\mu_\mathrm{CO}$ the \cob~atomic weight),  the obtained value if of less than 0.3~\kms~for a temperature of $T=30$~K. This way we assume that the observed line-width is mainly due to non-thermal processes. 

%Figure~\ref{low-high_fwhm} shows the \cob~FWHM integrated over the cluster area vs. the mean surface density of YSOs. The figure shows that clusters with high YSOs surface density are associated with less broad lines. While clusters with low YSOs surface density are spread over a more wide range of line widths. It makes sense to think that stars formed in more turbulent environments will inherit the parent cloud dynamics and be distributed in a more more spread way than stars formed in a less turbulent cloud. A more detail study is necessary, though, to confirm this idea.

\subsection{$N_{\textrm{\tiny{YSO}}}$ vs. dense cloud mass}\label{section_Ndense}
\citet*{lad10} found that the number of YSOs in a cluster is directly proportional to the dense cloud mass ($M_{0.8}$), which is the mass above a column density equivalent to $A_K \sim 0.8$~mag. This has important implications since establishes an empirical relation between the young stellar content and its parent molecular cloud. 

\begin{figure}
\begin{center}
\includegraphics[width=8cm]{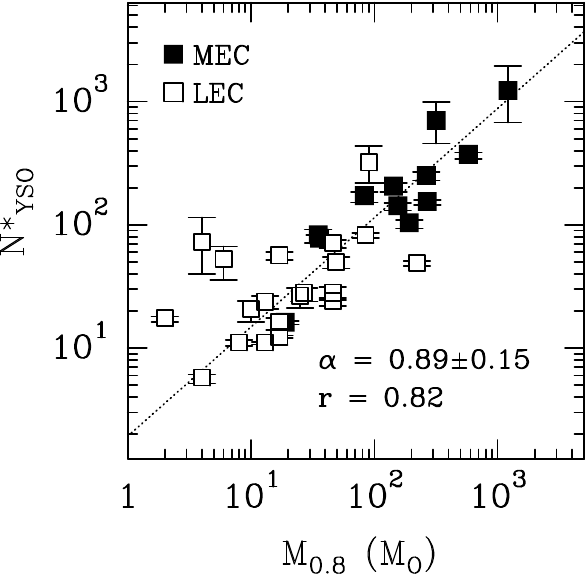}
\caption{Corrected number of cluster members and molecular mass above $A_K=0.8$~mag (M$_{0.8}$). The dotted line shows the best fit to the data using orthogonal regression. The error-bars are derived from the ${N^*}_{\mathrm{YSO}}$ values obtained with \ghm~in the range $-1.21\pm0.5$ (see Appendix~\ref{N_corrected}).}
\label{low-high_mass}
\end{center}
\end{figure}

To investigate this correlation among our cluster sample, we calculate $M_{0.8}$ by integrating the gas mass over the region inside the cluster convex hull area where $A_K > 0.8$~mag. For the number of YSOs, we use the total number of cluster members $N_{\textrm{\tiny{YSO}}}$. If well several cluster members are located in areas where $A_K < 0.8$, we assume that some of the young stars will move away from their birthplace \citep[][used all YSOs for which $A_K > 0.1$~mag and showed that their correlation does not depend on the cutoff magnitude for the accounted members]{lad10}. We find that the number of YSOs is proportional to ${M_{0.8}}^{\alpha}$ with an index $\alpha=0.6$. This is different from the index $\alpha=1$ obtained by \citet{lad10}. However, since our clusters are located further than the sample from \citet{lad10}, our results may be biased by the non-detection of faint cluster members. To correct this, we estimate the percentage of non-detected YSOs in each cluster (for a detailed explanation see Appendix~\ref{N_corrected}), and corrected the number of cluster members. After this, we find a correlation between the dense gas mass and the corrected number of cluster members (${N\textrm{*}_{\textrm{\tiny{YSO}}}}$) is given by (see Fig.~\ref{low-high_mass}):
\be
N\textrm{*}_{\textrm{\tiny{YSO}}} \propto {M_{0.8}}^{0.89 \pm 0.15}\label{eq-mass},
\ee
in better agreement with \citet{lad10}. The linear relation between the dense cloud mass and the number of stars suggest that the star formation rate (SFR) depends also linearly from the dense mass \citep{lad10}. %This correlation explains also the observed trend between the peak extinction and the number of YSOs (see Fig.~\ref{low-high_Ak}) since the more embedded the cluster is (higher peak $A_K$) the more dense mass it contains and hence it is more numerous.

%The different results obtained using the corrected number of cluster members remarks the importance of the detection or completeness limit in the analysis of such a regions. 

%The difference suggest that the percentage of detected cluster members decreases for the more numerous clusters. Figures~\ref{low-high_SD} and~\ref{low-high_Ak} show that the peak extinction as well as YSOs surface density increase with the number of cluster members. Since both peaks are in general located close to each other, clusters with more members will have more sources at higher extinctions than cluster with a few members. In other words, the more numerous clusters will have a higher percentage of non-detected members that less numerous clusters. 

%%%%%%%%%%%%%%%%%%%%%%%%%%%%%%%%%%%%%
%%%%%%%%%%%%%%%%%%%%%%%%%%%%%%%%%%%%%
%%%%%%%%%%%%%%%%%%%%%%%%%%%%%%%%%%%%%
%%%%%%%%%%%%%%%%%%%%%%%%%%%%%%%%%%%%%

\subsection{Star formation efficiency}
The star formation efficiency (SFE), is defined as the percentage of gas and dust mass converted into stars:
\be
\textrm{SFE} = \frac{M_{stars}}{M_{g+d}+M_{stars}}.
\ee
In many cases, the SFR is calculated over regions which are still forming stars. This means that there is still some molecular mass ($M_{g+d}$) that will be converted into star mass ($M_{stars}$). Since this is the case in the studied regions, the SFE will be then considered as a lower limit.

\begin{figure}
\begin{center}
\includegraphics[width=8cm]{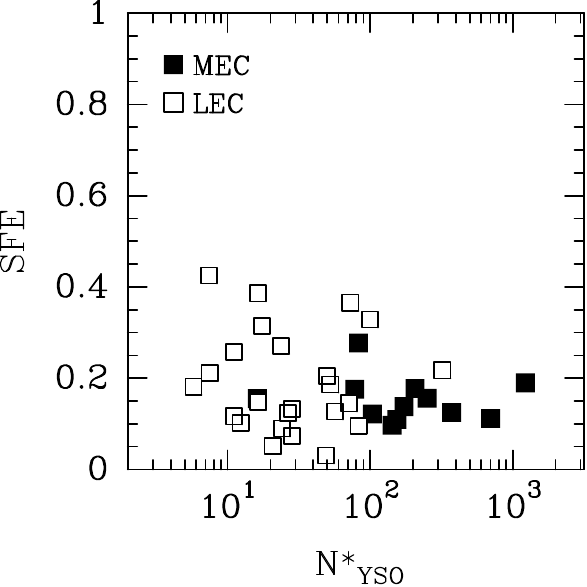}
\caption{Star formation efficiency (SFE) in embedded clusters. The chart shows the SFE for the corrected number of cluster members ($N\textrm{*}_{\mathrm{YSO}}$).}
\label{low-high_SFE}
\end{center}
\end{figure}

We use the cluster properties to estimate the SFE on a cluster size-scale. In this case, most of the low-density gas that surrounds the clusters is not taken into account since we calculate the cloud mass only from inside the cluster convex hull area. We use $M_{g+d}=\max(M_{A_K},M_{\sigma_{\mathrm{CO}}})$, $M_{stars}=0.5\times N\textrm{*}_{\textrm{\tiny{YSO}}}$ and find SFEs between 3 and 45\% with an average of $\sim20$\%. Those values are in agreement with the efficiencies needed to go from the core mass function (CMF) to the initial mass function (IMF) \citep*[e.g. 30\% in the Pipe nebula and 40\% in Aquila, from][respectively]{alv07,and10}. Also, similar values are obtained in cluster formation simulations from \citet{mas10} for a $10^3$~\msun~cloud. The SFE distribution over the corrected number of cluster members is shown in Figure~\ref{low-high_SFE}. There is not correlation between the star formation efficiencies and number of cluster members. This suggest that feedback processes, which may be particularly important in MEC, may start having an impact only in the later stages of the cluster evolution.
%(if we use $M_{stars}=1.0\times N_{\textrm{\tiny{YSO}}}$ those values change by less than 0.02). 

On a bigger spatial scale, we estimate the SFE for the whole regions, which is a better estimate of the SFE in molecular clouds. In this case, we calculate the cloud mass inside the field covered by all the IRAC and near-IR bands using our \cob~density maps since those are more sensitive to the less dense gas than the extinction maps. For the number of stars, we use ${N}_{\mathrm{YSO}}$ from Table~\ref{stars_detected} corrected by a factor 2 to roughly include the undetected YSOs. We find that the SFE in each region is between 4 and 8\%.
%%%%%%%%%%%%%%%%%%%%%%%%%%%%%%%%%%%%%
%%%%%%%%%%%%%%%%%%%%%%%%%%%%%%%%%%%%%
%%%%%%%%%%%%%%%%%%%%%%%%%%%%%%%%%%%%%
%%%%%%%%%%%%%%%%%%%%%%%%%%%%%%%%%%%%%

%%%%%%%%%%%%%%%%%%%%%%%%%%%%%%%%%%%%%
%%%%%%%%%%%%%%%%%%%%%%%%%%%%%%%%%%%%%
\begin{figure}
\begin{center}
\includegraphics[width=8cm]{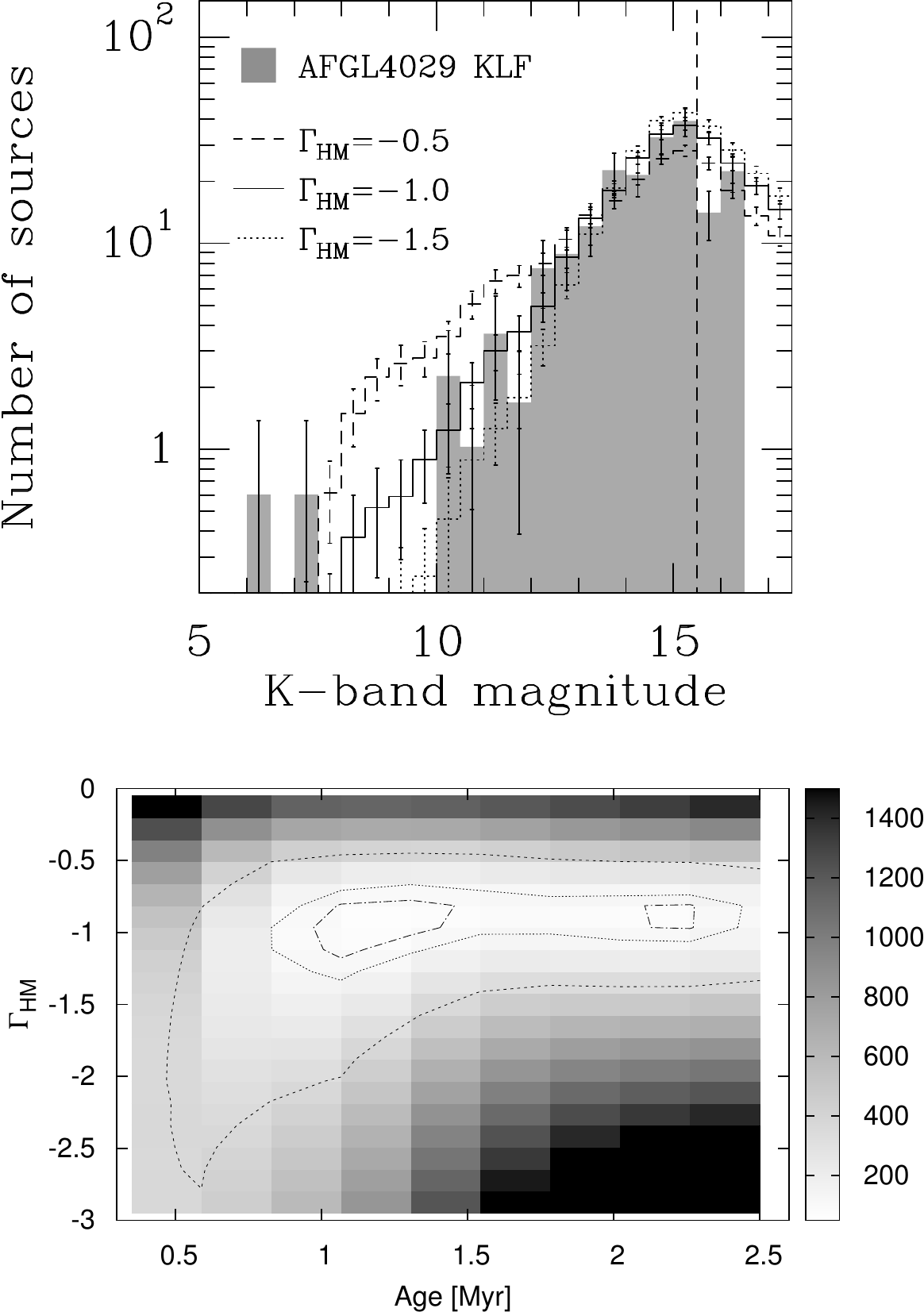}
%\plotone{fig111-eps-converted-to.pdf}
\caption{\textbf{Upper}: Field-substracted K-band luminosity function for cluster AFGL4029 (in grey) and for synthetic clusters (lines); the continuous line shows the synthetic cluster that better fits our data. All models have ages between 0.5 and 1.5 Myr. \textbf{Lower}: \ghm~and cluster-age parameter space for cluster AFGL4029. The contours are at 1, 3 and 5$\sigma$. The $\chi^2$ values are shown in the gray-scale.}
\label{imf_afgl4029}
\end{center}
\end{figure}
%%%%%%%%%%%%%%%%%%%%%%%%%%%%%%%%%%%%%
%%%%%%%%%%%%%%%%%%%%%%%%%%%%%%%%%%%%%

\subsection{The initial mass function (IMF) high-mass end}\label{section_IMF}
%Once formed, the evolution of stars is determined by their mass. This is why a complete understanding of the Initial Mass Function (IMF) is fundamental to understand the evolution of star clusters and to test theories of star formation. 
%On average, the value of $\Gamma$ is similar to -1.35. It was estimated first by \citet{sal55} over a range of masses lower than 10~M$_{\odot}$. In the case of massive stars, recent studies have revealed that the IMF is similar to the Salpeter IMF \citep{mas98,pfl06,sto06}, but the statistics are still poor since massive stars are rare and usually located far away from the Sun. In this section, we use the most numerous cluster from each region to study the high-mass end of the IMF slope (\ghm, $M>1$\msun).
Once formed, the evolution of stars is determined by their mass. This is why a complete understanding of the initial mass function (IMF) is fundamental to understand the evolution of star clusters as well as to constrain star formation theories. 
The IMF is defined as a power-law of the form:
\be
\frac{dN}{d\log m} = m^{\Gamma},
\ee
where \emph{N} is the number of stars of mass \emph{m} and $\Gamma$ is the mass function index or slope. 

We use synthetic cluster models to estimate the high-mass index of the IMF (\ghm) for the most numerous cluster in each region: AFGL4029, S235ABC, S252A, S255-2 and NGC7538. To derive \ghm, we compare the observed cluster K-band luminosity function (KLF, derived from all the detections in the K-band, see \S~\ref{section_imf}) with the KLF drawn from synthetic clusters having as variables the IMF slope and the range of ages. This way we constrain both the cluster age and the IMF index. Clusters IMF is in general described by several $\Gamma$~slopes for different mass ranges. \citet{mue02} found that the IMF for the Orion-Trapezium cluster is fitted by five slopes: $\Gamma_1=-1.21$ for $m~($\msun$)>0.6$, $\Gamma_2=-0.15$ for $0.6>m~($\msun$)>0.12$, $\Gamma_3=0.73$ for $0.12>m~($\msun$)>0.025$, $\Gamma_4=-0.5$ for $0.025>m~($\msun$)>0.027$ and $\Gamma_5=0$ for $m($\msun$)<0.017$. Since our observations detect sources of more than around 0.5-1~\msun, we vary the IMF slope only in the range $m>0.6$~\msun~and assume a Trapezium IMF for lower masses. Synthetic clusters are created using the Monte Carlo algorithm from \citet*{mue00}, which creates 100,000 stars with two main parameters as input: the IMF and age range. From those stars, we randomly select 2,000 to create a synthetic cluster. This cluster is later convolved with the observed clusters individual properties like extinction and IR-excess (for details see Appendix~\ref{section_imf}). Finally, we use a least squares fitting algorithm to compare the observed and synthetic KLF for different \ghm~and age range values (see Fig.~\ref{imf_afgl4029}).

We find that the five clusters have luminosity functions better fitted by \ghm~values between $-1.00$ and $-1.60$ and cluster ages of less than $\sim1.5$ Myr (see Table~\ref{gamma}). Clusters AFGL4029 and S235ABC have a top-heavy IMF. This may be of particular interest for AFGL4029 since this cluster is a good candidate for triggered star formation and some numerical models propose that triggering stars may trigger the formation of a more massive second generation of stars with a top-heavy IMF \citep{hok06}. On the other hand, cluster S255-2, which is also a candidate for triggered star formation, has a steeper IMF. We conclude that, if well the KLF is an useful method to estimate the IMF, it should be complemented with other observations like stars spectra to better constrain the cluster members ages and masses.

%%%%%%%%%%%%%%%%%%%%%%%%%%%%%%%%%%%%%
%%%%%%%%%%%%%%%%%%%%%%%%%%%%%%%%%%%%%
\begin{table}
\caption{Best-fited \ghm~and ages for selected clusters.}
\label{gamma}
\centering
\begin{tabular}{lrc} 
\hline
Cluster name & \ghm & Age [Myr] \\
\hline
AFGL4029  & $-1.0\pm0.3$ & $1.2\pm0.5$   \\
%                   & $-1.1\pm0.5$ & $1.5\pm0.5$   & yes \\ % bynaries
S235ABC    & $-1.0\pm0.4$ & $1.0\pm0.5$   \\ 
%                   & $-1.2\pm0.8$ & $1.5\pm0.5$   & yes \\ % bynaries
S252A        & $-1.2\pm0.5$ & $0.6\pm0.5$  \\ 
%                   & $-1.6\pm0.6$ & $0.5\pm0.4$   & yes \\ % bynaries
S255-2 \& S255N & $-1.6\pm0.3$ & $1.3\pm0.6$  \\ 
%                              & $-2.0\pm0.8$ & $1.5\pm0.5$   & yes \\ % bynaries
NGC7538    & $-1.3\pm0.5$ & $0.4\pm0.3$   \\
 %                  & $-1.3\pm0.6$ & $0.5\pm0.4$   & yes \\  % bynaries
\hline
\end{tabular}
\end{table}
%%%%%%%%%%%%%%%%%%%%%%%%%%%%%%%%%%%%%
%%%%%%%%%%%%%%%%%%%%%%%%%%%%%%%%%%%%%

\subsection{About triggered star formation}
We searched for trends among the physical properties of triggered cluster candidates and find not clear differences between them and clusters formed by "simple" gravitational collapse. Unless there is a clear age sequence like in cluster AFGL4029, it is difficult to estimate whether star formation around \hii~regions has been triggered or if it started before the arrival of the expansion wave. Even if this is the case, our observations show that once the clump becomes gravitationally unstable, the fragmentation processes and cluster dynamics rapidly erase the imprints of the triggering mechanism. This has already been suggested by the simulations from \citet*{dal13}. Finally, one of the key parameters missing in current studies about triggered star formation is the investigation of time causality between the expansion of the \hii~region and the age of the newly formed stars. 

\subsection{Properties of individual embedded clusters}
In this section we summarize the most important characteristics of the identified embedded clusters. We also search for individual sources of interest (not necessary inside clusters). The properties of clusters in region S254-S258 are explained in detail in \citet{cha08a}. All given coordinates are in the J2000 system.

\subsubsection{W5-east}\label{section_w5east_clusters}
\paragraph{G138.15+1.69}
This cluster is a good candidate for triggered star formation since it is located at the extreme of a pillar-like structure pointing in the direction of the ionizing star HD 18326 \citep{deh12}. Our dense-gas mass estimation ($M_{0.8}$) is a factor 2 higher than previous estimations from SCUBA data enclosing a similar area \citep{mor08}. The estimated \vlsr~is in agreement with previous \cob~observations from \citet{niw09}. 

\citet*{gin11} reported several outflows in the W5 region using CO 3--2 observations. We searched for their powering sources and find one IR-excess source located between the outflow \#25 red and blue lobes which is a good candidate (coordinates 3h00m56.16s +60d40m34.5s). However, this source is not identified as a HERSHEL point source by \citet{deh12} neither as a T-Tauri star by \citet{ogu02}. 

\citet{deh12} identified several YSOs towards G138.15+1.69. Of the 5 point-like sources detected in their study, 4 are classified as Class I following our classification scheme. Their fifth source is located inside the cluster convex hull area but is not detected in our observations. %Those are located at: 3h00m53.06s +60d40m37.5s, 3h00m55.32s +60d40m19.3s, 3h00m56.48s +60d40m11.8s and 3h00m56.5s +60d40m24.8s. 

We also identified a Class I source located at the top of a pillar-like structure between clusters G138.15+1.69 and AFGL4029, with coordinates: 3h01m47.04s +60d35m23.9s. 

\paragraph{AFGL4029}
The spatial distribution of Class I and Class II sources around AFGL4029 suggest a sequential star formation scenario. Figures~\ref{afgl4029irac_classes} and~\ref{w5_molecular} show that Class I sources are located preferentially east from the PAH ridge and are associated with the \cob~column density peak (most dense material), while Class II are located mainly to the west of the PAH front, closer to the ionizing star HD18326 and where the molecular material has already been removed. Assuming that Class I sources are younger than Class II, the observed star formation sequence makes this region a good candidate for triggered star formation; the ionizing front compresses and accumulates the molecular material as it expands, until exceeds its critical mass and collapses forming stars. Such a process will have the youngest stars located close to the expansion front, immersed in a pillar-like structure and preceded by older stars closer to the ionizing star. This is what we observe in AFGL4029 and in a minor scale in G138.15+1.69. However, the question if the expansion front compressed an already existing clump is still open. %Another question which may be related to the former one is why there are not massive stars among these western Class II population. 

We hypothesize that the Class II sources located up to $\sim$4~arc-minutes west from the cluster represent a previous period of star formation and are dynamically more evolved. However, this does not rule out the possibility that those sources are also part of cluster AFGL4029. Since they are outnumbered by the newly formed stars, the critical distance $d_c$ around the cluster is overweighted by the most embedded sources. Consequently, the western Class II sources will not be consider part of the cluster with our current cluster definition.

The dense-gas mass estimated for AFGL4029 is a factor 3 higher than previous calculations from \citet{mor08}. However, our estimations involve a larger area. As for cluster G138.15+1.69, we found a Class I source (3h01m31.30s +60d29m13.0s) associated with the outflow \#26 from \citet{gin11}. This source is also identified as the outflow counterpart by \citet{deh12}, corresponding to their HERSCHEL point source BRC14-c1.

\citet{deh12} identified several YSOs inside AFGL4029 convex hull area. Of the 13 point-like sources detected in their study, 4 are classified as Class II, 4 as Class I, 1 as YSO and 3 are not detected in our observations.

\paragraph{G138.32+1.51}
This cluster is located $\sim$3~arc-minutes south of AFGL4029 and it is associated with continuum emission at 850 microns \citep{mac11}. We found a Class I source (3h01m47.44s  +60d24m20.4s) located south-east of G138.32+1.51 which is associated with the outflow \#33 from \citet{gin11}. This source is also identified as the outflow counterpart by \citet{deh12}, corresponding to their HERSCHEL point source BRC14-a3.

\citet{deh12} identified one YSO inside the cluster convex hull that we classify as Class I source.

\paragraph{AFGL416}
The bipolar PAH structure observed in this cluster is due to the interaction between the stellar winds from massive stars and the ISM \citep*{fel87}. Figure~\ref{afgl4029_124} shows that the radio emission peaks between the lobes. On the other hand, Figure~\ref{w5_molecular} shows that the \cob~column density between the lobes increases from the center to the east and west sides. All of this implies that massive stars in AFGL416 are located along the east-west filament. Those stars generate strong stellar winds that sweep out the molecular material more efficiently along the north-south direction due to the lower column density, creating the observed bipolar morphology.

Our observations suggests that the interaction between molecular material associated with cluster AFGL416 and the ionizing front from star HD18326 is weak. 

\citet{gin11} found 2 outflows east of AFGL416 for which we have not detected powering source candidates.

\citet{deh12} identified several YSOs inside the cluster convex hull area. Of the 5 point-like sources detected in their study, 2 are classified as Class I, 1 as YSO and 2 are not detected in our observations.

\subsubsection{S235}

\paragraph{G173.51+2.79}
This cluster is located north-west from the \hii~region S235. It is also called S235 North-West  \citep{kir08}, FSR 784 \citep*{fro07} and Koposov 7 \citep*{kop08}. \citet{dew11} found 3 YSOs in this cluster, all of them are also identified as YSOs in our observations. 

Cluster G173.51+2.79, as well as clusters \textbf{G173.62+2.88} and \textbf{G173.67+2.87}, are located in the outer parts of the \hii~region S235 and have been proposed as triggered star formation candidates. However, unlike clusters AFGL4029 and G138.15+1.69, these do not seem to be located in pillar-like structures. \citet{kir08} observed \cob~and CS emission towards the region S235 and found good agreement between the distribution of molecular material associated with these three clusters and what is expected in a triggered or induced star formation scenario. 

\paragraph{G173.63+2.69}
Cluster located south-west of the \hii~region S235 and together with S235AB and S235C, is associated with molecular material at a different \vlsr~than the rest of the clusters surrounding the \hii~region S235 (see Table~\ref{cloud}).

\paragraph{S235C}
This cluster is probably part of cluster S235AB since it is located inside of it and it has a similar \vlsr. We identify a Class I source candidate to be the ionizing source of the UC~\hii~region (05h40m51.41s +35d38m30.0s, see Figure~\ref{s235_124}). There is another Class I source at about 1 arc-minute east from S235C (05h40m57.7s +35d38m19.1s) which shows an arc-shaped PAH structure pointing towards S235C. 

\paragraph{S235AB}
This cluster has the highest YSOs surface density in our sample ($10^3$ stars per pc$^{-2}$) and it has a different \vlsr~than the rest of clusters located around the \hii~region S235 (see Fig.~\ref{s235_molecular} and Table~\ref{cloud}). We identified two Class I sources as candidates for powering the \hii~regions S235A and S235B with coordinates 05h40m52.58s +35d42m18.6s and 05h40m52.39s +35d41m29.4s respectively. We also identified 75 out of the 77 YSOs found by \citet{dew11} as cluster members. Clusters BDSB71 to BDSB73 \citep{bic03b} are included in S235AB.

\paragraph{G173.66+2.78}
This cluster may be associated to cluster S235 due to its location and similar \vlsr. We identified all the YSOs found by \citet{dew11} as cluster members.

\paragraph{S235}
Also called S235 Central \citep{kir08}, this cluster is apparently located inside the \hii~region S235. The projected distance from the cluster to the ionizing star BD+351201 is $\sim1$~pc. At this distance, we may expect to observe signatures of interaction between the ionizing front and the molecular material associated with the cluster (like in AFGL4029). The absence of such signatures suggest that the cluster is either located in another plane compared to the ionizing star, or not associated directly with the molecular material at this location. Since the ionizing star has a \vlsr~of $-18$~\kms~\citep{kir08} and the \cob~ material associated with S235 has a \vlsr~of $-19.8$~\kms, we hypothesize that the cluster is located in a different plane than the \hii~region \citep{kir08}. 

The sources IRS1 and IRS2 are identified as cluster members and classified as Class I. The 9 YSOs found by \citet{dew11} are also identified as cluster members. The cluster CBB2 \citep{cam11} is part of cluster S235.

\paragraph{G173.62+2.88 (East 2)}
Located north-east from cluster S235. We identify 18 out of the 19 YSOs found by \citet{dew11} as cluster members. \citet{dew11} found 3 Spitzer-IRAC sources without NIR counterpart: e2s1, e2s2 and e2s3. Of those, we detect e2s2 and e2s3 in the K-band. All of them are classified as Class I. 

\paragraph{G173.67+2.87 (East 1)}
We identify all the YSOs found by \citet{dew11} as cluster members. \citet{dew11} identified 4 Spitzer-IRAC sources without NIR counterparts: e1s1, e1s2, e1s3 and e1s4. We detect e1s2 and e1s3 in the H and K bands and e1s4 in the J, H and K bands. All of them (e1s1 to e1s4) are classified as Class I. The source e1s1 is also associated with extended emission seen in the K-band and at 4.5 microns, signs of outflow activity.

\subsubsection{S252}

\paragraph{G189.79+0.29}
Located to the west of cluster S252A, it is associated with molecular material at \vlsr~$=10$~\kms (S252A has \vlsr~of $8.3$~\kms).

\paragraph{G189.84+0.29}
It may be associated with cluster S252A due to its similar \vlsr~(see Table~\ref{cloud}).

\paragraph{G189.95+0.22}
This cluster has a \vlsr~of $5.8$~\kms, suggesting that it is located in a different plane in the sky than the rest of the clusters in the region.

\paragraph{S252A}
It harbors the compact-\hii~region S252A on its north-west side. The ionizing source of this compact-\hii~region is classified as Class I (06h08m32.05s +20d39m18.3s). The extension of this cluster agrees well with the sub-millimeter emission detected by \citet{tej06}.  The derived IMF slope is also in agreement with estimations by \citet{jos12} of $-1.33$. However, they cover a different mass range (from 0.3 to 2.5 \msun) and use a different methodology. 

Both the \cob~column density map and the extinction map show that the molecular material decreases towards the center of the compact-\hii~region. This was also noticed by \citet{tej06} in their sub-millimeter observations. Moreover, they find that the peak emission at 850 microns lays $\sim1$~arc-minute east from the center of \hii~region, close to its edge. The sub-millimeter peak coincides with methanol and water maser emission and also with a Class I source (06h08m35.33s +20d39m06.9s) detected in the IRAC bands 3 and 4 and in the MIPS map. This source also seems to be associated with emission from shocked gas seen as a green blob in Figure~\ref{s252_124}. Whether the formation of new stars at the sub-millimeter peak was triggered or not by the expansion of the contiguous \hii~region is still an open question.

\paragraph{G189.94+0.33}
Cluster located on the PAH ridge, south-east of S252A and G189.84+0.29. Figure~\ref{s252_124} shows some extended radio emission at the east side of the PAH ridge. This emission may be due to the interaction of the ionizing front with the material along the ridge. The 24 microns emission observed right at the west side of the radio emission and the rapid increase of the column density evidence how the molecular cloud is being compressed by the expansion front. G189.94+0.33 may be also a candidate for triggered star formation.

\paragraph{S252C}
This cluster has a $\sim2$~pc size lobe on the east side as shown in Figure~\ref{s252_124} and in the MIPS map. This lobe is probably created by strong stellar winds emitted by the ionizing source located close to the cluster center. The ionizing source of the compact-\hii~region is classified as Class II. 

\paragraph{G189.95+0.54}
Cluster located $\sim5$~arc-minutes south-east from S252C. Its \vlsr~suggest that it may be placed in a different plane in the sky than other clusters in the region.

\paragraph{S252E}
Similarly to cluster S252C, the dust emission around cluster S252E shows lobes on its east and west side. The east lobe is not completely covered by our IRAC map. The candidate ionizing source of the compact-\hii~region S252E is classified as Class II.

S252E is located at a projected distance of about 3~pc from the ionizing star of the \hii~region NGC2175. At this distance, we would expect to observe signs of interaction between the molecular material and the ionization front like a pillar. The absence of such a morphology suggest that this cluster is located in another plane compared to the ionizing star.\\
\\
The \hii~region \textbf{S252B} is located between clusters S252A and S252C. We find no cluster around the ionizing source (classified as Class II) and propose three alternatives to explain these observations: isolated star formation, runaway star or non-detected cluster members. The \cob~column-density map shows density enhancements on the west side of S252C and east side of S252A which are likely produced by the expansion of the \hii~region S252B. This means that the ionizing star of S252B is not a foreground or background source and that it could have been expelled from nearby embedded clusters due to dynamical interaction with other YSOs. Alternatively, this case may be similar to \hii~regions S255 and S257 in the S254-S258 complex. Those \hii~regions have a similar size than S252B and are powered by stars with similar spectral type. However, \citet[][and this paper]{cha08a} identified some YSOs around the ionizing star in the \hii~region S257 suggesting that there is likely a cluster associated to that source. Those cluster members are difficult to detect due to the brightness of the extended emission inside the \hii~region, which may be the case for S252B. Furthermore, early disk evaporation due to a nearby ionizing star will make those cluster members difficult to detect at mid-IR wavelengths. Finally, we cannot leave aside the possibility that the ionizing source of \hii~region S252B was formed in isolation.

% The estimated age of the \hii~region S252B is of around 1~Myr (assuming an ionizing star with spectral type B0.0V). The column density map shows that there is no molecular material inside S252B (Figures~\ref{s235_13coden} and \ref{s252_slides} respectively). This is consistent with the distribution of molecular material in other \hii~regions of similar ages. 

\subsubsection{S254-S258}
We investigate this region using the YSOs identified by \citet{cha08a} and adding the sources detected from MIPS observations. In general the number of cluster members as well as cluster sizes are very similar to the previous results from \citet{cha08a}. We find two main differences: former cluster S258 in now divided in clusters S258 and G192.70+0.03, and the number of identified members for cluster G192.63+0.00 decreased by a factor of two. We believe that the last is due to the proximity of cluster S255-2; the YSOs in G192.63+0.00 are outnumbered by the dense population of S255-2 which decreases the value of the critical distance $d_c$ used for both clusters. This way, the most dispersed population of G192.63+0.00 is not included as part of the cluster by our MST algorithm.

The new distance to this region \citep[1.6~kpc, ][]{ryg10} has an effect in the derived YSOs surface densities and clusters molecular mass, changing the values given in \citet{cha08a} by a factor of 2 or 3. However, this does not change the conclusions drawn from cluster to cluster comparison in the region since those factors are the same for all clusters.

\subsubsection{NGC7538}\label{section_NGC7538}

\paragraph{G111.44+0.79}
This cluster is located west of the \hii~region NGC7538, along a dusty filament connecting also clusters G111.48+0.80, G111.49+0.81 and NGC7538. It is also associated with several clumps detected at 850 microns \citep{rei05}.

\paragraph{G111.44+0.75}
Cluster located $\sim1$ arc-minute south of G111.44+0.79 and associated with several sub-millimeter clumps \citep{rei05}. This cluster is located along a different filament than G111.44+0.79 that is also connected with cluster NGC7538. 

\paragraph{G111.48+0.80 and G111.49+0.81}
These clusters are located between the \hii~region NGC7538 and cluster G111.44+0.79. They are associated with the same dust filament as G111.44+0.79 as well as with sub-millimeter clumps \citep{rei05}. 
 
\paragraph{G111.47+0.75}
Cluster located along the same dusty filament as G111.44+0.75. It is associated with sub-millimeter emission \citep{rei05} and with a water maser \citep[H1, from][]{kam90}. This maser is located less than 1 arcsecond from a Class I source with coordinates 23h13d22.2s +61d25m44.1s.

\paragraph{NGC7538}
Located along the south side of \hii~region NGC7538 (cluster NGC7538S is part of cluster NGC7538). NGC7538 is the largest and most massive of all the identified clusters in the five studied regions. However, since this region is also the furthest one, the cluster could be composed of several sub-structures that are not clearly separated by the MST algorithm. This could explain the cluster difference in the peak $\Sigma_{\mathrm{YSO}}$ value with respect to the other clusters trend in Figure~\ref{low-high_SD}. 

Sources IRS1 to 4, IRS9 and IRS11 are classified as Class I. Some sources with spectral type of early B-star \citep{pug10} and the far-infrared source NGC7538S are members of cluster NGC7538. The \cob~density map shows that the column density increases rapidly from the center to the south and north sides of the \hii~region. The densest area in the region is located around sources IRS1-IRS3. It is a very active site of star formation and it seems to be the joint point of two dusty filaments (A and B, see Fig.~\ref{ngc7538_molecular}).

\paragraph{G111.57+0.81}
This cluster is located east from the \hii~region NGC7538. It contains two Class I sources located at the tip of pillar-like structures and associated with the sub-millimeter clumps SMM~61 and SMM~57 from \citet{rei05} (their coordinates are 23h14d01.7s +61d30m15.7s and 23h13d58.5s +61d30m51.5s respectively). 

%\paragraph{G111.61+0.77}
%Cluster located east from G111.57+0.81. There is no evidence of interaction with the ionizing front of the \hii~region.

\paragraph{G111.59+0.64 and G111.65+0.63}
These clusters are located 3 to 5 arc-minutes south-east from the \hii~region NGC7538. They are associated with molecular material at \vlsr~of $-51$~\kms, suggesting that they belong to a different plane in the sky than the rest of the clusters in the region (see Fig.~\ref{ngc7538_molecular} and Table~\ref{cloud}).

\section{Conclusions}\label{section_conclusions3}
We performed a multi-wavelength study on 5 regions of massive star formation: W5-east, S235, S252, S254-S258 and NGC7538. Spitzer-IRAC/MIPS and NIR observations were used to classify the stellar population. While the molecular content was studied using \coa, \cob~observations and extinction maps.

We found in total 3021 YSOs, including 539 Class I and 1186 Class II. A minimum spanning tree algorithm was used to identify YSO clusters based on the characteristic separation of their members. A total of 41 embedded clusters were found, 15 of which have not being identified before.

The Class I sources are spatially correlated with the most dense molecular material. They are also located in regions with higher YSOs surface density and are distributed more hierarchically than Class II. All this agrees well with the picture where stars are formed in dense and fractally arranged dust filaments. Then, dynamical interactions rearrange the YSOs in a more centrally condensed distribution. 

We find that the mean separation between cluster members is smaller than the cluster Jeans length in most cases. This difference is more evident in the case of MEC. In addition, the \cob~line width of clusters associated molecular material shows that the clusters are turbulent. This agrees with a scenario in which fragmentation is likely driven by turbulence. Though magnetic fields may also play an important role as a support against gravity.

Between 30 and 50\% of the total number of YSOs are not included in clusters. This percentage of scattered population depends on the used cluster finding algorithm. We propose that between 10 and 20\% of the scattered population in the studied regions corresponds, indeed, to cluster members. 

We compared the physical properties of embedded clusters associated with high-mass stars and clusters with no evidence of harboring massive stars. We find no systematical differences in the correlations derived for both samples. In all cases, the MEC seem to be an extrapolation of the LEC.

The correlation between the clusters dense mass and the number of cluster members is investigated. We find that this correlation is close to linear, in agreement with previous findings from \citet{lad10}. We also find that the star formation efficiency is rather constant along the clusters mass range. In average, the estimated SFE agrees well with the mass factor between the cores mass function and the initial mass function.

The spatial distribution of Class I and Class II sources in clusters AFGL4029 and G138.15+1.69 suggests a sequential star formation which moves in the same direction as the ionization front. This is in agreement with previous estimation of the YSOs ages around these clusters and supports the hypothesis that both clusters were created in a triggered star formation scenario \citep[see][and references there in]{cha11}. 

We classify a total of 24 OB-type stars as either Class I or Class II sources. The IR-excess emitted by those sources may be due to a dusty structure around them. Those sources are good candidates for future high resolution studies aimed to search for disks in high-mass stars. The presence of a rotating structure around massive stars suggest that they are form via accretion.

%Finally, we prove the importance of accounting for the completeness limit in the observations. In order to improve the cross-checking between observational and theoretical results, we propose to include results for a few completeness limits in future models. 

%We used synthetic clusters to model the high-mass end of the IMF. We estimated the IMF slope by comparing the synthetic KLF with the observed KLF in five embedded massive clusters. We found that the best fit correspond to IMF slopes of the order of \ghm$\sim$ -1.1. This value is slightly flatter than a Salpeter IMF. We found that the value of \ghm~depends on the cluster age as well as the inclusion of binary systems in the models. The highest \ghm~values correspond to clusters AFGL4029 and S255-2. These clusters are good candidates for triggered star formation. We think that it is important to investigate the possible effects on the IMF due to triggered star formation.

\section*{Acknowledgments}
This work is based in part on observations made with the Spitzer Space Telescope, which is operated by the Jet Propulsion Laboratory, Caltech, under a contract with NASA. Support for this work was provided by NASA through a contract issued by JPL/Caltech. We also thank NOAO for their student thesis support. The Five College Radio Astronomy Observatory was supported by NSF grant AST 0540852. Chris Brunt is supported by an RCUK Fellowship at the University of Exeter, UK. This work is based in part on the IRAC post-BCD processing software "IRACProc" developed by Mike Schuster, Massimo Marengo and Brian Patten at the Smithsonian Astrophysical Observatory. This research used the facilities of the Canadian Astronomy Data Centre operated by the National Research Council of Canada with the support of the Canadian Space Agency. This research has made use of the NASA/ IPAC Infrared Science Archive, which is operated by the Jet Propulsion Laboratory, California Institute of Technology, under contract with the National Aeronautics and Space Administration. We thank the Spanish MINECO for funding support from grants CSD2009-00038, AYA2009-07304 and AYA2012-32032.
%%%%%%%%%%%%%%%%%%%%%%%%%%%%%%%%%%%%%
%%%%%%%%%%%%%%%%%%%%%%%%%%%%%%%%%%%%%



\appendix

\section{Background contamination estimates}\label{background}
The detection of distant galaxies in our mid-IR data is expected due to the relative transparency of the molecular clouds at these wavelengths. Since galaxies and YSOs have similar IR colors, it is important to find a way to distinguish between both to avoid including galaxies as part of our YSOs population. A simple method  would be to remove all the sources located in the galaxies-dominated areas (GDA) in the color-magnitude diagrams \citep[e.g.][]{cha08a,gut09}. However, the number of background galaxies will decrease in highly embedded regions due to the dust extinction. Therefore, we complement the GDA with our extinction maps and for each YSO candidate, we calculate the extinction along the line of sight. Then, we apply this extinction to the galaxies in the color-magnitude diagram to emulate the galaxies color as if they were affected by the same extinction as the YSO candidate. Finally, we define the GDA in the color-magnitude diagram and confirm the candidate as a YSO if it is located outside the GDA.

%%%%%%%%%%%%%%%%%%%%%%%%%%%%%%%%%%%%%
%%%%%%%%%%%%%%%%%%%%%%%%%%%%%%%%%%%%%
\begin{figure}
\begin{center}
\includegraphics[width=8cm]{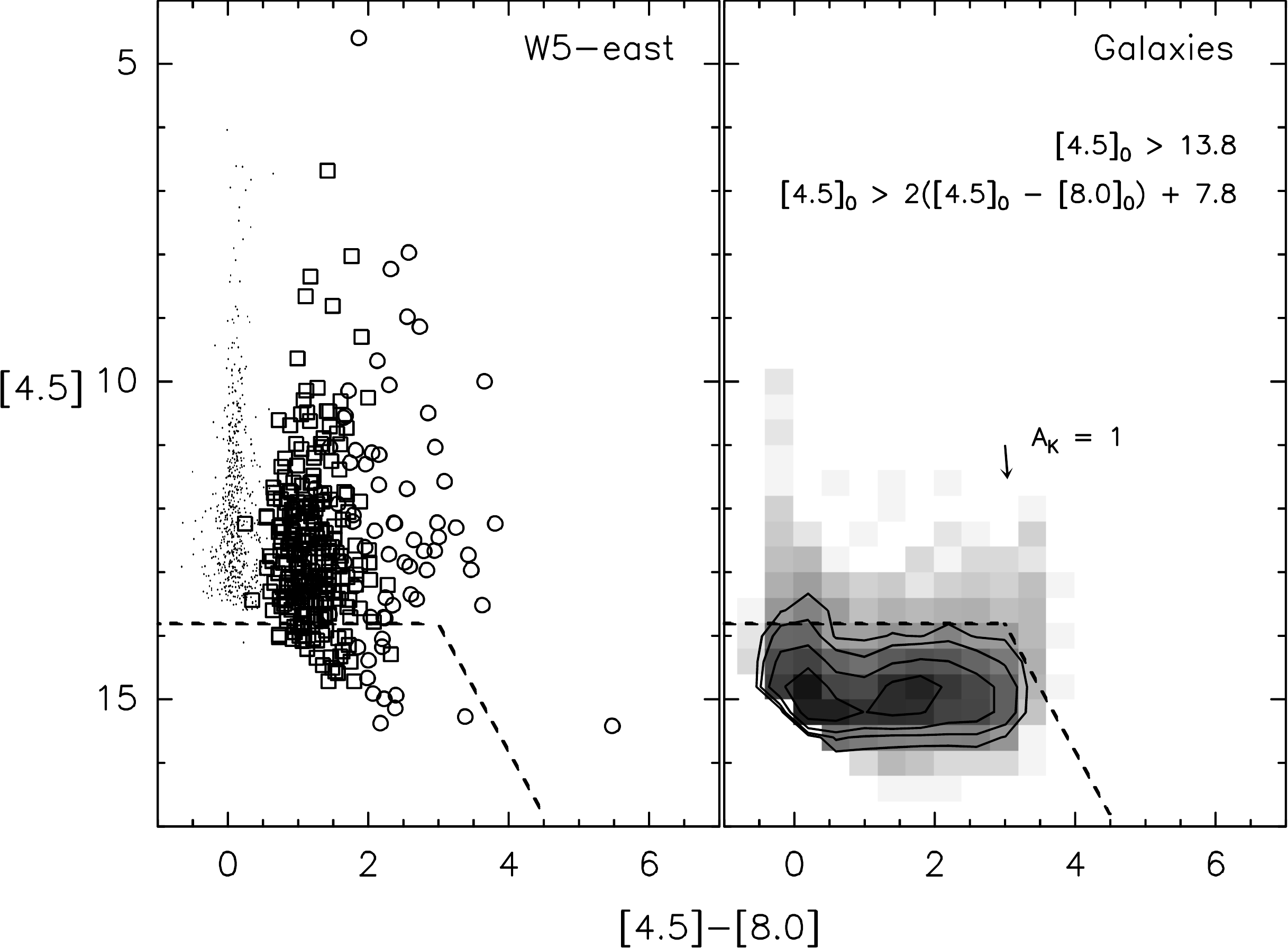}
%\epsscale{0.5}
%\plotone{figA1b-eps-converted-to.pdf}
%\plotone{figA1c-eps-converted-to.pdf}
\caption{Color-magnitude diagram used to remove background contamination in region W5-east. This diagram is used to eliminate contaminants detected in the four IRAC bands (K vs. K$-[4.5]]$ and J vs. J$-$K diagrams are used for the other YSOs classification schemes). Dashed lines enclose the galaxy dominated areas (GDA) in each chart. The chart on the left shows all the detected sources. Candidate YSOs are labeled as circles (Class I) and squares (Class II). Other detections are shown as dots. The chart on the right shows the galaxies from the IRAC Shallow survey for $A_K=0$. Contours begin at 0.5 galaxy per mag$^{-2}$. The arrow shows the extinction vector for $A_K=1$. For each YSO candidate $i$ with associated extinction $A_{K_i}$, the GDA moves along the extinction vector to reach the YSO extinction. Then, the candidate is removed if it is located inside the GDA.}
\label{image_background}
\end{center}
\end{figure}
%%%%%%%%%%%%%%%%%%%%%%%%%%%%%%%%%%%%%
%%%%%%%%%%%%%%%%%%%%%%%%%%%%%%%%%%%%%

We use different color-magnitude diagrams for the different YSOs: $[4.5]$ vs. $[4.5]-[8.0]$ for Class I and Class II candidates, K vs. K$-[4.5]$ for IR-excess candidates and J vs. J$-$K for sources detected only in the near-IR bands. We use NIR and Spitzer-IRAC galaxy observations from the IRAC Shallow Survey \citep{eis04}, which covers 8.5 square degrees of the sky in the NOAO Deep Wide-Field Survey in Bo\"otes and used \textbf{3$\times$30 second frames} (deeper than our IRAC observations). Figure~\ref{image_background} shows as an example the GDA in the $[4.5]$ vs. $[4.5]-[8.0]$ color-magnitude diagram for W5-east at $A_K=0$. In total we identified 11 Class I and 29 Class II sources as background contamination in W5-east, 11 Class I and 2 Class II in S235, 15 Class I and 5 Class II in S252, 1 Class I and 1 Class II in S254-S258 and 7 Class I and 9 Class II in NGC7538.

Unresolved asymptotic giant branch (AGB) stars will have colors similar to YSOs in the near and mid-IR wavelengths. We estimate the amount of AGB contaminants over the observed fields using the AGB stars distribution derived by \citet*{jac02} and \citet{rob08}. We find that the number of AGB stars for each region FOV is of the order of 1. This is due to the location of the studied regions towards the outer parts of Galaxy. Since AGB \textbf{star contamination is} negligible with respect to the number of YSOs we did not perform a further correction in this matter.

\section{Comparing YSO detections with previous publications in W5-east and S235}\label{comparison_other_studies}
\citet{ogu02} and \citet{nak08} identified 114 stars with H$\alpha$ emission in region W5-east over the same FOV as in our observations. Of those, we identify 84 (74\%) as YSOs: 3 as Class I, 65 as Class II and 16 as having IR-excess. In the same region, and using IRAC+NIR data, \citet{koe08} found 292 YSOs (including Class I, Class II and transition disks following their classification scheme). Of those, we identified 251 (86\%): 36 Class I, 191 Class II and 24 IR-excess sources. In region S235, \citet{dew11} detected 230 YSOs using also IRAC+NIR data. Of those, we detected 220 (96\%); 46 Class I, 165 Class II and 9 sources with IR-excess.

\begin{figure}
\begin{center}
\includegraphics[width=8cm]{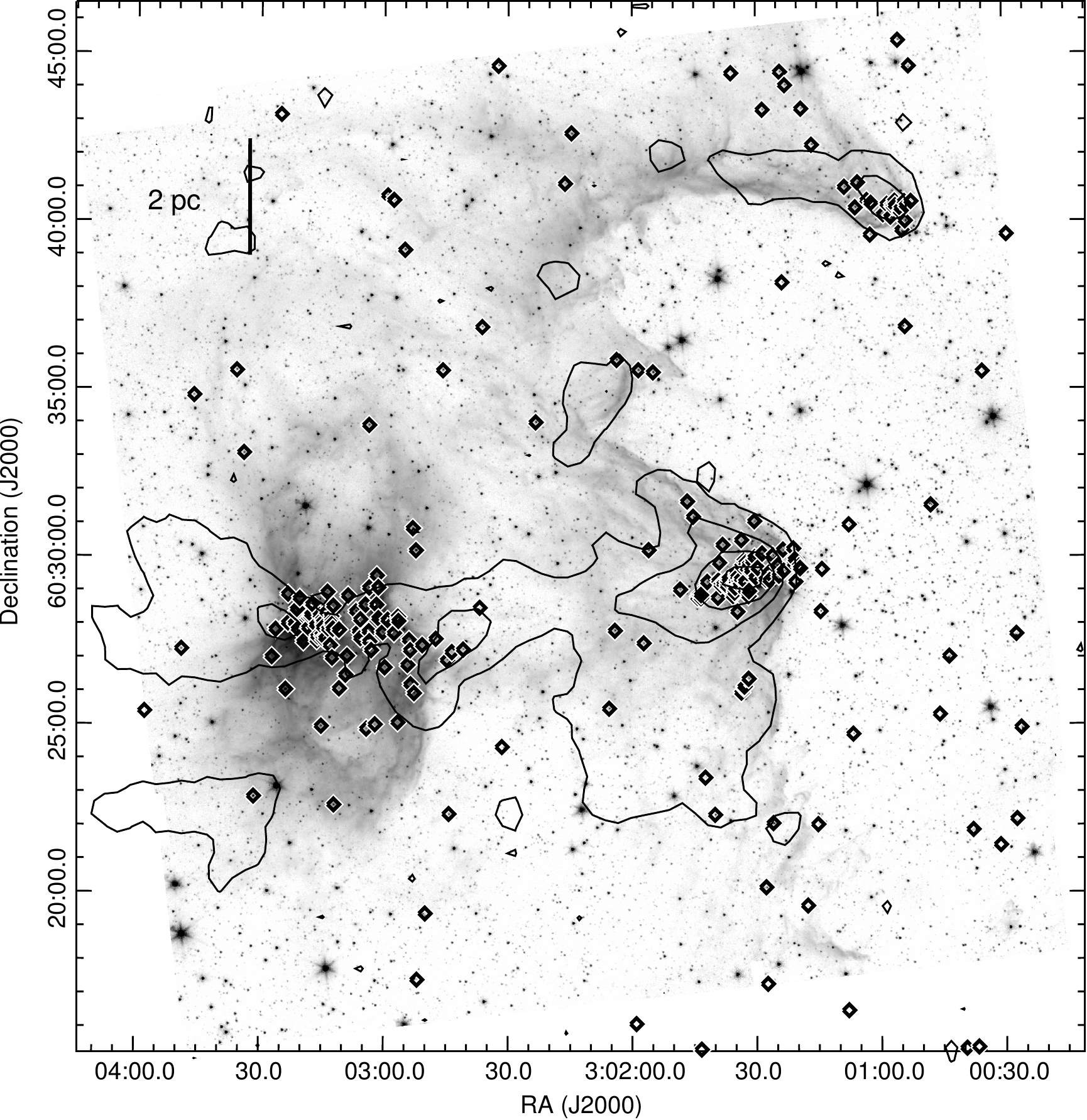}
\caption{Distribution of YSOs identified in this work that were not identified as either Class I or Class II by \citet{koe08} in region W5-east. The contours show the \cob~surface density distribution. Contours are between $4\times 10^{15}$ and $4\times 10^{16}$~cm$^{-2}$.}
\label{additional_w5east}
\end{center}
\end{figure}

\begin{figure}
\begin{center}
\includegraphics[width=8cm]{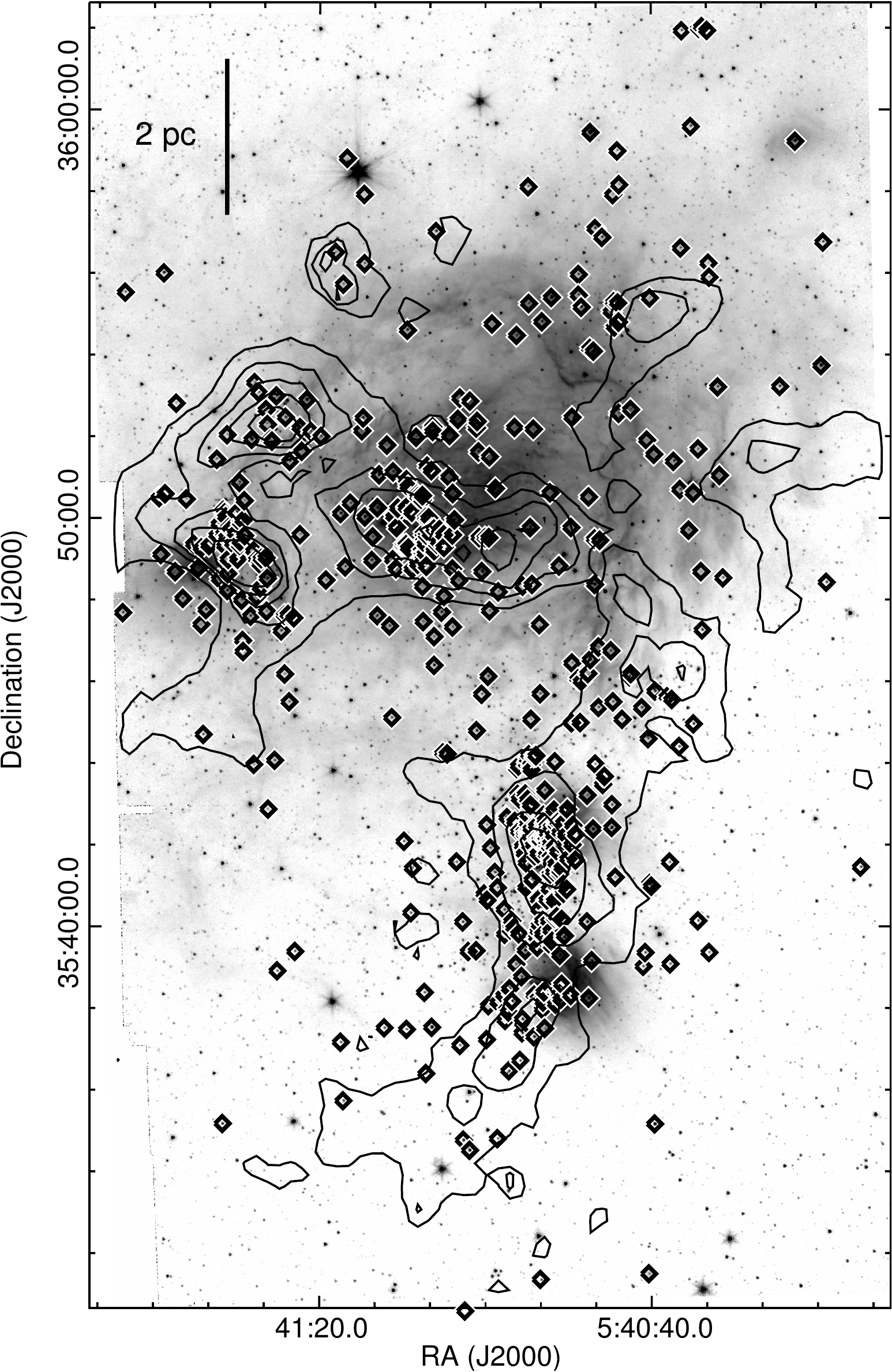}
\caption{Distribution of YSOs identified in this work that were not identified by \citet{dew11} in region S235. The contours show the \cob~surface density distribution. Contours are between $4\times 10^{15}$ and $1\times 10^{16}$~cm$^{-2}$.}
\label{additional_s235}
\end{center}
\end{figure}

The longer integration times and better spatial resolution of our NIR observations has a direct impact in the number of identified YSOs. In region W5-east, we identify 478 YSOs, while \citet{koe08} identified 292 over the same FOV. In region S235, we identify 630 YSOs while \citet{dew11} identified 230 in the same FOV. Approximately 60\% of the new YSOs are detected using NIR observations (66 and 59\% in W5-east and S235 respectively). Figures~\ref{additional_w5east} and~\ref{additional_s235} show the distribution of the YSOs that were not identified by \citet{koe08} and \citet{dew11} in W5-east and S235 respectively. In both regions, those sources are distributed in groups whose location correlates well with the molecular material and embedded clusters. If they were background galaxies, they would be distributed homogeneously over the field. We conclude that this population is mainly composed \textbf{of} YSOs. 

%%%%%%%%%%%%%%%%%%%%%%%%%%%%%%%%%%%%%
%%%%%%%%%%%%%%%%%%%%%%%%%%%%%%%%%%%%%
%%%%%%%%%%%%%%%%%%%%%%%%%%%%%%%%%%%%%
%%%%%%%%%%%%%%%%%%%%%%%%%%%%%%%%%%%%%

%%%%%%%%%%%%%%%%%%%%%%%%%%%%%%%%%%%%%
%%%%%%%%%%%%%%%%%%%%%%%%%%%%%%%%%%%%%
%%%%%%%%%%%%%%%%%%%%%%%%%%%%%%%%%%%%%
%%%%%%%%%%%%%%%%%%%%%%%%%%%%%%%%%%%%%

\section{Estimating the non-detected cluster population}\label{N_corrected}
We identify the young stellar population in regions located at a few kpc from the Sun. At those distances, our completeness limit allows to detect stars down to $\sim0.5-1$~\msun~and we would like to estimate the percentage of sources below this limit. For example, if we assume that the embedded clusters follow an Orion IMF \citep{mue02}, the percentage of non-detected members would be up to 50\%. Now, if we account for the cluster extinction, this percentage may be even higher. In the following, we explain the algorithm used to estimate the percentage of non-detected cluster members.

The completeness limit will vary from one cluster to another due to their different properties like extinction, extended emission brightness and separation between cluster members. To account for these differences we calculated the 90\% completeness limit for each cluster individually. This is done by adding synthetic stars over the cluster K-band image and then counting the percentage of synthetic stars recovered. The completeness limit (CL$_\mathrm{K}$) corresponds to the magnitude where this percentage goes below 90\%. After this, we calculate the percentage of synthetic stars ($\kappa$) with K magnitudes fainter than CL$_\mathrm{K}$ from a synthetic cluster having similar properties as the observed clusters (see Appendix~\ref{section_imf}) and estimate the corrected number of cluster members ($N\textrm{*}_{\mathrm{YSO}}$) from:
\be
N\textrm{*}_{\mathrm{YSO}} = \frac{N_{\mathrm{YSO}}/(1-IR)}{1-\kappa}.
\ee

Where $IR$ is the percentage of cluster members without IR-excess (see~\S~\ref{appendix_IRexcess}). One important assumption when creating the synthetic cluster is the assumed IMF. In our case, we used the Orion-Trapezium IMF from \citet{mue02}. However, in \S~\ref{section_IMF}, we show that the high-mass IMF slope (\ghm, see Table~\ref{gamma}) for some clusters differs from the Orion-Trapezium value (\ghm$=-1.21$). This difference becomes the main source of error when estimating percentage of detected sources $\kappa$ and the corrected number of cluster members $N\textrm{*}_{\mathrm{YSO}}$. We account for this by calculating $\kappa$ for \ghm$=-1.00$ and \ghm$=-1.60$ (which encloses the observed \ghm~values from Table~\ref{gamma}). We find that $\kappa$ varies \textbf{by on} average 10\%, which translates into a factor of less than 3 for $N\textrm{*}_{\mathrm{YSO}}$. The error bars in Figure~\ref{low-high_mass} are derived from those calculations.

%Moreover, if we use the corrected number of cluster members (see Appendix~\ref{N_corrected}) and assume a mean mass of 0.5~\msun~per source, the average ratio increases to $4.9\pm2$ ($4.2\pm2$ for LEC and $6.5\pm2$ for MEC).

\section{Adding cluster properties to synthetic cluster}\label{section_imf}
We use synthetic clusters to estimate the KLF and the percentage of non-detected cluster members. Those synthetic clusters are created using the Montecarlo code described in \citet{mue00}. This code allows to convolve synthetic clusters with ad-hoc extinction and/or IR-excess distribution functions in order to reproduce individual cluster properties. For the extinction distribution we use values from our extinction maps.

% and it consist of $10^5$ stars for which certain physical properties like mass, and near-IR magnitudes are predefined. We used a range of ages between 0.5 and 1.5~Myr and masses between 0.02 and 31~\msun~with a stars mass distribution given by the Trapezium-Orion IMF \citep{mue02}. Then, for each cluster, we add the mean K-band extinction to the synthetic K-band magnitude and calculate the percentage of synthetic stars ($\kappa$) that are fainter than the cluster CL$_\mathrm{K}$.

%The synthetic luminosity function is calculated from the average of 200 synthetic clusters. 

\subsection{IR-excess}\label{appendix_IRexcess}
%K-band extinction towards the observed sources was calculated by derredening cluster members (sources classified as Class I, Class II and with IR-excess) to the CTTS loci from \citet{mey97} on the H-K vs. J-H color diagram. We used the extinction law from \citet{ind05}. 
The IR-excess of the observed sources, given by their H-K color, was calculated assuming that affects only the K-band and that the intrinsic H-K color of cluster members is 0.3. We estimate the IR-excess by first derredening the YSOs to the CTTS loci from \citet*{mey97} transformed to the K-4.5 vs. H-K color diagram \citep[using the extinction law from][]{fla07}, and then moving them to their intrinsic H-K color. Then, the IR-excess distribution function is convolved with the synthetic cluster K-band magnitudes.

In addition, we use the ratio between the number of YSOs and the number of field subtracted K-band detections inside the convex hull area (see~\S~\ref{appendix_field}) to estimate the percentage of cluster members without IR-excess.

 %The total number of members was estimated by subtracting a control-field surface density of stars from the cluster surface density of stars in the K-band mosaic. Figure~\ref{modelKLF} shows the comparison of a KLF without corrections, corrected by IR-excess and corrected by extinction and IR-excess for a synthetic cluster with age between 0.5-1.5~Myr.

%\subsection{Including binaries}
%The range of binary systems in massive stars runs from 15 to 80\% \citep{gar01}. Since we are not able to resolve close binary systems, we created two sets of synthetic clusters: without binaries and with a 50\% binary fraction. Binary systems were created by randomly adding another star from the original 100,000 to 1,000 (50\%) of the stars in the synthetic cluster. 

\subsection{Field contamination in KLF}\label{appendix_field}
The KLF is derived from all the detections in the K-band in an area enclosing the cluster. Evidently, not all those detections are cluster members and hence the observed luminosity function is contaminated by the presence of field stars. We remove this contamination by subtracting a control field KLF from the cluster KLF. In order to do this properly, we added the extinction due to the molecular cloud to the field stars. This is done by convolving the control field with the extinction distribution of the cluster members. Once the control field is convolved, we construct the KLF, normalize it by the area and subtract it from the cluster KLF. %Figure~\ref{observedKLF} shows the observed and field subtracted KLF for cluster AFGL4029.

%\newpage
\section{Gaussian decomposition of \cob~spectrum in embedded clusters}\label{section_gaussian}
%%%%%%%%%%%%%%%%%%%%%%%%%%%%%%%%%%%%%
%%%%%%%%%%%%%%%%%%%%%%%%%%%%%%%%%%%%%
\begin{figure}
\begin{center}
\includegraphics[width=8cm]{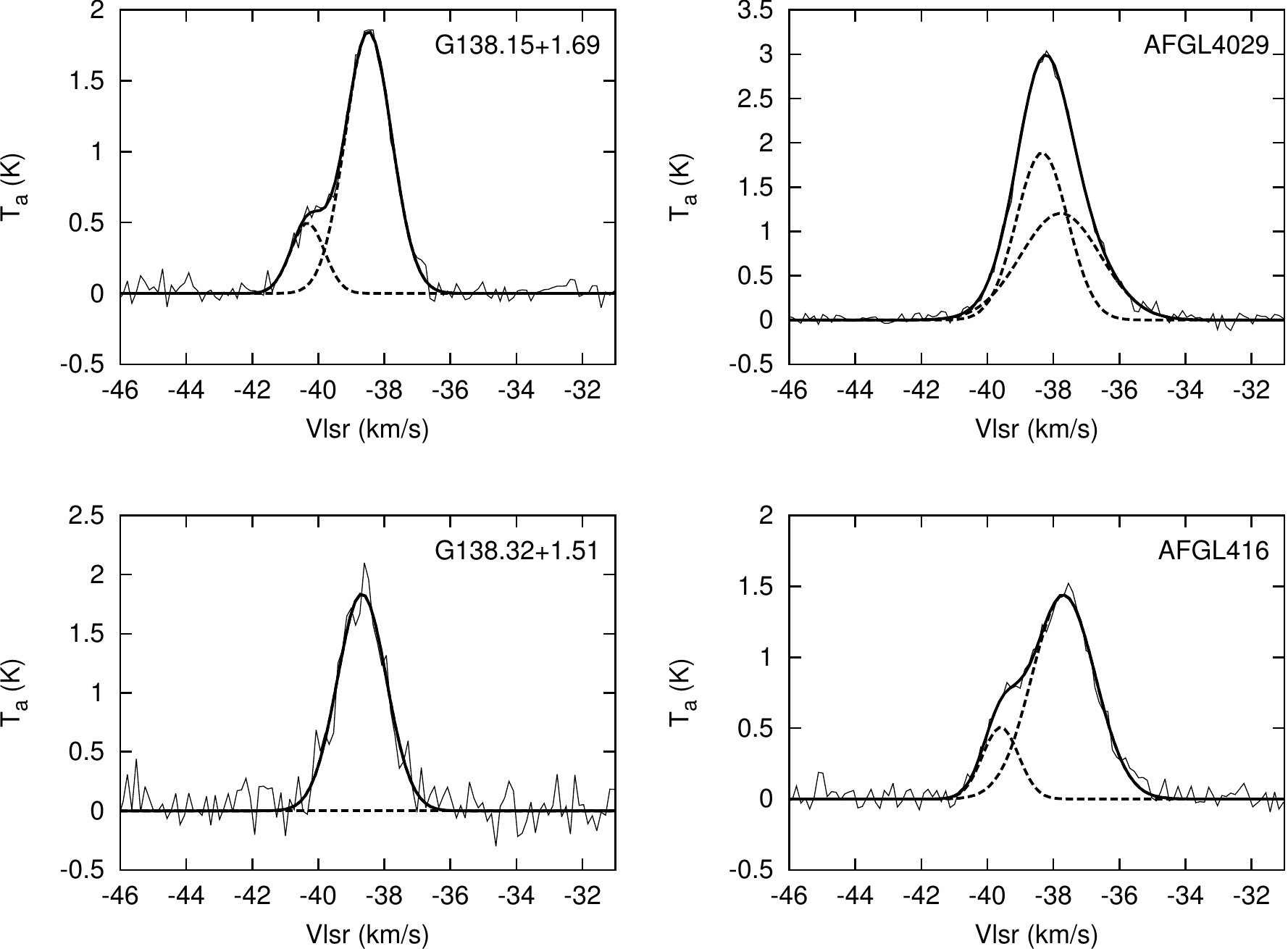}
\caption{Integrated spectrum over the corresponding convex hull area for embedded clusters in region W5-east. The spectrum is decomposed in several Gaussian components. The continuous line shows the components added. Different components are shown by segmented lines. Parameters are summarized in Table~\ref{table_vlsr}.}
\label{image_spectrum_w5}
\end{center}
\end{figure}

\begin{figure}
\begin{center}
\includegraphics[width=8cm]{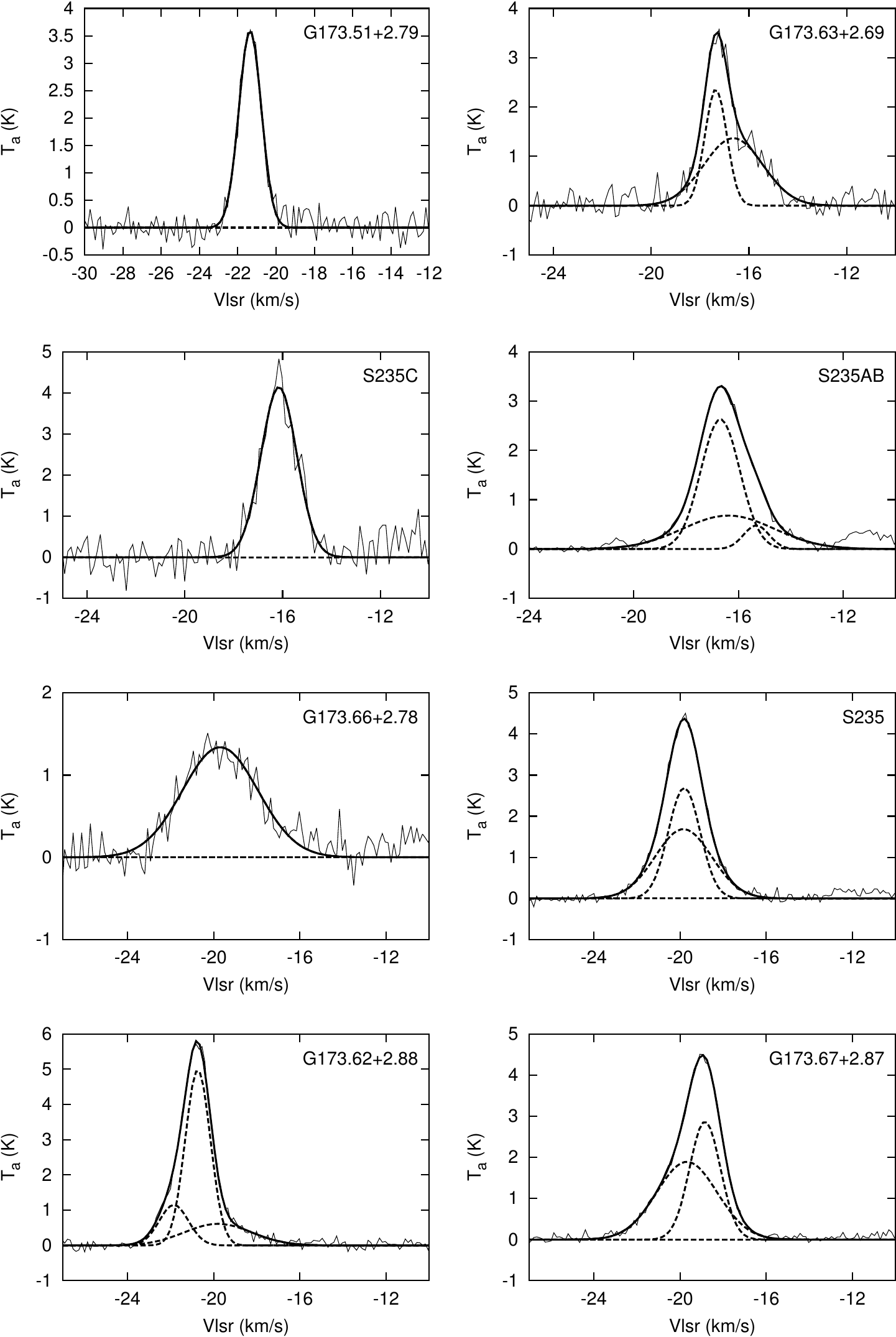}
\caption{Same as Figure~\ref{image_spectrum_w5} but for region S235.}
\label{image_spectrum_s235}
\end{center}
\end{figure}

\begin{figure}
\begin{center}
\includegraphics[width=8cm]{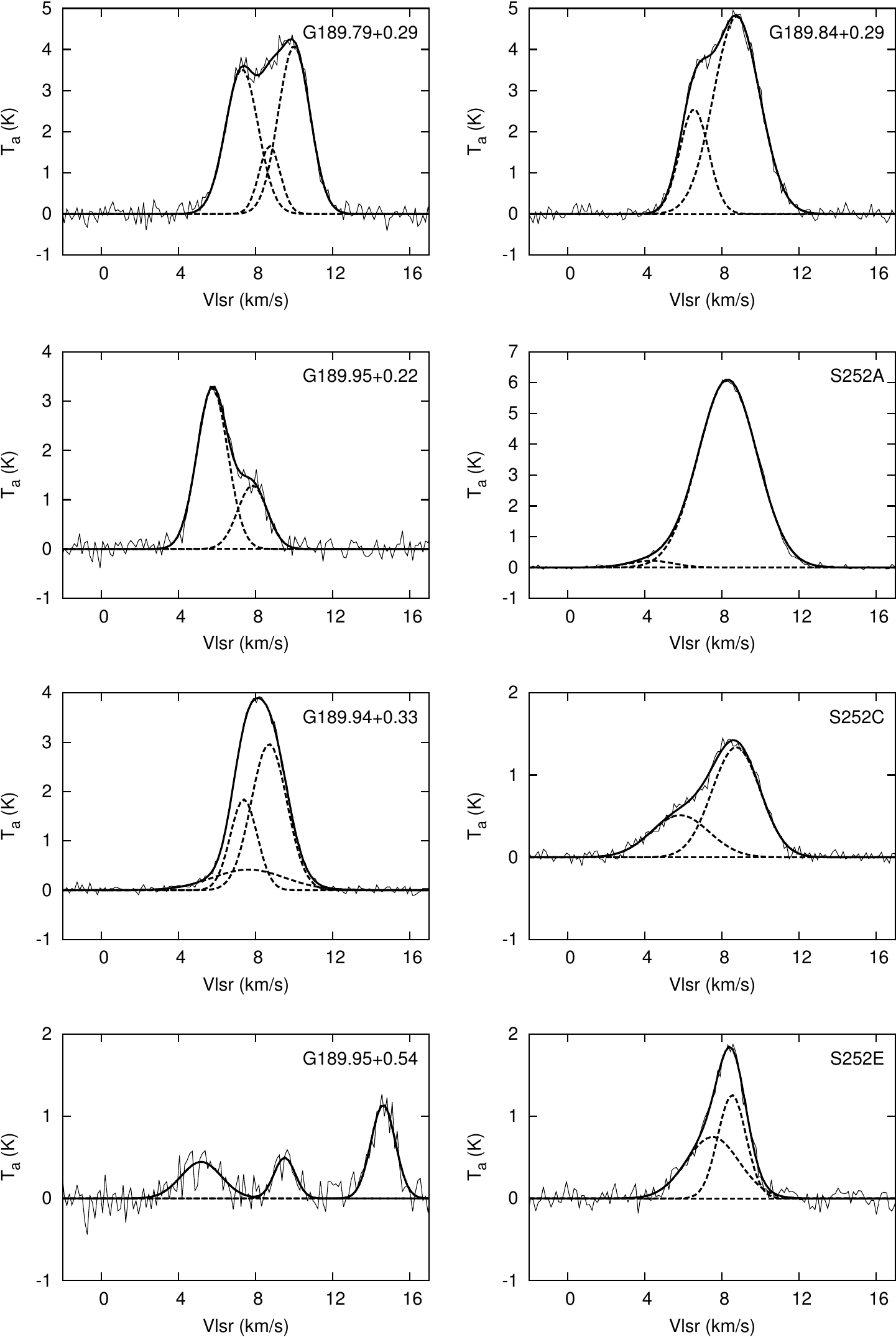}
\caption{Same as Figure~\ref{image_spectrum_w5} but for region S252.}
\label{image_spectrum_s252}
\end{center}
\end{figure}

\begin{figure}
\begin{center}
\includegraphics[width=8cm]{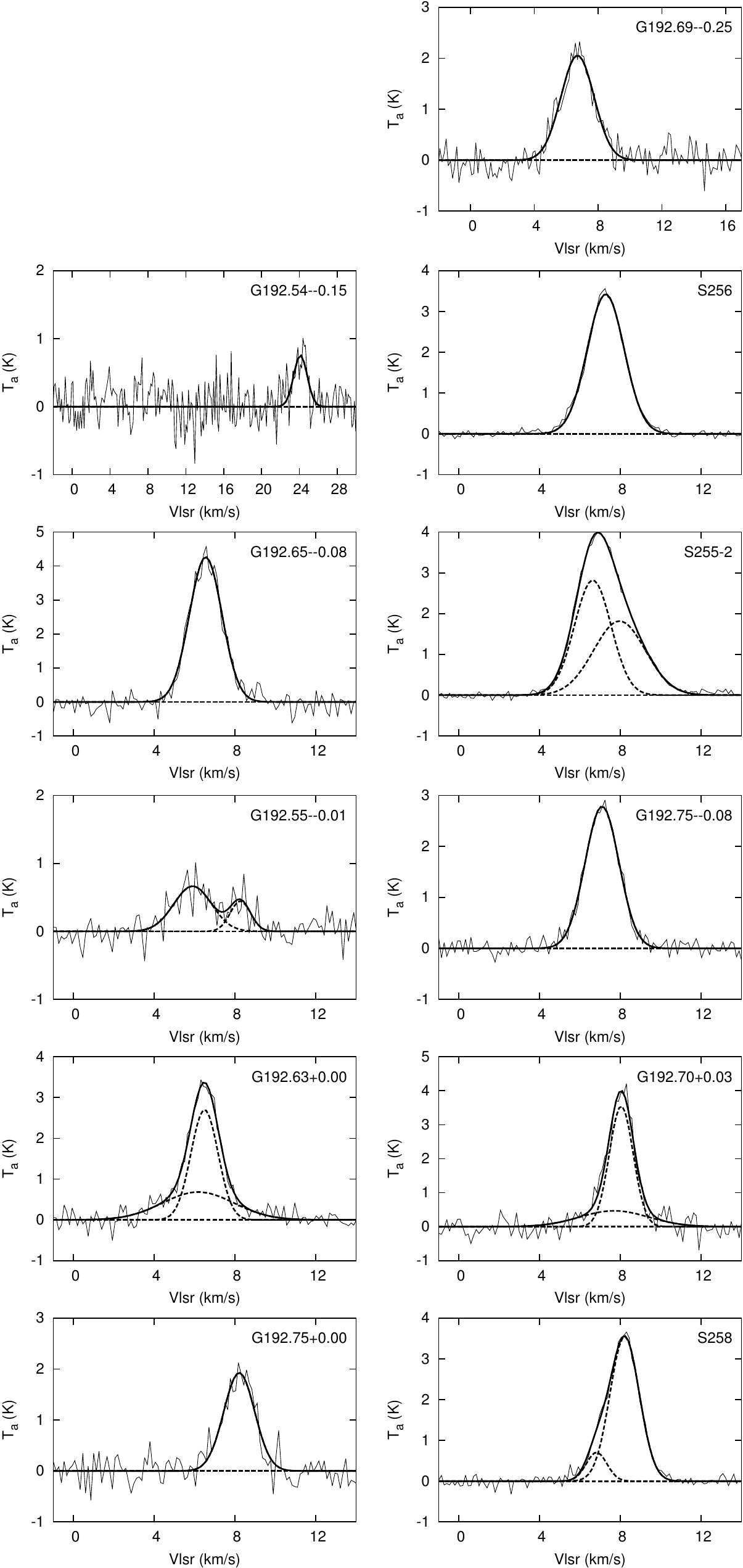}
\caption{Same as Figure~\ref{image_spectrum_w5} but for region S254-S258.}
\label{image_spectrum_s255}
\end{center}
\end{figure}

\begin{figure}
\begin{center}
\includegraphics[width=8cm]{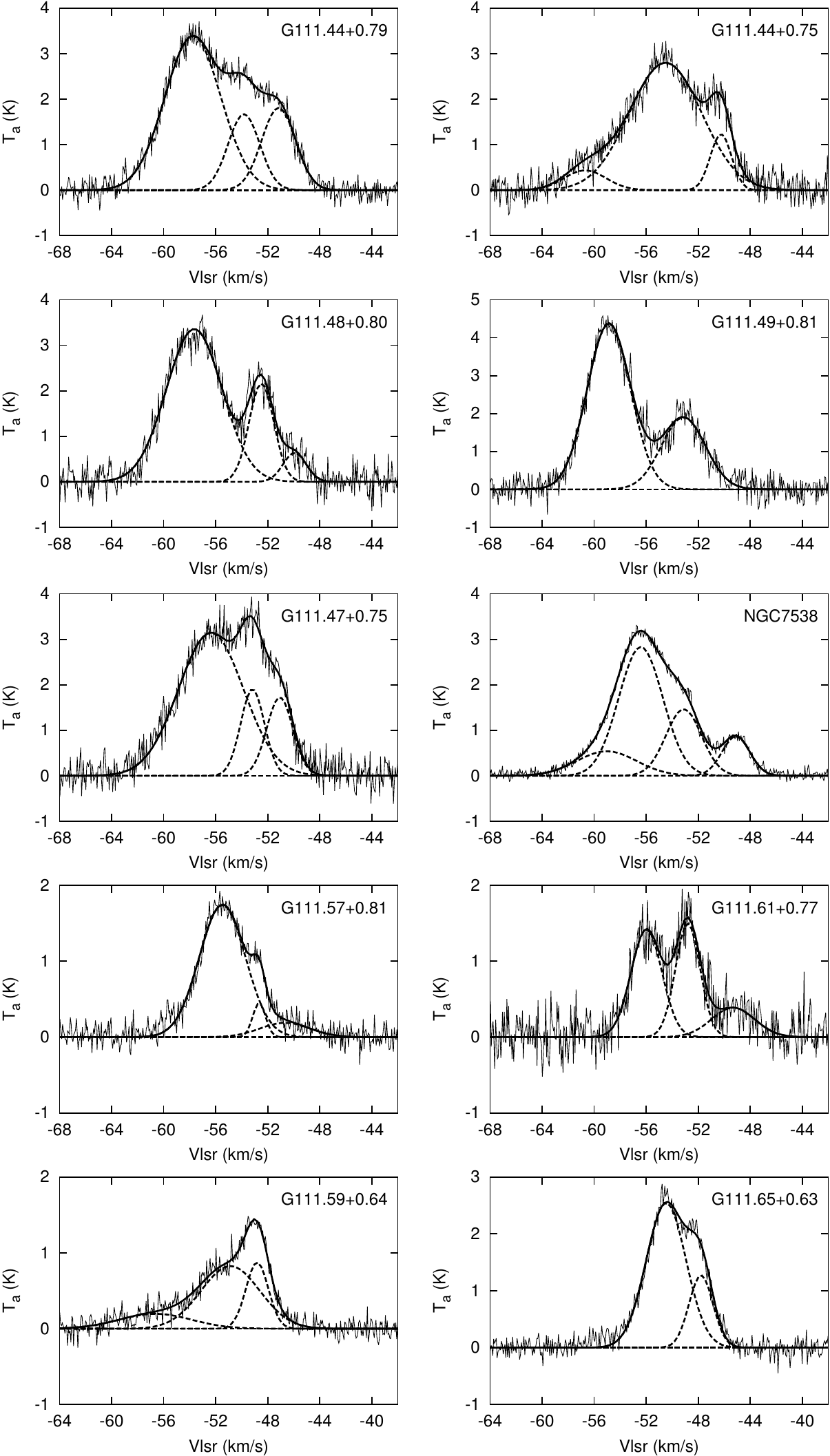}
\caption{Same as Figure~\ref{image_spectrum_w5} but for region NGC7538.}
\label{image_spectrum_ngc7538}
\end{center}
\end{figure}
%%%%%%%%%%%%%%%%%%%%%%%%%%%%%%%%%%%%%
%%%%%%%%%%%%%%%%%%%%%%%%%%%%%%%%%%%%%

\begin{landscape}
\begin{table}
\begin{minipage}{115mm}
\caption{Clusters \cob~spectrum decomposition}
\label{table_vlsr}
\begin{tabular}{lcccccccccccc} 
Name & T$_1$ & \vlsr$_1$ & FWHM$_1$ & T$_2$ & \vlsr$_2$ & FWHM$_2$ & T$_3$ & \vlsr$_3$ & FWHM$_3$ & T$_4$ & \vlsr$_4$ & FWHM$_4$ \\
& (K) & (\kms) & (\kms) & (K) & (\kms) & (\kms) & (K) & (\kms) & (\kms) & (K) & (\kms) & (\kms) \\
\hline                                                                  
G138.15+1.69 & 1.85 & -38.5 & 1.65 & 0.50 & -40.3 & 1.20 & - & - & - & - & - & - \\ 
AFGL4029 & 1.89 & -38.3 & 1.81 & 1.20 & -37.8 & 2.84 & - & - & - & - & - & - \\ 
G138.32+1.51 & 1.84 & -38.7 & 1.74 & - & - & - & - & - & - & - & - & - \\ 
AFGL416 & 1.43 & -37.7 & 2.26 & 0.51 & -39.6 & 1.29 & - & - & - & - & - & - \\ 
\\
G173.51+2.79 & 3.60 & -21.4 & 1.39 & - & - & - & - & - & - & - & - & - \\ 
G173.63+2.69 & 2.36 & -17.4 & 1.06 & 1.37 & -16.6 & 2.84 & - & - & - & - & - & - \\ 
S235C & 4.13 & -16.2 & 1.72 & - & - & - & - & - & - & - & - & - \\ 
S235AB & 2.63 & -16.7 & 1.76 & 0.68 & -16.4 & 4.11 & 0.48 & -15.3 & 1.13 & - & - & - \\ 
G173.66+2.78 & 1.34 & -19.7 & 4.14 & - & - & - & - & - & - & - & - & - \\ 
S235 & 2.69 & -19.8 & 1.74 & 1.69 & -19.8 & 3.08 & - & - & - & - & - & - \\ 
G173.62+2.88 & 4.89 & -20.7 & 1.36 & 1.15 & -21.8 & 1.57 & 0.61 & -19.8 & 3.67 & - & - & - \\ 
G173.67+2.87 & 2.87 & -18.8 & 1.69 & 1.89 & -19.7 & 3.29 & - & - & - & - & - & - \\ 
\\
G189.79+0.29 & 4.07 & 10.0 & 1.93 & 3.53 & 7.3 & 1.97 & 1.65 & 8.7 & 1.29 & - & - & - \\ 
G189.84+0.29 & 4.80 & 8.8 & 2.82 & 2.55 & 6.6 & 1.76 & - & - & - & - & - & - \\ 
G189.95+0.22 & 3.27 & 5.8 & 1.88 & 1.29 & 7.8 & 1.79 & - & - & - & - & - & - \\ 
S252A & 6.10 & 8.3 & 3.57 & 0.22 & 4.3 & 2.75 & - & - & - & - & - & - \\ 
G189.94+0.33 & 2.96 & 8.7 & 2.16 & 1.84 & 7.4 & 1.65 & 0.42 & 7.6 & 4.42 & - & - & - \\ 
S252C & 1.34 & 8.7 & 2.94 & 0.52 & 5.8 & 3.53 & - & - & - & - & - & - \\ 
G189.95+0.54 & 0.50 & 9.5 & 1.27 & 1.14 & 14.6 & 1.48 & 0.44 & 5.2 & 2.44 & - & - & - \\ 
S252E & 1.26 & 8.5 & 1.72 & 0.75 & 7.5 & 3.20 & - & - & - & - & - & - \\ 
\\
G192.69--0.25 & 2.05 & 6.7 & 2.44 & - & - & - & - & - & - & - & - & - \\ 
G192.54--0.15 & 0.76 & 24.1 & 1.67 & - & - & - & - & - & - & - & - & - \\ 
S256 & 3.42 & 7.3 & 2.16 & - & - & - & - & - & - & - & - & - \\ 
G192.65--0.08 & 4.21 & 6.5 & 1.90 & - & - & - & - & - & - & - & - & - \\ 
S255-2 \& S255N & 2.82 & 6.6 & 2.14 & 1.81 & 8.0 & 3.01 & - & - & - & - & - & - \\ 
G192.55--0.01 & 0.67 & 5.9 & 2.19 & 0.45 & 8.3 & 1.18 & - & - & - & - & - & - \\ 
G192.75--0.08 & 2.79 & 7.1 & 1.97 & - & - & - & - & - & - & - & - & - \\ 
G192.63+0.00 & 2.70 & 6.5 & 1.57 & 0.68 & 6.2 & 4.16 & - & - & - & - & - & - \\ 
G192.70+0.03 & 3.55 & 8.0 & 1.39 & 0.47 & 7.7 & 4.00 & - & - & - & - & - & - \\ 
G192.75+0.00 & 1.93 & 8.2 & 1.81 & - & - & - & - & - & - & - & - & - \\ 
S258 & 3.55 & 8.2 & 1.72 & 0.70 & 6.8 & 1.20 & - & - & - & - & - & - \\ 
\\
G111.44+0.79 & 3.38 & -57.8 & 5.05 & 1.81 & -51.2 & 3.06 & 1.67 & -53.8 & 2.87 & - & - & - \\ 
G111.44+0.75 & 2.80 & -54.5 & 6.53 & 1.23 & -50.3 & 1.90 & 0.44 & -60.7 & 3.78 & - & - & - \\ 
G111.48+0.80 & 3.35 & -57.6 & 5.03 & 2.16 & -52.5 & 2.26 & 0.63 & -49.9 & 2.09 & - & - & - \\ 
G111.49+0.81 & 4.37 & -58.9 & 3.92 & 1.90 & -53.1 & 3.97 & - & - & - & - & - & - \\ 
G111.47+0.75 & 3.14 & -56.4 & 5.88 & 1.90 & -53.2 & 2.14 & 1.73 & -51.1 & 2.35 & - & - & - \\ 
NGC7538 & 2.82 & -56.4 & 4.00 & 1.46 & -53.1 & 3.22 & 0.87 & -49.1 & 2.51 & 0.54 & -59.0 & 5.38 \\ 
G111.57+0.81 & 1.74 & -55.5 & 4.16 & 0.45 & -52.7 & 1.15 & 0.19 & -50.7 & 4.49 & - & - & - \\ 
G111.61+0.77 & 1.50 & -52.8 & 2.23 & 1.42 & -56.0 & 2.77 & 0.38 & -49.3 & 3.90 & - & - & - \\ 
G111.59+0.64 & 0.87 & -48.8 & 2.02 & 0.82 & -50.9 & 5.31 & 0.20 & -56.6 & 6.51 & - & - & - \\ 
G111.65+0.63 & 2.54 & -50.5 & 3.60 & 1.27 & -47.8 & 2.16 & - & - & - & - & - & - \\ 
\hline
\end{tabular}
\end{minipage}
\end{table}
\end{landscape}
%%%%%%%%%%%%%%%%%%%%%%%%%%%%%%%%%%%%%%%%%%%%%%%%
%%%%%%%%%%%%%%%%%%%%%%%%%%%%%%%%%%%%%%%%%%%%%%%%
%\begin{table*}
%\caption{}
%\label{table_vlsr}
%\vbox to220mm{\vfil Landscape Table E1 goes here
%\vfil}
%\end{table*}
%%%%%%%%%%%%%%%%%%%%%%%%%%%%%%%%%%%%%%%%%%%%%%%%
%%%%%%%%%%%%%%%%%%%%%%%%%%%%%%%%%%%%%%%%%%%%%%%%

\clearpage

%Since our observations are complete only for stars of mass $m > 1~\textrm{M}_{\odot}$, we compare the model and observed KLF down to a magnitude of 15.5 in the K-band. This magnitude corresponds to the photosphere of a 1~M$_{\odot}$ model star at a distance of 2.8~kpc (distance to the furthest cluster). Model and observed KLF were normalized by matching the number of stars brighter than the cutoff magnitude. Then, model and observed KLF were compared by calculating the $\chi^2$ value:
%\be
%\chi^2 = \sum_{i=1}^{n}(N_O - N_M)^2,
%\ee
%where $N_O$ and $N_M$ are the number of observed and model stars per bin in the KLF respectively. 

%We study the shape of the model KLF depending on the synthetic cluster properties. The age range of the clusters changes the position of the KLF peak, shifting it to fainter magnitudes at old ages. The value of \ghm~changes the slope of the KLF, a flat \ghm~corresponds to a flat KLF. The inclusion of binaries shifts the peak of the model KLF to brighter magnitudes, thus the best fitted model with binaries will have in general an older age than the best fitted model without binaries. The variation of the model KLF depending on the age, value of \ghm, and the inclusion of binaries is shown in Figure~\ref{klf2}.

%%%%%%%%%%%%%%%%%%%%%%%%%%%%%%


\begin{thebibliography}{99}
\bibitem[\protect\citeauthoryear{Allen et al.}{2004}]{all04} Allen L. et al., 2004, ApJS, 154, 363
\bibitem[\protect\citeauthoryear{Alves, Lada \& Lombardi}{Alves et al.}{2007}]{alv07} Alves J., Lada C., Lombardi M., 2007, A\&A, 462, 17
\bibitem[\protect\citeauthoryear{Andr\'e et al.}{2010}]{and10} Andr\'e Ph. et al., 2010, A\&A, 518, 102
\bibitem[\protect\citeauthoryear{Balog et al.}{2004}]{bal04} Balog Z., Kenyon S., Lada E., Barsony M., Vink\'o J., G\'aspa\'r A., 2004, AJ, 128, 2942 
\bibitem[\protect\citeauthoryear{Barriault \& Joncas}{2007}]{bar07} Barriault L., Joncas G., 2007, ApJ, 667, 257 
\bibitem[\protect\citeauthoryear{Battinelli}{1991}]{bat91} Battinelli P., 1991, A\&A, 244, 69 
\bibitem[\protect\citeauthoryear{Becker \& Fenkart}{1971}]{bec71} Becker W., Fenkart R., 1971, A\&AS, 4, 241 
\bibitem[\protect\citeauthoryear{Beichman}{1979}]{bei79} Beichman C. A., 1979, PhD thesis Hawaii Univ., Honolulu
\bibitem[\protect\citeauthoryear{Bica, Dutra \& Barbuy}{Bica et al.}{2003}]{bic03a} Bica E., Dutra C., Barbuy B., 2003, A\&A, 397, 177
\bibitem[\protect\citeauthoryear{Bica et al.}{2003}]{bic03b} Bica E., Dutra C., Soares J., Barbuy B., 2003, A\&A, 404, 223
\bibitem[\protect\citeauthoryear{Bonnell, Vine \& Bate,}{Bonnell et al.}{2004}]{bon04} Bonnell I. A., Vine S. R., Bate M. R., 2004, MNRAS, 349, 735
\bibitem[\protect\citeauthoryear{Bonnell, Clarke \& Bate}{Bonnell et al.}{2008}]{bon08} Bonnell I. A., Clarke C. J., Bate M. R., 2008, MNRAS, 389, 1556
\bibitem[\protect\citeauthoryear{Bressert et al.}{2012}]{bre12} Bressert E. et al., 2012, A\&A, 542, 49
\bibitem[\protect\citeauthoryear{Brunt}{2004}]{bru04} Brunt C., 2004, in Clemens D., Shah R., Brainerd T., eds, Proc. of ASP Conference 317. Milky Way Surveys: The Structure and Evolution of our Galaxy, p. 79
\bibitem[\protect\citeauthoryear{Camargo, Bonatto \& Bica}{Camargo et al.}{2011}]{cam11} Camargo D., Bonatto C., Bica E., 2011, MNRAS, 416, 1522 
\bibitem[\protect\citeauthoryear{Cardelli, Clayton \& Mathis}{Cardelli et al.}{1989}]{car89} Cardelli J., Clayton C., Mathis J., 1989, ApJ, 345, 245 
\bibitem[\protect\citeauthoryear{Carpenter et al.}{1993}]{car93} Carpenter J., Snell R., Schloerb F., 1995, ApJ, 450, 201 
\bibitem[\protect\citeauthoryear{Carpenter et al.}{1995}]{car95} Carpenter J., Snell R., Schloerb F., Skrutskie M., 1993, ApJ, 407, 657 
\bibitem[\protect\citeauthoryear{Carpenter, Heyer \& Snell}{Carpenter et al.}{2000}]{car00} Carpenter J., Heyer M., Snell R., 2000, ApJS, 130, 381 
\bibitem[\protect\citeauthoryear{Cartwright \& Whitworth}{2004}]{car04} Cartwright A., Whitworth A., 2004, MNRAS, 348, 589 
\bibitem[\protect\citeauthoryear{Cesaroni et al.}{2005}]{ces05} Cesaroni R., Neri R., Olmi L., Testi L., Walmsley C. M., Hofner P., 2005, A\&A, 434, 1039 
\bibitem[\protect\citeauthoryear{Chauhan et al.}{2011}]{cha11} Chauhan N., Pandey A., Ogura K., Jose J., Ojha D., Samal M., Mito H., 2011, MNRAS, 415, 1202
\bibitem[\protect\citeauthoryear{Chavarr\'{\i}a et al.}{2008}]{cha08a} Chavarr\'{\i}a L., Allen L., Hora J., Brunt C., Fazio G., 2008, ApJ, 682, 445 
\bibitem[\protect\citeauthoryear{Chavarr\'{\i}a et al.}{2010}]{cha10} Chavarr\'{\i}a L., Mardones D., Garay G., Escala A., Bronfman L., Lizano S., 2010, ApJ, 710, 583 
\bibitem[\protect\citeauthoryear{Chavarr\'{\i}a-K et al.}{1989}]{cha89} Chavarr\'{\i}a-K C., Leitherer C., de Lara E., S\'anchez O., Zickgraf F., 1989, A\&A, 215, 51 
\bibitem[\protect\citeauthoryear{Churchwell, Walmsley \& Cesaroni}{Churchwell et al.}{1990}]{chu90} Churchwell E., Walmsley C., Cesaroni R., 1990, A\&AS, 83, 119 
\bibitem[\protect\citeauthoryear{Condon et al.}{1998}]{con98} Condon J., Cotton W., Greisen E., Yin Q., Perley R., Taylor G., Broderick J., 1998, AJ, 115, 1693
\bibitem[\protect\citeauthoryear{Crampton, Georgelin \& Georgelin}{Crampton et al.}{1978}]{cra78} Crampton D., Georgelin Y. M., Georgelin Y. P., 1978, A\&A 66, 1
\bibitem[\protect\citeauthoryear{Cyganowski et al.}{2008}]{cyg08} Cyganowski C. et al., 2008, AJ, 136, 2391 
\bibitem[\protect\citeauthoryear{Dale, Ercolano \& Bonnell}{Dale et al.}{2013}]{dal13} Dale J., Ercolano B., Bonnell I., 2013, MNRAS, 431, 1062
\bibitem[\protect\citeauthoryear{Deharveng et al.}{1997}]{deh97} Deharveng L., Zavagno A., Cruz-Gonz\'alez I., Salas L., Carrasco L., 1997, A\&A, 317, 459
\bibitem[\protect\citeauthoryear{Deharveng et al.}{2008}]{deh08} Deharveng L., Lefloch B., Kurtz S., Nadeau D., Pomar\`es M., Caplan J., Zavagno A., 2008, A\&A, 482, 585
\bibitem[\protect\citeauthoryear{Deharveng et al.}{2012}]{deh12} Deharveng L. et al., 2012, A\&A, 546, 74
\bibitem[\protect\citeauthoryear{Dewangan \& Anandarao}{2011}]{dew11} Dewangan L., Anandarao B., 2011, MNRAS, 414, 1526 
\bibitem[\protect\citeauthoryear{Dickman}{1978}]{dic78} Dickman R., 1978, ApJS, 37, 407
\bibitem[\protect\citeauthoryear{Eisenhardt et al.}{2004}]{eis04} Eisenhardt P. et al., 2004, ApJS, 154, 48
\bibitem[\protect\citeauthoryear{Erickson et al.}{1999}]{eri99} Erickson N., Grosslein R., Erickson R., Weinreb S., 1999, IEEE Trans. Microwave Theory Tech., 47, 2212
\bibitem[\protect\citeauthoryear{Evans \& Blair}{1981}]{eva81} Evans N., Blair G., 1981, ApJ, 246, 394 
\bibitem[\protect\citeauthoryear{Fazio et al.}{2004}]{faz04} Fazio G. G. et al., 2004, ApJS, 154, 10
\bibitem[\protect\citeauthoryear{Felli, Habing \& Israel}{Felli et al.}{1977}]{fel77} Felli M., Habing H., Israel F., 1977, A\&A, 59, 43 
\bibitem[\protect\citeauthoryear{Felli, Hjellming \& Cesaroni}{Felli et al.}{1987}]{fel87} Felli M., Hjellming R., Cesaroni R., 1987, A\&A, 182, 313 
\bibitem[\protect\citeauthoryear{Felli et al.}{2006}]{fel06} Felli M., Massi F., Robberto M., Cesaroni R., 2006, A\&A, 453, 911 
\bibitem[\protect\citeauthoryear{Flaherty et al.}{2007}]{fla07} Flaherty K., Pipher J., Megeath T., Winston E., Gutermuth R., Muzerolle J., Allen L., Fazio G., 2007, ApJ, 663, 1069
\bibitem[\protect\citeauthoryear{Frerking, Langer \& Wilson}{Frerking et al.}{1982}]{fre82} Frerking M., Langer W., Wilson R. W., 1982, ApJ, 262, 590
\bibitem[\protect\citeauthoryear{Froebich, Scholz \& Raftery}{Froebich et al.}{2007}]{fro07} Froebich D., Scholz A., Raftery C., 2007, MNRAS, 374, 399
\bibitem[\protect\citeauthoryear{Garay et al.}{2007}]{gar07} Garay G., Mardones D., Brooks K., Videla L., Contreras Y., 2007, ApJ, 666, 309
\bibitem[\protect\citeauthoryear{Georgelin, Georgelin \& Roux}{Georgelin et al.}{1973}]{geo73} Georgelin Y. M., Georgelin Y. P., Roux S., 1973, A\&A, 25, 337 
\bibitem[\protect\citeauthoryear{Ginsburg, Bally \& Williams}{Ginsburg et al.}{2011}]{gin11} Ginsburg A., Bally J., Williams J. P., 2011, MNRAS, 418, 2121
\bibitem[\protect\citeauthoryear{Grasladen \& Carrasco}{1975}]{gra75} Grasladen G., Carrasco L., 1975, A\&A, 43, 259 
\bibitem[\protect\citeauthoryear{Gutermuth et al.}{2008}]{gut08} Gutermuth R. et al., 2008, ApJ, 674, 336
\bibitem[\protect\citeauthoryear{Gutermuth et al.}{2009}]{gut09} Gutermuth R., Megeath T., Myers P., Allen L., Pipher J., Fazio G., 2009, ApJS, 184, 18
\bibitem[\protect\citeauthoryear{Herpin et al.}{2012}]{her12} Herpin F. et al., 2012, A\&A, 542, 76
\bibitem[\protect\citeauthoryear{Heyer et al.}{1989}]{hey89} Heyer M., Snell R., Morgan J., Schloerb F., 1989, ApJ, 346, 220
\bibitem[\protect\citeauthoryear{Heyer et al.}{1998}]{hey98} Heyer M., Brunt C., Snell R., Howe J., Schloerb F., Carpenter J., 1998, ApJS, 115, 241
\bibitem[\protect\citeauthoryear{Hillwig et al.}{2006}]{hil06} Hillwig T., Gies D., Bagnuolo W., Huang W., McSwain M., Wingert D., 2006, ApJ, 639, 1069 
\bibitem[\protect\citeauthoryear{Hiltner}{1956}]{hil56} Hiltner W. A., 1956, ApJ, 120, 454 
\bibitem[\protect\citeauthoryear{Hora et al.}{2004}]{hor04} Hora J. et al., 2004, in Mather J., ed., Proc. of the SPIE: Optical, Infrared, and Millimeter Space Telescopes, Vol. 5487, p. 77
\bibitem[\protect\citeauthoryear{Hosokawa \& Inutsuka}{2006}]{hok06}Hosokawa T., Inutsuka S., 2006, ApJ, 648, 131
\bibitem[\protect\citeauthoryear{Hunter \& Massey}{1990}]{hun90} Hunter D., Massey P., 1990, AJ, 99, 846
\bibitem[\protect\citeauthoryear{Indebetouw et al.}{2005}]{ind05} Indebetouw R. et al., 2005, ApJ, 619, 931
\bibitem[\protect\citeauthoryear{Israel}{1977}]{isr77} Israel F., 1977, A\&A, 59, 27 
\bibitem[\protect\citeauthoryear{Israel \& Felli}{1978}]{isr78} Israel F., Felli M., 1978, A\&A, 63, 325 
\bibitem[\protect\citeauthoryear{Israel et al.}{2003}]{isr03} Israel F. et al., 2003, A\&A, 401, 99
\bibitem[\protect\citeauthoryear{Jackson, Ivezi\'c \& Knapp}{Jackson et al.}{2002}]{jac02} Jackson T., Ivezi\'c Z., Knapp G., 2002, MNRAS, 337, 749
\bibitem[\protect\citeauthoryear{Jose et al.}{2012}]{jos12} Jose J. et al., 2012, MNRAS, 424, 2486
\bibitem[\protect\citeauthoryear{Kameya et al.}{1990}]{kam90} Kameya O., Morita K-I., Kawabe R., Ishiguro M., 1990, ApJ, 355, 562 
\bibitem[\protect\citeauthoryear{Kirsanova et al.}{2008}]{kir08} Kirsanova M., Sobolev A., Thomasson M., Wiebe D., Johansson L., Seleznev A., 2008, MNRAS, 388, 729 
\bibitem[\protect\citeauthoryear{Krassner, Pipher \& Sharpless}{Krassner et al.}{1979}]{kra79} Krassner J., Pipher J., Sharpless S., 1979, A\&A, 77, 302 
\bibitem[\protect\citeauthoryear{Koenig et al.}{2008}]{koe08} Koenig X., Allen L., Gutermuth R., Hora J., Brunt C., Muzerolle J., 2008, ApJ, 688, 1142 
\bibitem[\protect\citeauthoryear{Koposov, Glushkova \& Zolotukhin}{Koposov et al.}{2008}]{kop08} Koposov S., Glushkova E., Zolotukhin I., 2008, A\&A, 486, 771 
\bibitem[\protect\citeauthoryear{Kraus et al.}{2006}]{kra06} Kraus S. et al., 2006, A\&A, 455, 521 
\bibitem[\protect\citeauthoryear{Kumar et al.}{2007}]{kum07} Kumar M., Davis C., Grave J., Ferreira B., Froebrich D., 2007, MNRAS, 374, 54
\bibitem[\protect\citeauthoryear{Kurtz, Churchwell \& Wood}{Kurtz et al.}{1994}]{kur94} Kurtz S., Churchwell E., Wood D., 1994, ApJS, 91, 659 
\bibitem[\protect\citeauthoryear{Kutner \& Ulich}{1981}]{kut81} Kutner M. L., Ulich B. L., 1981, ApJ, 250, 341
\bibitem[\protect\citeauthoryear{Lada \& Wooden}{1979}]{lad79} Lada C., Wooden D., 1979, ApJ, 232, 158 
\bibitem[\protect\citeauthoryear{Lada}{1987}]{lad87} Lada C., 1987, IAU Symp 115 Star-forming Regions, ed. C. J. Lada (Dordrecht: Reidel ), 1 
\bibitem[\protect\citeauthoryear{Lada \& Lada}{2003}]{lad03} Lada C., Lada E., 2003, ARA\&A, 41, 57
\bibitem[\protect\citeauthoryear{Lada, Lombardi \& Alves}{Lada et al.}{2010}]{lad10} Lada C., Lombardi M., Alves J., 2010, ApJ, 724, 687
\bibitem[\protect\citeauthoryear{MacKenzie et al.}{2011}]{mac11} MacKenzie, T. et al., 2011, MNRAS, 415, 1950
\bibitem[\protect\citeauthoryear{Mampaso et al.}{1987}]{mam87} Mampaso A., Pismis P., Vilchez J., Phillips J., 1987, RMxAA, 14, 474 
\bibitem[\protect\citeauthoryear{Maschberger et al.}{2010}]{mas10} Maschberger Th., Clarke C., Bonnell I., Kroupa P., 2010, MNRAS, 404, 1061
\bibitem[\protect\citeauthoryear{McKee \& Tan}{2003}]{mck03} McKee Ch., Tan J., 2003, ApJ, 585, 850 
\bibitem[\protect\citeauthoryear{Meyer, Calvet \& Hillendrand}{Meyer et al.}{1997}]{mey97} Meyer M., Calvet N., Hillendrand L., 1997, AJ, 114, 288 
\bibitem[\protect\citeauthoryear{Minkowski}{1946}]{min46} Minkowski R., 1946, PASP, 58, 305 
\bibitem[\protect\citeauthoryear{Mirabel et al.}{1987}]{mir87} Mirabel I., Ruiz A., Rodr\'{\i}guez L., Cant\'o J., 1987, ApJ, 318, 729 
\bibitem[\protect\citeauthoryear{Morgan et al.}{2008}]{mor08} Morgan L., Thompson M., Urquhart J., White G., 2008, A\&A, 477, 577 
\bibitem[\protect\citeauthoryear{Moscadelli et al.}{2009}]{mos09} Moscadelli L., Reid M., Menten K., Brunthaler A., Zheng X., Xu Y., 2009, ApJ, 693, 406 
\bibitem[\protect\citeauthoryear{Muench, Lada \& Lada}{Muench et al.}{2000}]{mue00} Muench A., Lada E., Lada C., 2000, ApJ, 533, 358 
\bibitem[\protect\citeauthoryear{Muench et al.}{2002}]{mue02} Muench A., Lada E., Lada C., Alves J., 2002, ApJ, 573, 366 
\bibitem[\protect\citeauthoryear{Muench et al.}{2007}]{mue07} Muench A., Lada C., Luhman K., Muzerolle J., Young E., 2007, ApJ, 134, 411 
\bibitem[\protect\citeauthoryear{Nakano et al.}{2008}]{nak08} Nakano M., Sugitani K., Niwa T., Itoh Y., Watanabe M., 2008, PASJ, 60, 739 
\bibitem[\protect\citeauthoryear{Niwa et al.}{2009}]{niw09} Niwa T., Tachihara K., Itoh Y., Oasa Y., Sunada K., Sugitani K., Mukai T., 2009, A\&A, 500, 1119 
\bibitem[\protect\citeauthoryear{Noriega-Crespo et al.}{2004}]{nor04} Noriega-Crespo A. et al., 2004., ApJS, 154, 352
\bibitem[\protect\citeauthoryear{Oey et al.}{2013}]{oey13} Oey M., Lamb J., Kushner C., Pellegrini E., Graus A., 2013, ApJ, 768, 66 
\bibitem[\protect\citeauthoryear{Ogura, Sugitani \& Pickles}{Ogura et al.}{2002}]{ogu02} Ogura K., Sugitani K., Pickles A., 2002, AJ, 123, 2597 
\bibitem[\protect\citeauthoryear{Ojha et al.}{2004}]{ojh04} Ojha D., Ghosh S., Kulkarni V., Testi L., Verma R., Vig S., 2004, A\&A, 415, 1039 
\bibitem[\protect\citeauthoryear{Ojha et al.}{2011}]{ojh11} Ojha D. et al., 2011, ApJ, 738, 156 
\bibitem[\protect\citeauthoryear{Olofsson}{1983}]{olo83} Olofsson G., 1983, A\&A, 120, 1 
\bibitem[\protect\citeauthoryear{Patten et al.}{2006}]{pat06} Patten B. et al., 2006, ApJ, 651, 502
\bibitem[\protect\citeauthoryear{Pestalozzi, Elitzur \& Conway}{2009}]{pes09} Pestalozzi M., Elitzur M., Conway J., 2009, A\&A, 501, 999
%\bibitem[\protect\citeauthoryear{Pismis}{1970}]{pis70} Pismis P., 1970, Bolet\'{\i}n de los Observatorios de Tonantzintla y Tacubaya Vol. 5, pp. 219-227
\bibitem[\protect\citeauthoryear{Price \& Walker}{1976}]{pri76} Price S. D., Walker R. G., 1976, in Interim Report Air Force Geophysics Lab., Hanscom AFB, MA. Optical Physics Div. The AFGL four color infrared sky survey: Catalog of observations at 4.2, 11.0, 19.8 and 27.4 micrometers
\bibitem[\protect\citeauthoryear{Puga et al.}{2010}]{pug10} Puga E. et al., 2010, A\&A, 517, 2
\bibitem[\protect\citeauthoryear{Ray et al.}{1990}]{ray90} Ray T., Poetzel R., Solf J., Mundt R., 1990, ApJ, 357, 45
\bibitem[\protect\citeauthoryear{Reid \& Wilson}{2005}]{rei05} Reid M., Wilson C., 2005, ApJ, 625, 891
\bibitem[\protect\citeauthoryear{Reid et al.}{2009}]{rei09} Reid M., Menten K., Brunthaler A., Zheng X., Moscadelli L., Xu Y., 2009, ApJ, 693, 397
\bibitem[\protect\citeauthoryear{Robitaille et al.}{2008}]{rob08} Robitaille T. et al., 2008, AJ, 136, 2413
\bibitem[\protect\citeauthoryear{Rygl et al.}{2010}]{ryg10} Rygl K., Bunthaler A., Reid M., Menten K., van Langevelde H., Xu Y., 2010, A\&A, 511, 2
\bibitem[\protect\citeauthoryear{Sandell et al.}{2009}]{san09} Sandell G., Goss W., Wright M., Corder S., 2009, ApJ, 699, 31
\bibitem[\protect\citeauthoryear{Sandell \& Wright}{2010}]{san10} Sandell G., Wright M., 2010, ApJ, 715, 919
\bibitem[\protect\citeauthoryear{Schmeja \& Klessen}{2006}]{schm06} Schmeja S., Klessen R., 2006, A\&A, 449, 151
\bibitem[\protect\citeauthoryear{Schmeja, Kumar \& Ferreira}{Schmeja et al.}{2008}]{sch08} Schmeja S., Kumar M., Ferreira B., 2008, MNRAS, 389, 1209
\bibitem[\protect\citeauthoryear{Schmeja}{2011}]{sch11} Schmeja S., 2011, AN, 332, 172
\bibitem[\protect\citeauthoryear{Schmidt et al.}{2009}]{sch09} Schmidt W., Federrath C., Hupp M., Kern S., Niemeyer J., 2009, A\&A, 494, 127
\bibitem[\protect\citeauthoryear{Schuster, Marengo \& Patten}{2006}]{sch06} Schuster M., Marengo M., Patten B., 2006, in Silva D., Doxsey R., eds, Proc. of the SPIE: Observatory Operations: Strategies, Processes, and Systems, Vol. 6270, p. 65
\bibitem[\protect\citeauthoryear{Sharpless}{1959}]{sha59} Sharpless S., 1959, ApJS, 4, 257
\bibitem[\protect\citeauthoryear{Snell et al.}{1988}]{sne88} Snell R., Huang Y., Dickman R., Claussen M., 1988, ApJ, 325, 853
\bibitem[\protect\citeauthoryear{Tej et al.}{2006}]{tej06} Tej A., Ojha D., Ghosh S., Kulkarni V., Verma R., Vig S., Prabhu T., 2006, A\&A, 452, 203
\bibitem[\protect\citeauthoryear{Thompson, Thronson \& Campbell}{Thompson et al.}{1983}]{tho83} Thompson R., Thronson H., Campbell B., 1983, ApJ, 266, 614
\bibitem[\protect\citeauthoryear{Troland \& Crutcher}{2008}]{tro08} Troland T., Crutcher R., 2008, ApJ, 680, 457
\bibitem[\protect\citeauthoryear{Vlemmings}{2008}]{vle08} Vlemmings W. H., 2008, A\&A, 484, 773
\bibitem[\protect\citeauthoryear{Wang et al.}{2011}]{wan11} Wang Y. et al., 2011, A\&A, 527, 32
\bibitem[\protect\citeauthoryear{Werner et al.}{1979}]{wer79} Werner M., Becklin E., Gatley I., Matthews K., Neugebauer G., Wynn-Williams C., 1979, MNRAS, 188, 463
\bibitem[\protect\citeauthoryear{Winston et al.}{2007}]{win07} Winston E. et al., 2007, ApJ, 669, 493 
\bibitem[\protect\citeauthoryear{Wouterloot \& Brand}{1989}]{wou89} Wouterloot J., Brand J., 1989, A\&A, 80, 149
\bibitem[\protect\citeauthoryear{Wynn-Williams, Beckling \& Neugebauer}{Wynn-Williams et al.}{1974}]{wyn74} Wynn-Williams C., Beckling E., Neugebauer G., 1974, ApJ, 187, 473
\bibitem[\protect\citeauthoryear{Xu et al.}{2006}]{xu06} Xu Y. et al., AJ, 2006, 132, 20
\bibitem[\protect\citeauthoryear{Zapata, Rodr\'{\i}guez \& Kurtz}{Zapata et al.}{2001}]{zap01} Zapata L., Rodr\'{\i}guez L., Kurtz S., RMxAA, 2001, 37, 83
\bibitem[\protect\citeauthoryear{Zinnecker \& Yorke}{2007}]{zin07} Zinnecker H., Yorke H., 2007, ARA\&A, 45, 481
\end{thebibliography}
\end{document}